\documentclass[review]{elsarticle}
\usepackage{graphicx}
\usepackage{amsmath}
\usepackage{enumerate}

\usepackage{natbib}
\setcitestyle{authoryear,open={(},close={)}}

\journal{----}

\graphicspath{Figures}             
\DeclareGraphicsExtensions{.pdf,.jpeg,.png,.eps}

\usepackage{subfigure}
\usepackage{multirow}
\usepackage{setspace}
\usepackage{epstopdf}
\usepackage{bbm}
\usepackage{geometry}
\usepackage{graphicx}
\usepackage{epstopdf}
\usepackage{graphicx}
\usepackage{enumerate}       
\usepackage[latin1]{inputenc}
\usepackage{graphics,amsthm,amsfonts}
\usepackage{indentfirst}       
\usepackage{amssymb,amsmath,graphicx,color}
\usepackage{comment}
\usepackage{appendix}
\usepackage{array}
\usepackage{multirow}
\usepackage{appendix}
\usepackage{chngcntr} 
\usepackage{float}
\usepackage{setspace}
\usepackage{etoolbox}

\theoremstyle{plain}

\theoremstyle{definition}

\theoremstyle{remark}

\newcommand{\bmu}{\mbox{\boldmath $\mu$}}

\newcommand{\balpha}{\mbox{\boldmath $\alpha$}}
\newcommand{\bbeta}{\mbox{\boldmath $\beta$}}

\newcommand{\bE}{\mbox{\boldmath $E$}}

\newcommand{\bY}{\mbox{\boldmath $Y$}}

\newcommand{\bM}{\mbox{\boldmath $M$}}

\newcommand{\bB}{\mbox{\boldmath $B$}}

\begin{document}
\doublespacing
	
	\begin{frontmatter}

	\title{Bayesian Dynamic Estimation of Mortality Schedules in Small Areas}
	\author[1,2]{Guilherme Lopes de Oliveira \corref{mycorrespondingauthor}}
	\address[1]{Departamento de Computa\c c\~ao, Centro Federal de Educa\c c\~ao Tecnol\'ogica de Minas Gerais, Brazil}
	\cortext[mycorrespondingauthor]{Corresponding author. Departamento de Computa\c c\~ao, CEFET-MG, Av. Amazonas, 7675, Nova Gameleira, CEP 30510-000, Belo Horizonte, Minas Gerais, Brazil. E-mail: guilhermeoliveira@cefetmg.br}
		
	\author[2]{Rosangela Helena Loschi}
	\address[2]{Departamento de Estatística, Universidade Federal de Minas Gerais, Brazil}
	\author[3]{Renato Martins Assun\c c\~ao}
	\address[3]{Departamento de Ci\^encia da Computa\c c\~ao, Universidade Federal de Minas Gerais, Brazil}
\begin{abstract}

The determination of the shapes of mortality curves, the estimation and projection of mortality patterns over time, and the investigation of differences in mortality patterns across different small underdeveloped populations have received special attention in recent years. 
The challenges involved in this type of problems are the common sparsity and the unstable behavior of observed death counts in small areas (populations). 
These features impose many difficulties in the estimation of reasonable mortality schedules.
In this chapter, we present a discussion about this problem and we introduce the use of relational Bayesian dynamic  models for estimating and smoothing mortality schedules by age and sex.
Preliminary results are presented, including a comparison with a methodology recently proposed in the literature.
The analyzes are based on simulated data as well as mortality data observed in some Brazilian municipalities.	
\end{abstract}

\end{frontmatter}

\section{Introduction}
\label{intro}

The mortality rate, life expectancy and other indicators of longevity are of fundamental importance to measure the health  and well-being conditions of human populations.
Methods for describing mortality patterns are common in demography, but the   understanding of mortality evolution plays an important role in many other fields, such as actuarial science, epidemiology and genetics.

Demographic studies often use data related to the entire population. 
Because of this mortality studies are commonly performed at an aggregate level. More specifically, mortality data are usually available as the number of deaths and the population exposure to risk in a particular population (e.g., country, state, municipality, counties) and age group (e.g., sequential 5-year age intervals).
In such context, the focus is to estimate the mortality rate, i.e. deaths per population at risk and age in each population. Typically, the age-specific mortality rates are calculated separately by sex or other characteristics, providing the so-called \textit{mortality schedules}.

\begin{figure}[htb!]
	\centering
	\subfigure
	{
		\includegraphics[scale=0.325]{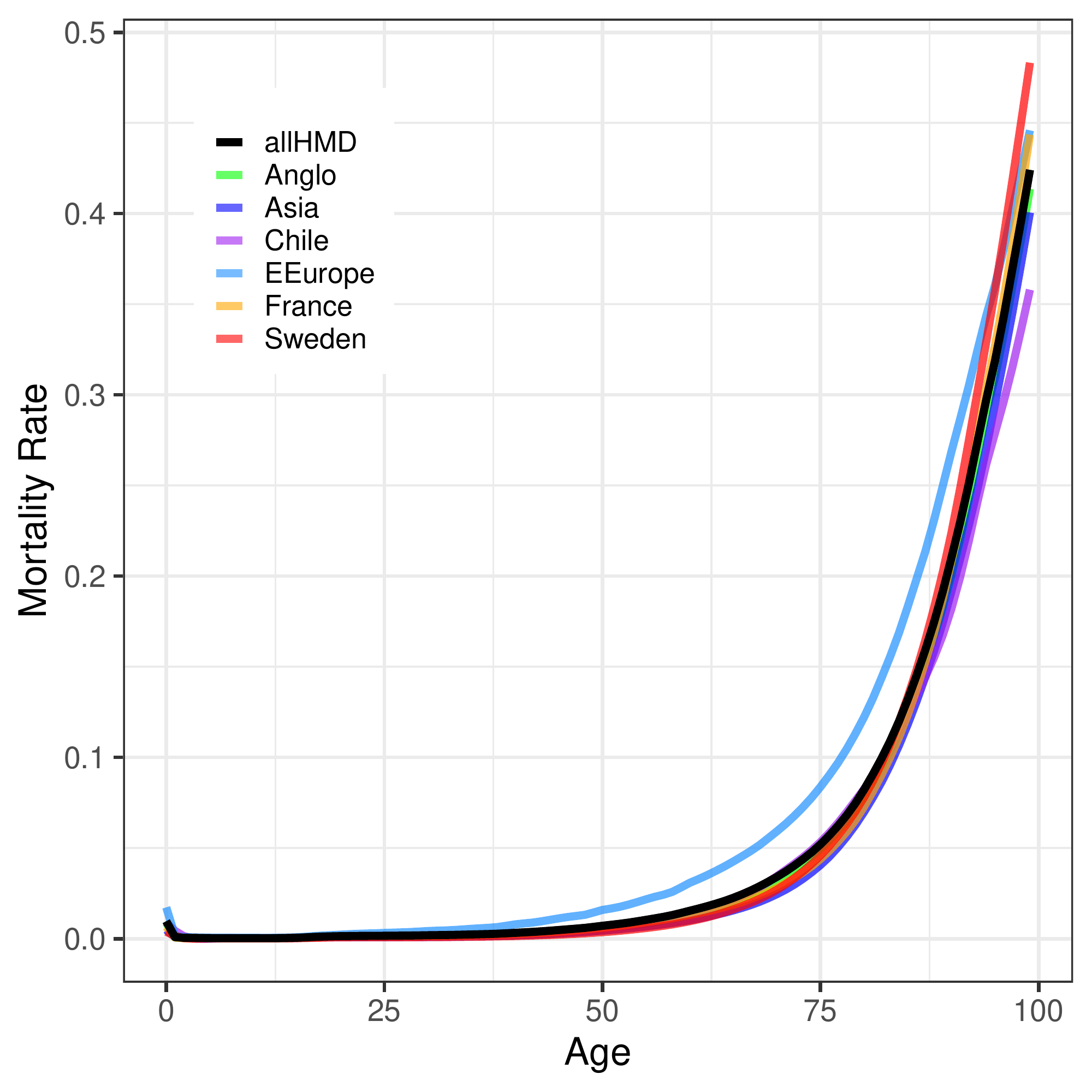}
	}
	\subfigure
	{
		\includegraphics[scale=0.325]{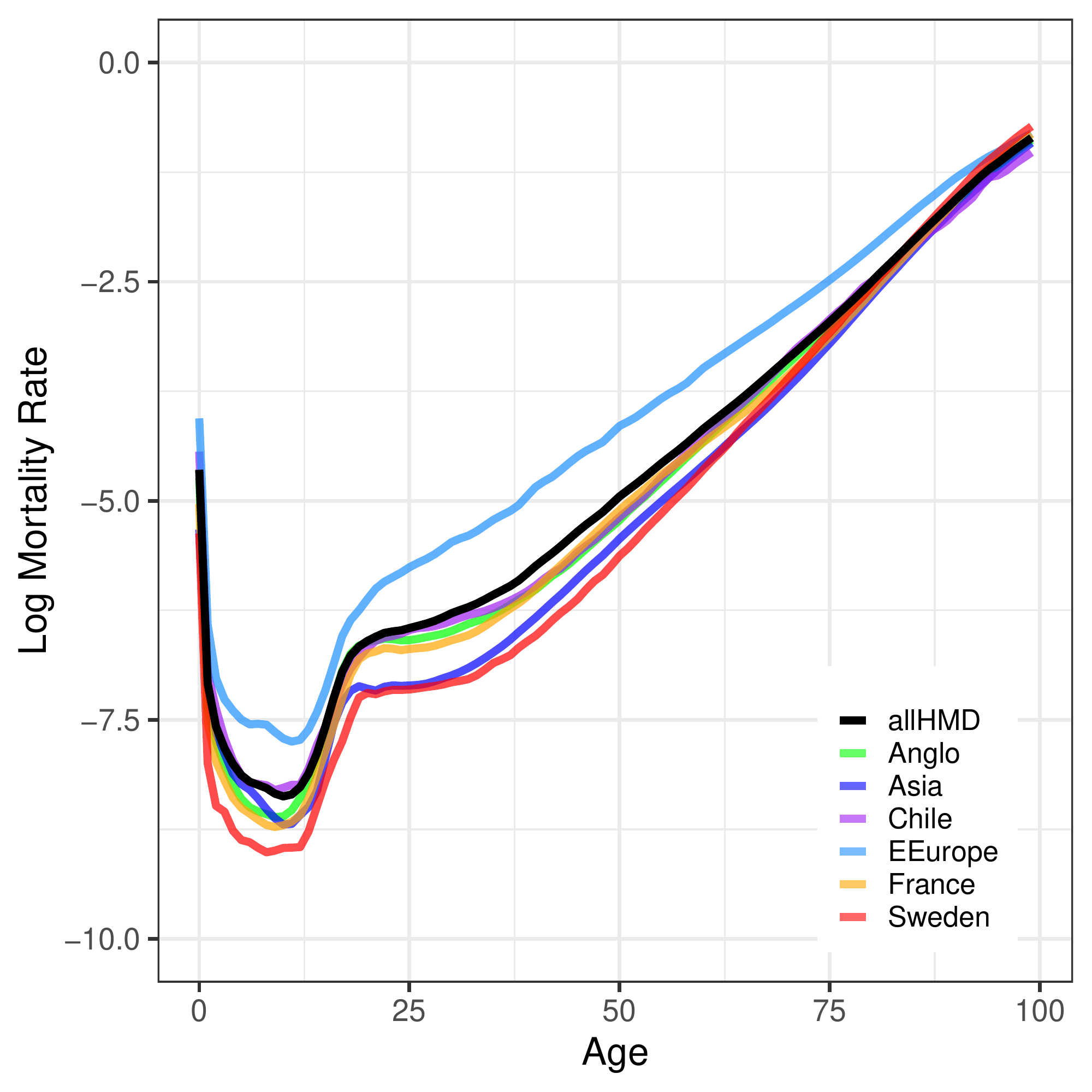}
	}
	\caption{Seven mortality schedules by 1-year age intervals extracted from the life tables available in the Human Mortality Database: all countries,  Chile,  Sweden, France, Eastern European countries in HMD,  Anglophone countries and  Asian countries. Original scale (left) and logarithmic scale (right). Source: \cite{HMD2015} through \cite{Gonzaga2016}'s website {http://topals-mortality.schmert.net}.}
	\label{fig:HMDexamples}
\end{figure}

In human populations, mortality rate curves generally display regular patterns. 
The left panel of Figure \ref{fig:HMDexamples} illustrates the usual shape for human mortality schedules considering data selected from the life tables available in the Human Mortality Database (HMD) \citep{HMDprotocol2020}.
The curves were obtained by fitting spline functions to the observed values of mortality rates by 1-year age intervals for each selected life table. Spline smoothing is considered in order to clarify the underlying shape. 
It can be seen that, in general: (i) the mortality rate tends to decay quickly after birth; (ii) then,  it apparently remains stable until age 35; and (iii) an exponential growth can be noticed from there.
Indeed, with basis on historical evidence, the previous features are commonly  observed for a large range of human mortality curves. 
Alternatively, mortality curves are often analyzed on the logarithmic scale (right panel of Figure \ref{fig:HMDexamples}).
In this case, the typical variations of mortality rates across ages can be better noted. 	
In particular: (1) the mortality is relatively high in the first years of life; (2) then, it shows a steep decline between birth until a regular minimum around age 10; (3) the mortality curve typically reveals an "accident hump'' at young adult ages (around, roughly, ages 15 to 30); (4) above age 30 the curve is fairly linear (in the log scale); and (5) at the oldest ages (say, above age 90) there is often a deceleration of mortality.  

When dealing with large populations, specially in developed countries, the usual shape for the mortality curve is easily identified in general. That is the case for most life tables available in the HMD (e.g., see curves in Figure \ref{fig:HMDexamples}).
However, that is not true when analyzing data from populations with poor data quality as, for instance, in developing countries that present incomplete coverage of the vital registration systems as well as errors in age declaration for both population and death counts.

Besides the data registration issues, the production of life tables is even more difficult when focusing in small populations. In this cases, the observed rates are often highly erratic and may have a great amount of null death counts (numerator of mortality rates) or even the lack of individuals exposure to risk in some age intervals (denominator of mortality rates).

\begin{figure}[htb!]
	\centering
	\subfigure
	{
		\includegraphics[scale=0.225]{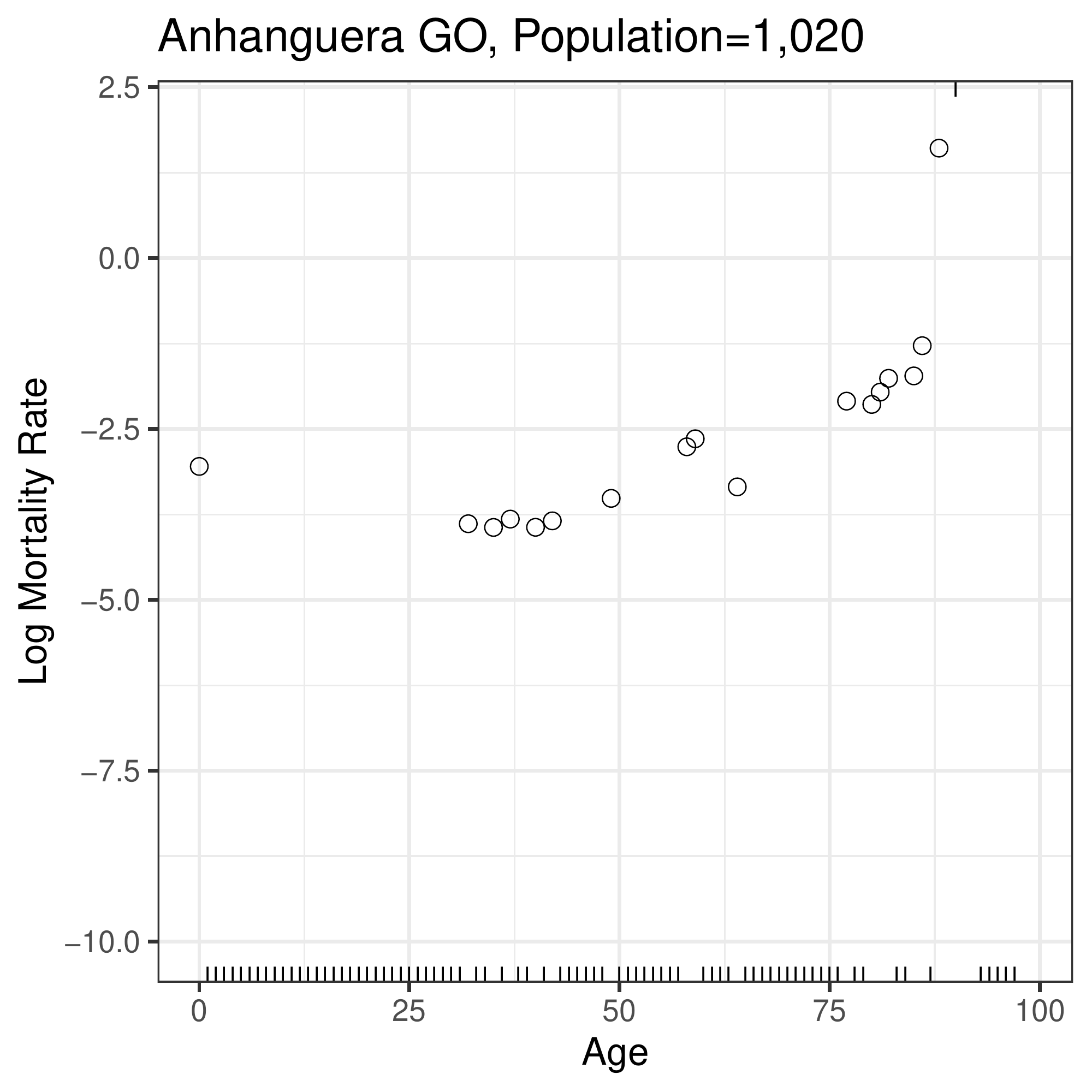}
	}
	\subfigure
	{
		\includegraphics[scale=0.225]{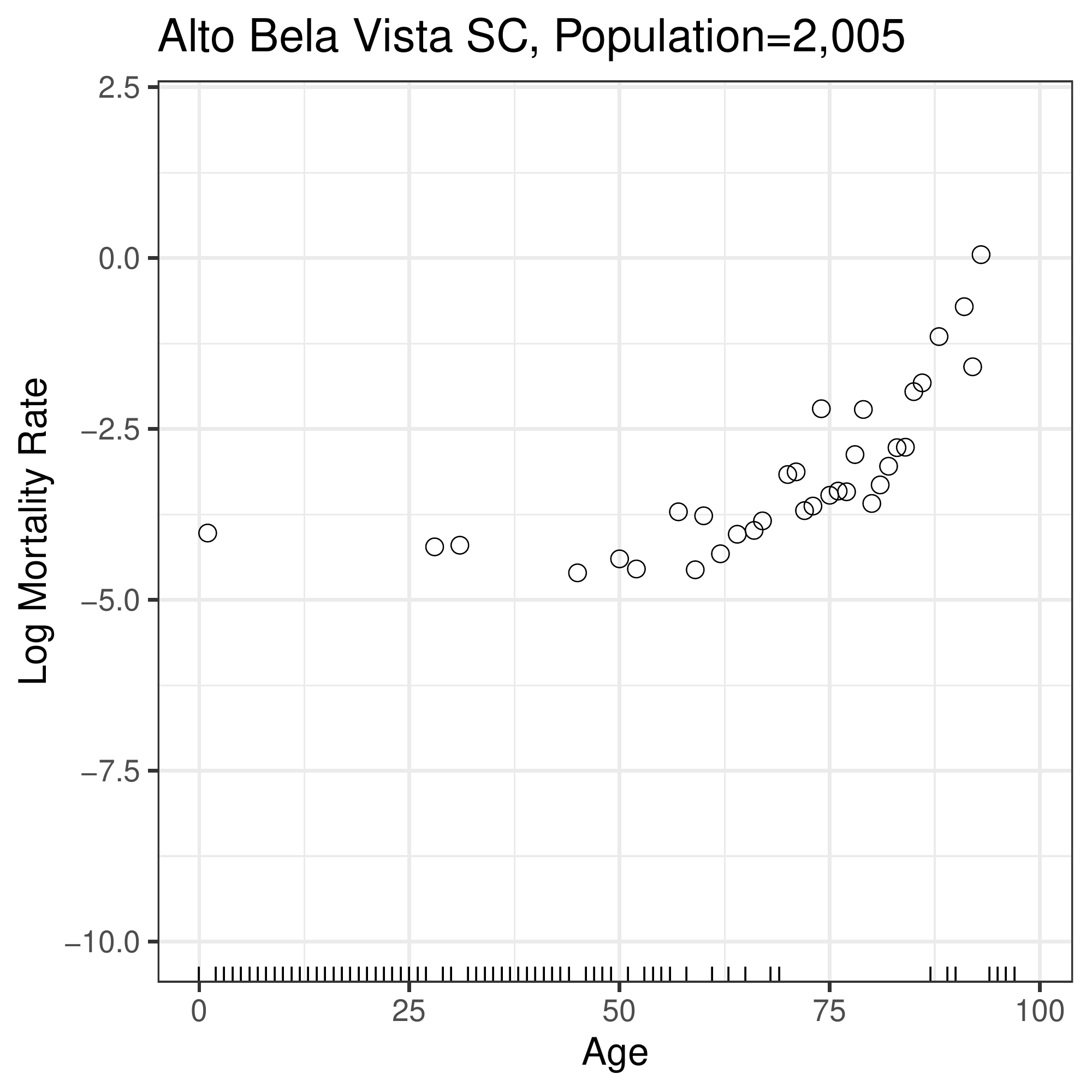}
	}\subfigure
	{
		\includegraphics[scale=0.225]{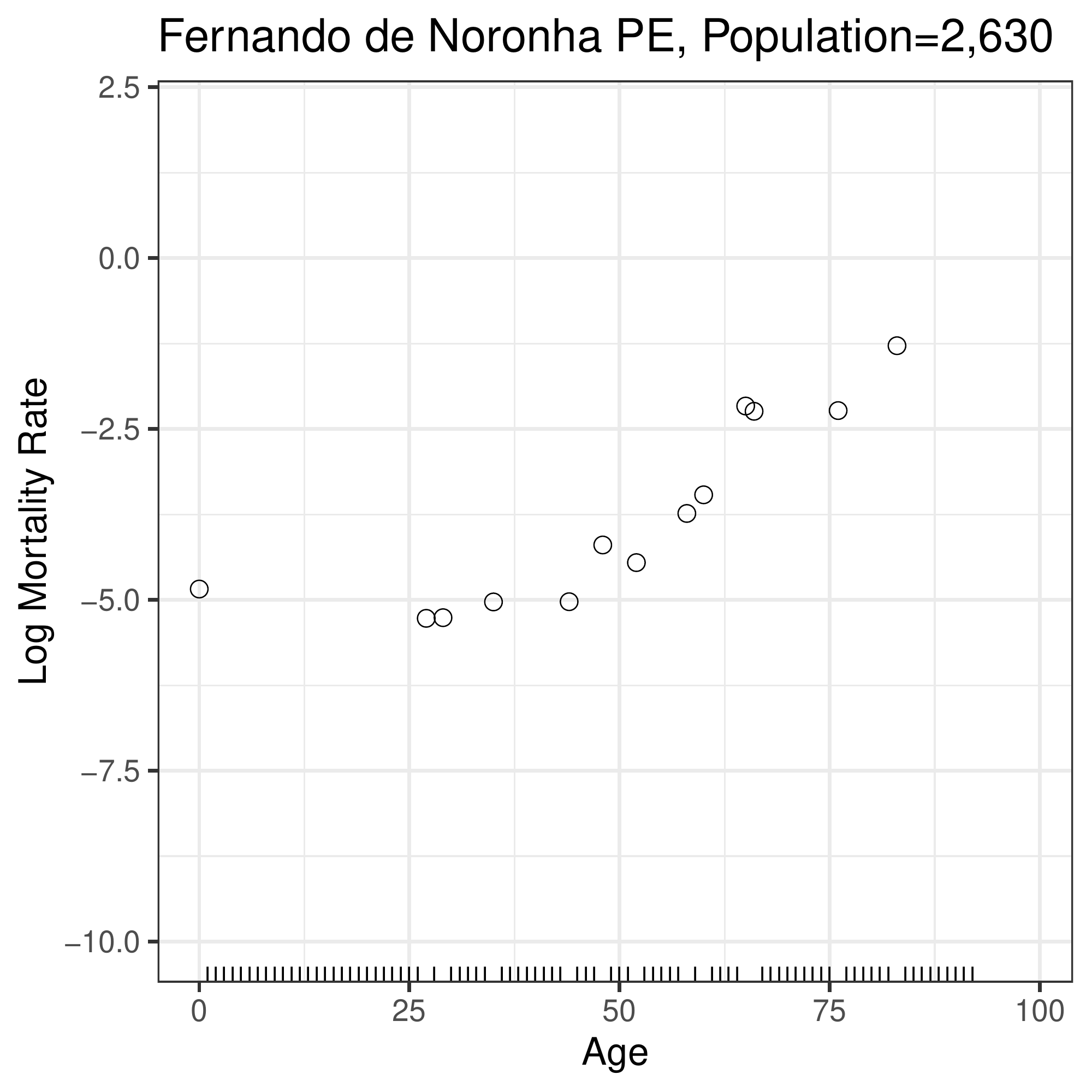}
	} \\
	\subfigure
	{
		\includegraphics[scale=0.225]{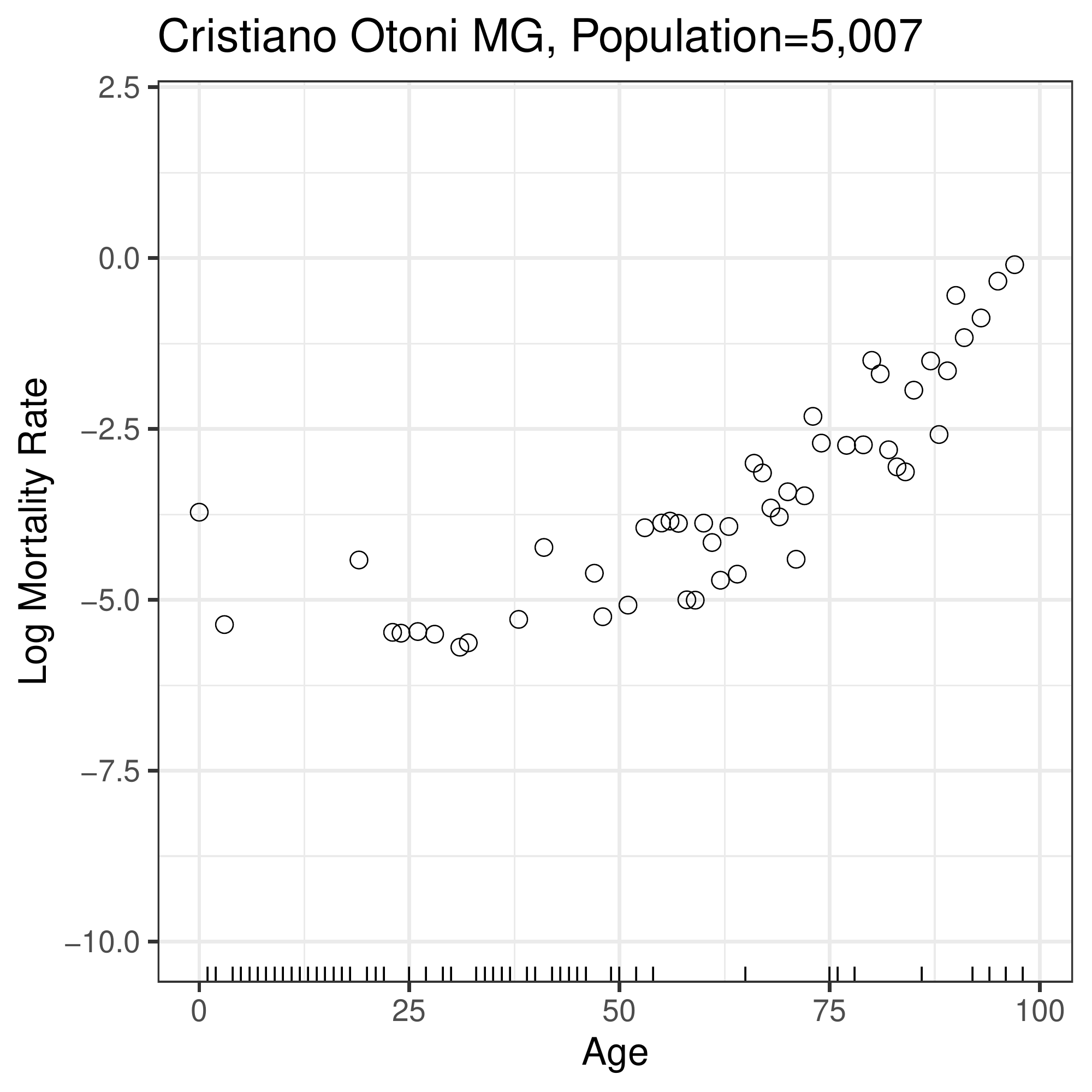}
	}
	\subfigure
	{
		\includegraphics[scale=0.225]{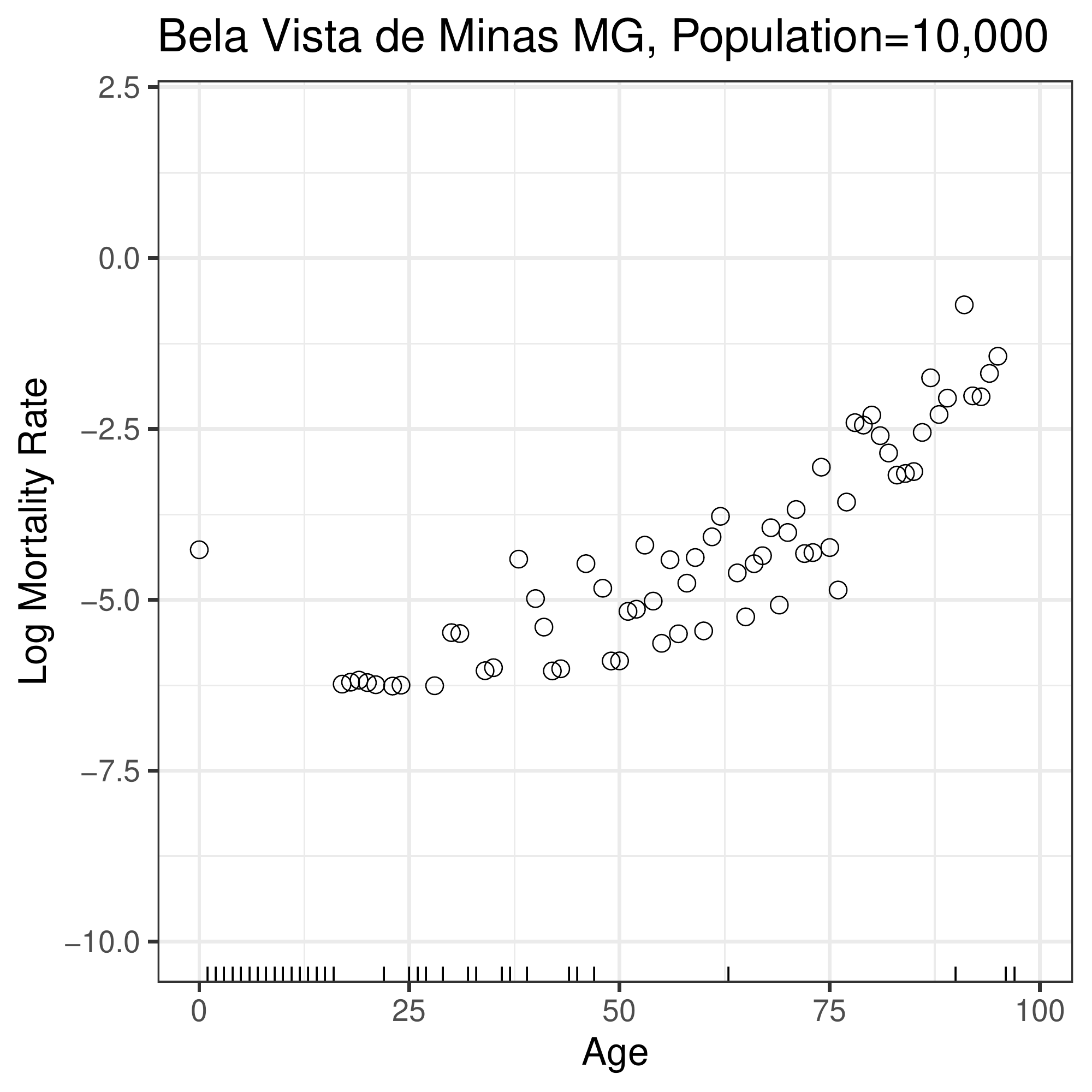}
	}\subfigure
	{
		\includegraphics[scale=0.225]{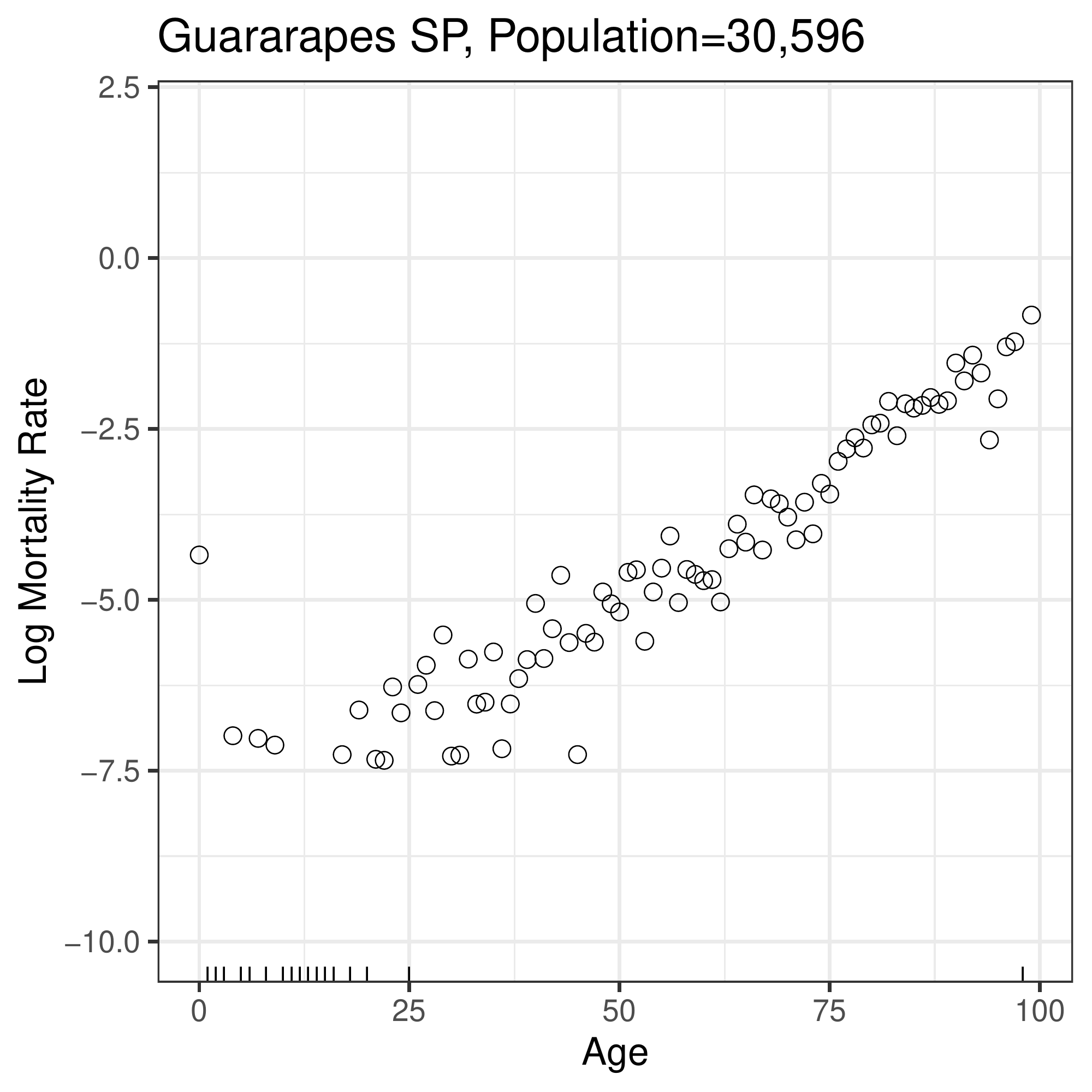}
	} \\
	\subfigure
	{
		\includegraphics[scale=0.225]{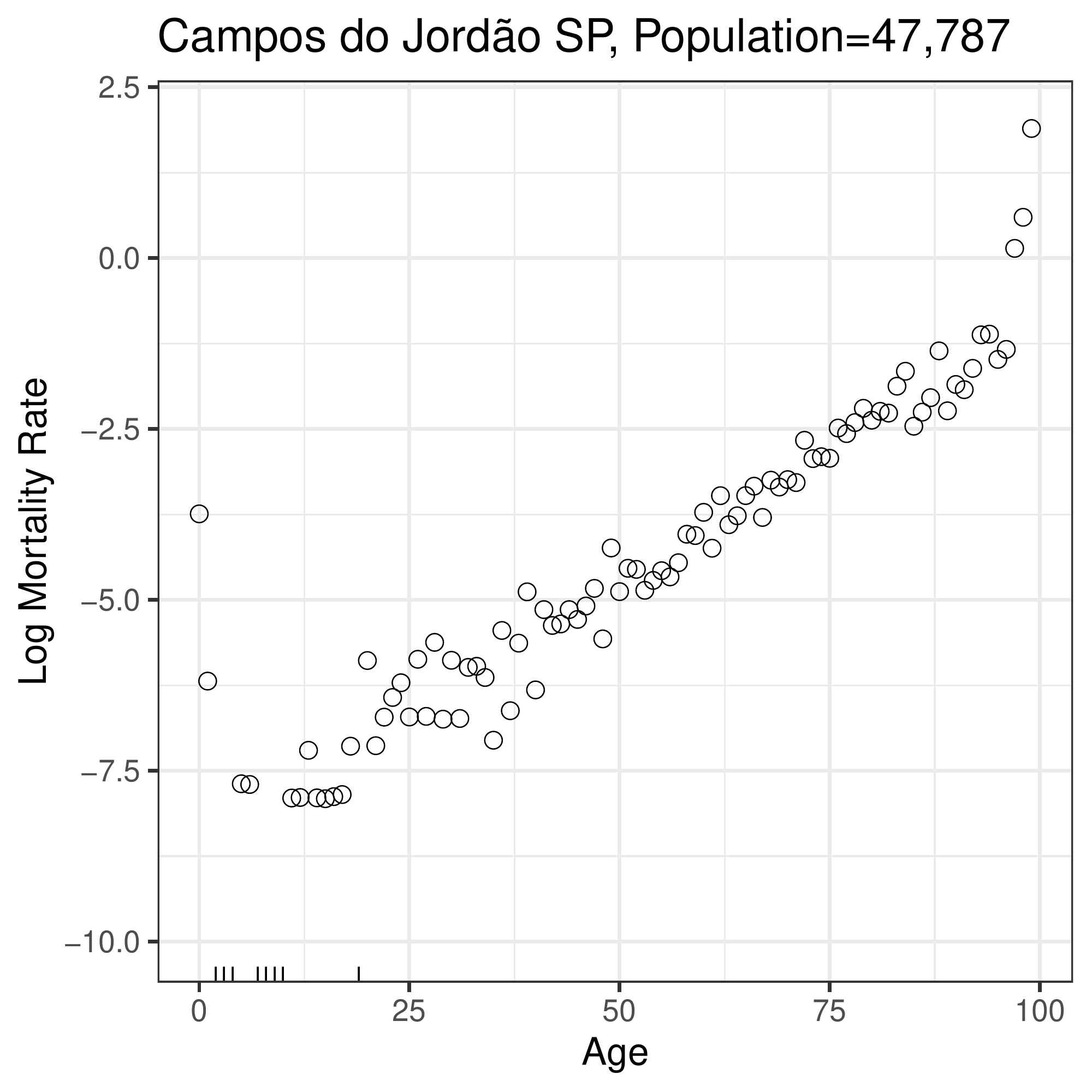}
	}
	\subfigure
	{
		\includegraphics[scale=0.225]{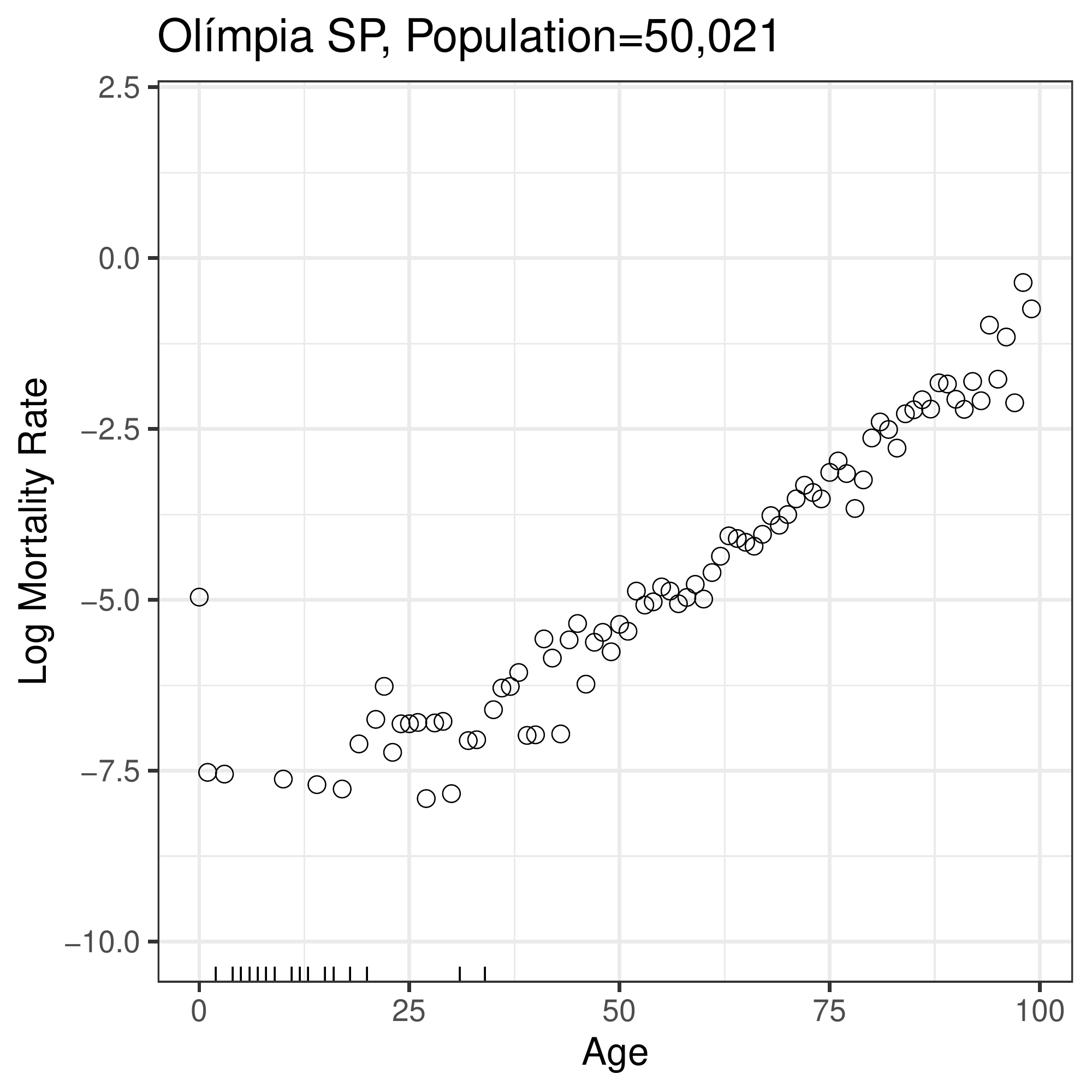}
	}\subfigure
	{
		\includegraphics[scale=0.225]{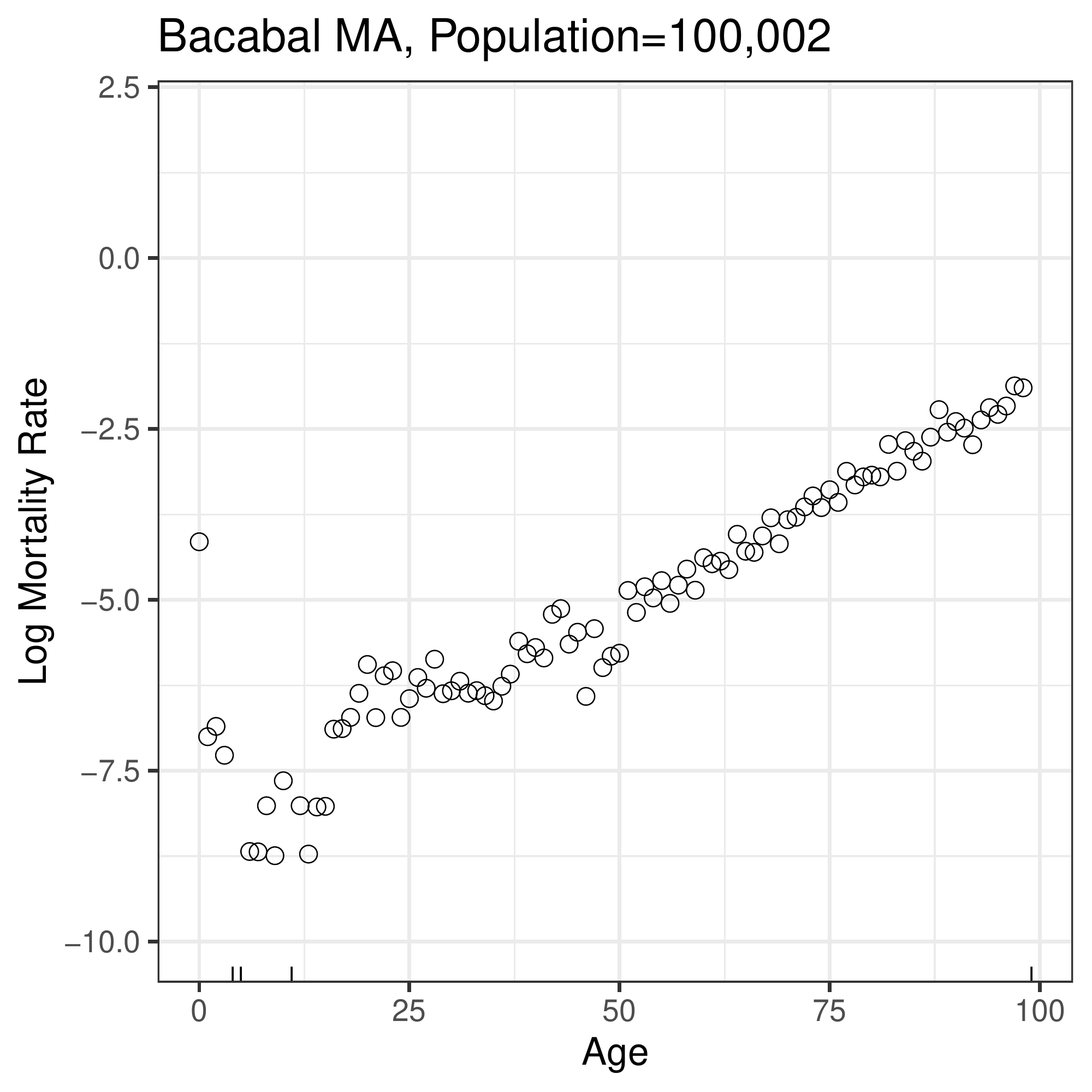}
	} \\
	\subfigure
	{
		\includegraphics[scale=0.225]{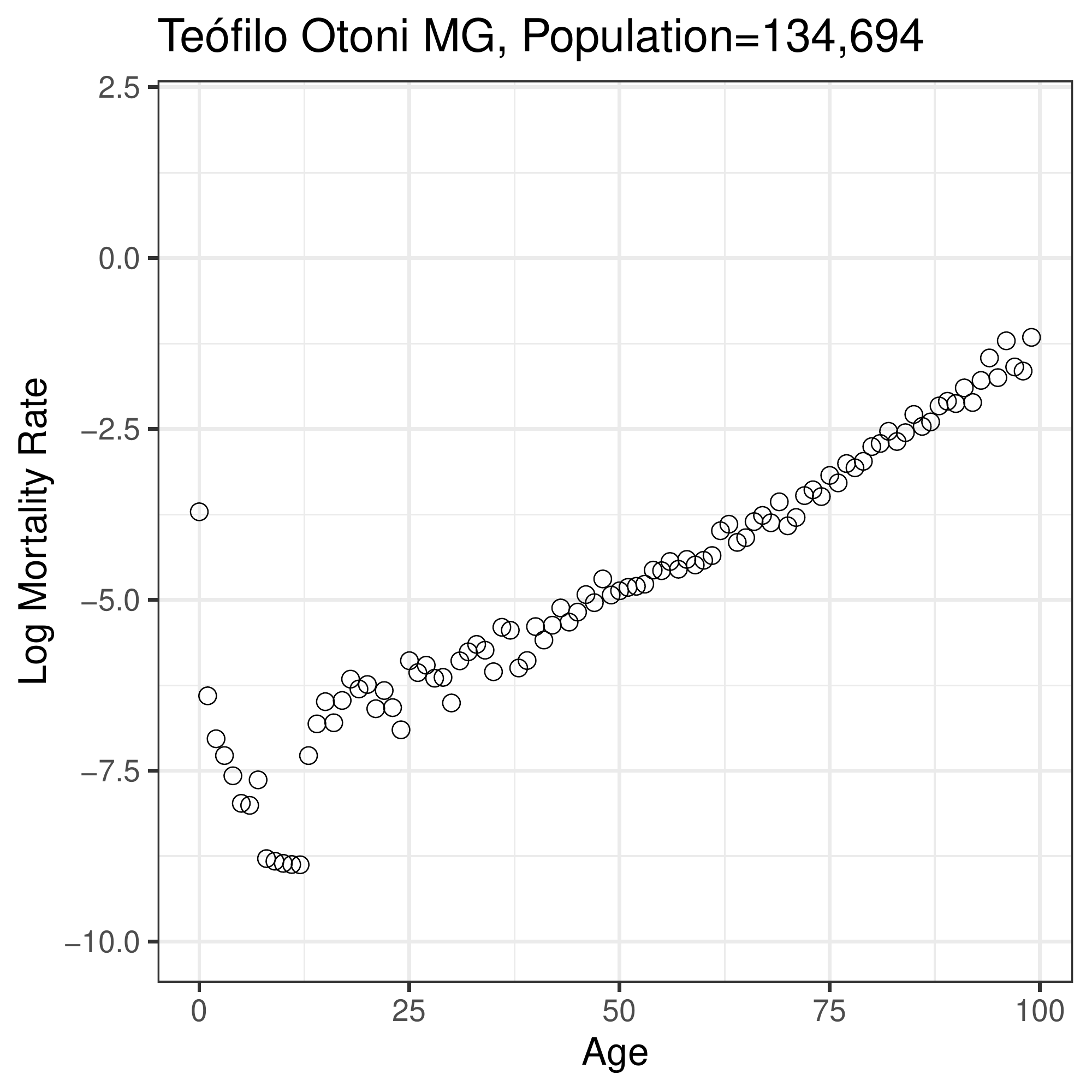}
	}
	\subfigure
	{
		\includegraphics[scale=0.225]{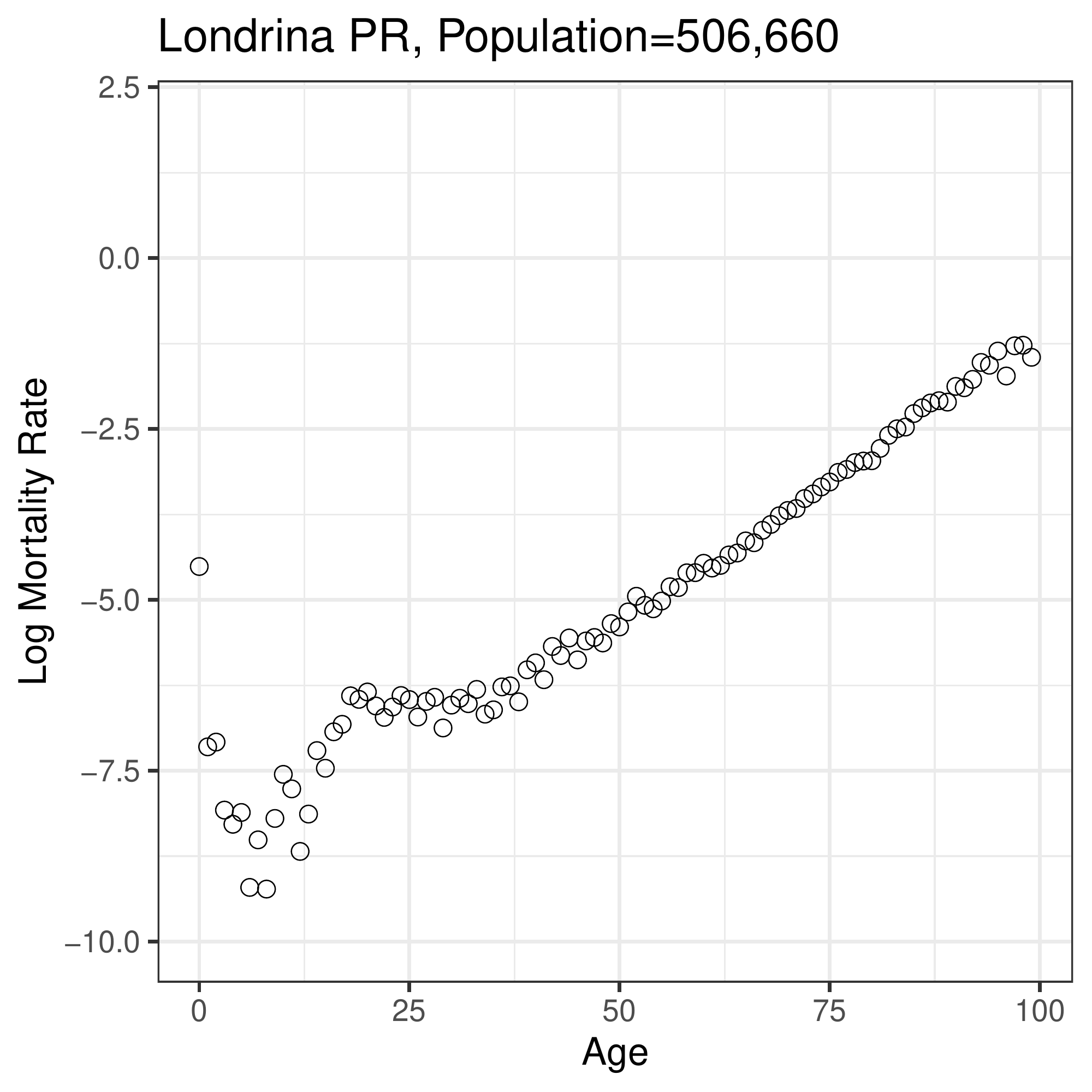}
	}\subfigure
	{
		\includegraphics[scale=0.225]{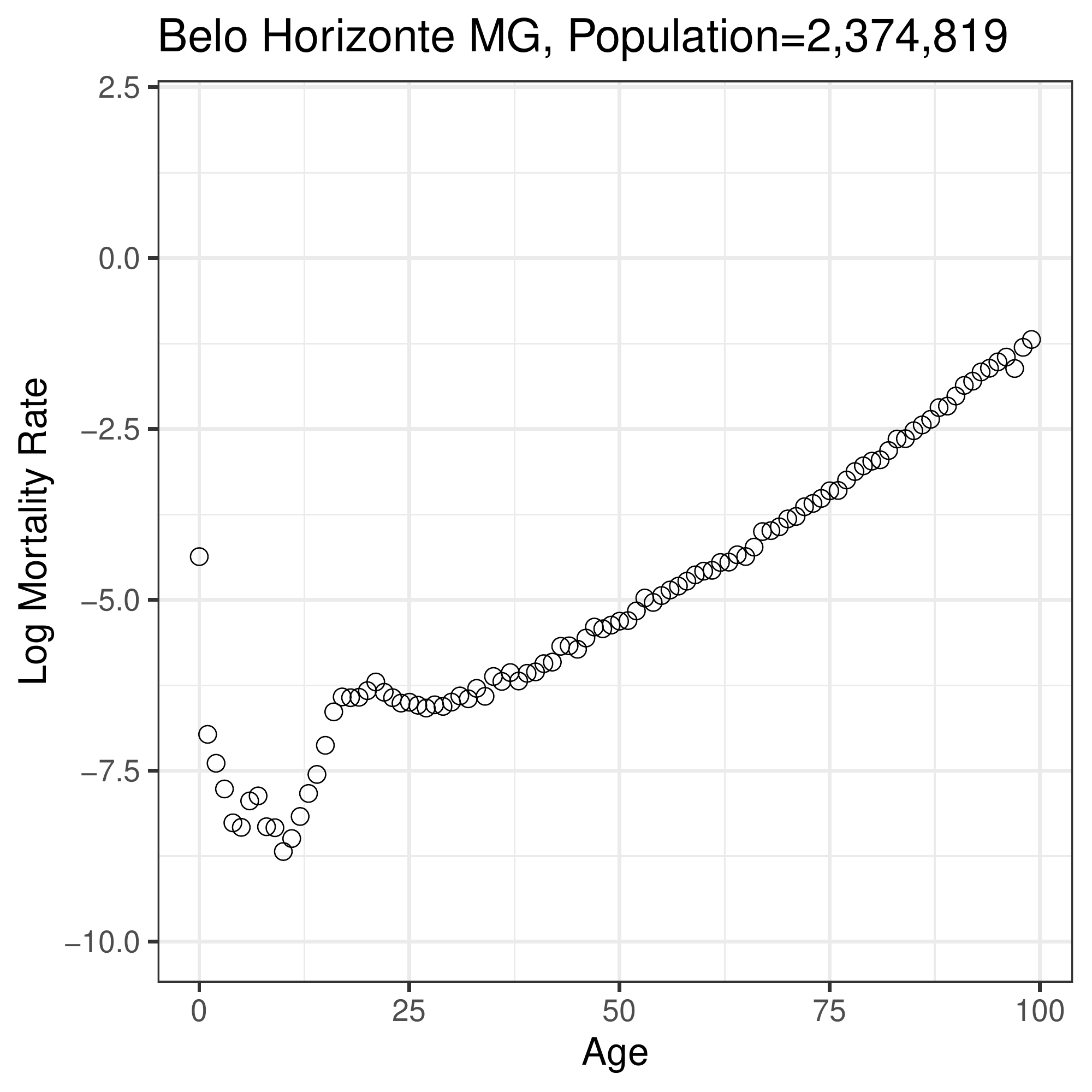}
	} \\
	\caption{Observed mortality schedules (in the log scale) for selected Brazilian municipalities from 2009 to 2011, both sexes. Open circles represent the log-mortality rate for each single-year of age. Tick marks on the horizontal axis represent ages with no reported deaths or ages with no population at risk, which makes impossible to calculate the mortality rate. Data sources: IBGE (2010) and Brazilian Ministry of Health (http://www.datasus.gov.br).}
	\label{fig:mortality_curves_Brazil}
\end{figure}

In addition to the presence of extreme values, the mortality rates observed in small populations tend to present high variability across ages and sexes (male and female). 
Such data characteristics make it difficult to identify the true underlying mortality pattern.
As an example, Figure \ref{fig:mortality_curves_Brazil} displays the log-mortality rates observed for some out of the 5,565 Brazilian municipalities. 
There is a higher variability in the smallest populations due to the greater occurrence of low and null counts, as a consequence of having a small number of individuals at risk in some ages.
According to the national census of 2010, for 45\% of the Brazilian municipalities the population is smaller than 10,000 inhabitants and for 90\% of them it is smaller than 50,000.

\clearpage
Figure \ref{fig:mortality_curves_Brazil} also shows that it is possible to identify the shape of usual mortality schedules for municipalities in which population is greater than 100,000 inhabitants. 
It also shows that data noise is smaller for large populations.
For small populations, besides the great variability, there are several age intervals with zero deaths count, specially at infant and young ages, making cumbersome the identification of the shape of the mortality schedules.

Naturally,  the sparsity in observed data becomes even more severe when subnational groups are disaggregated by age and sex, usual procedure in life table estimation.
Reliable measurements and comparative analysis of mortality levels, age patterns and sex differences for regional populations  help to better understand health status at local levels and to guide policy definitions and changes in the targets for public investments.
It may also help in the appropriate derivation of life expectancy and other measures used to perform population projections. 

Some important goals of mortality modeling include describing the shape of mortality curves, estimating and projecting mortality patterns over time, and investigating
differences in mortality patterns across different populations \citep{Gompertz1825,Brass1971,Heligman1980,Coale83,LeeCarter1992,Dellaportas2001,Dowd2011,Li2014,Wilmoth2012,Lima2016,AlexopoulosDellaportas2019}.
Because mortality schedules generally display regular patterns,  smoothing approaches are a natural choice to analyze changes in mortality rates, age-structure decompositions and the construction of continuous life tables \citep{Kashiwagi1992,Alexander2017}.
In this context, a common approach involves spline smoothing functions (see, e.g., \citet{Currie2004,DeBeer2012,Camarda2012,Gonzaga2016,Alexander2018} and references there in).

For developed countries, where annual population updates tends to be available and vital registration systems have good quality, researchers have recently made important advances in statistical modeling and smoothness of complete mortality schedules in small areas (see \cite{Lima2016} and the references there in). 

For less developed countries, the identification of discrepancies in mortality patterns across large regions (such as the states of a country) may not be a complicated task, especially because the populations size tend to be large.
In the presence of abundant population, even simple parametric demographic models may produce good estimates for the target mortality rates. 
However, studies on complete age- and sex-specific mortality schedules at subnational levels are rare in developing countries due to lack of updated information.
Because of that, demographers and statistical epidemiologists have proposed  a voluminous literature
for estimation of partial mortality schedules in developing countries, especially infant and child mortality indicators  \citep{Souza2010,Walker2012,Silva2013,Alkema2014,You2015}. 
Some studies with focus in adult and old-age mortality levels in such areas can also be found \citep{Kannisto88,Timaeus1991,Hill2009}.
Many methods rely on indirect information from surveys or censuses to adequately address the volatility of these estimates due to data quality issues regarding vital registration systems.

As noted by \cite{Lima2016}, a prominent and promising modeling approach for estimation of complete mortality schedules in less developed small populations involves the combination of statistical models with formal demography methods.
The incorporation of demographic knowledge into statistical modeling frameworks is intended to ensures that estimates
and projections have plausible patterns across ages.

Many authors have evaluated the efficacy of using parametric modeling structures combined with empirical regularities observed in mortality schedules obtained from external trustful sources \citep{Brass1971,Coale83,Murray2003,Wilmoth2012,Alexander2017,Gonzaga2016,Clark2019}. 
Some authors refers to these sort of methods as relational models. 
The idea behind relational models is that complete age patterns in mortality, while inherently non-linear, exhibit strong similarities across different populations.
Thus, patterns observed in high-quality data (e.g., mortality schedules in Figure \ref{fig:HMDexamples}) are used as a standard basis for producing estimates of mortality in populations where observed data are sparse or have poor quality, as in the small populations illustrated in Figure \ref{fig:mortality_curves_Brazil}.

In this context of ''borrowing strength'' from an external mortality standard, recently, some authors have used data available in the Human Mortality Database \citep{HMDprotocol2020} to build more flexible systems of life tables. 
For instance, the Bayesian hierarchical relational model proposed by \cite{Alexander2017} ensures a relatively smooth trend in mortality over time at the same time of sharing information across geographic areas.
As an alternative, the approach of \cite{Gonzaga2016} linearly relates a mortality standard and penalized  spline offsets to smooth mortality curves in small areas (populations).

The main goal of this work is to  propose alternative models to estimate complete mortality schedules, specially in small underdeveloped populations. 
In order to estimate and smooth the associated mortality curves,  we propose some Bayesian hierarchical dynamic models jointly with an underlying functional form that captures regularities in age patterns of mortality. 
The dynamic terms relates mortality rates across age intervals as an attempt to relatively smooth trend in the complete mortality curve. 
In turn, the use of a mortality standard is intended to penalize departures from the usual characteristic shapes of the target curves.
We consider simulated and real datasets for different population sizes in order to exploit the advantages and drawbacks of the proposed methods. 
Competing approaches are compared in terms of adequacy and quality of smoothed mortality estimates. 

\section{Estimating and Smoothing Mortality Curves}
\label{model:mortality_curves_models}

For a given population $i$ and age $x$ assume that the total number of deaths is $Y_{i,x}$ and the total exposures individuals is $E_{i,x}$, $i=1,\dots,n$ and $x=0,1,\dots,A$. 
Assume that the expected mortality rate is given by the product of exposures $E_{i,x}$ and an unknown parameter $\theta_{i,x}$ which represents the mortality risk in each population $i$ and age $x$.  
A fully empirical (naive) estimate for the risk can be obtained by computing the observed death rates at the respective population and age $M_{i,x} = Y_{i,x}/E_{i,x}$ for all $i,x$.
In terms of probabilistic modeling, a Poisson distribution with mean $E_{i,x}\theta_{i,x}$ is an usual choice to model the observed count $Y_{i,x}$ in which case an appropriate inferential method can be used to estimate the unknown parameter $\theta_{i,x}$.

To simplify the presentation, mortality data are commonly prepared as rectangular arrays. For each age in particular population, we have the number of deaths, the number of total exposures and the observed mortality rates arranged in $n\times A$ matrices $\bY$, $\bE$ and $\bM$, respectively. The rows are indexed by population and the columns are indexed by age. Without loss of generality, in this work we consider $A=100$ corresponding to ages $0,1,2,\dots,98,99$.

We aim to estimate a smoothed mortality curve for each population of interest. 
For doing that, we must estimate the related mortality rates at each age.
We approach this problem by using a regression structure for the mortality rates in which the covariate corresponds to  a mortality standard curve. Such a covariate is used  to inform about the usual pattern observed for human mortality curves (see discussion in Section \ref{intro}).
In order to "borrow strength'' from the relation of mortality rates in subsequent age intervals, we treat the sequence of age-specific mortality rates as a time series.
Then, a dynamic structure across them is imposed through the model coefficients, thus providing a smoothed solution.   Such a modeling framework  is proposed in Section \ref{sec:dynamicPoisson}.

In our analysis, the standard mortality curve corresponds to the a mortality schedule calculated with basis in all life tables available in the Human Mortality Database in 2015 \citep{HMD2015}.
The information was obtained from the supplementary materials that were made available by \cite{Gonzaga2016} who also used such a standard in their model definition ({http://topals-mortality.schmert.net}). 
The standard mortality schedule is available separately for males and females. 
The standard schedule obtained by tacking the mean between the two sex-specific schedules is the ''allHMD'' curve displayed in Figure \ref{fig:HMDexamples}.

In the following subsections we present the models we consider to analyze the datasets of interest.

\subsection{Dynamic Poisson Model}\label{sec:dynamicPoisson}

Assume that the total number of deaths for population $i$ and age $x$ has the Poisson distribution
$$Y_{i,x}\sim Poisson(E_{i,x}\theta_{i,x}),$$ 
where $E_{i,x}$ is an offset corresponding to the total exposure individuals and $\theta_{i,x}$ denotes the mortality risk, $i=1,,\dots,n$ and $x=0,1,\dots,A$. 
Consider that 
$$\log(\theta_{i,x}) = \beta_{i,x} + \mu_{{i}}S_{i,x},$$
where $S_{i,x}$ is the mortality standard obtained from the HMD and $\beta_{i,x}$ and $\mu_{i,x}$ are the regression parameters for population $i$ and age $x$. 
We assume the same standard for the $n$ populations, that is, $S_{i,x}=S_{x}$, but a different standard can be applied for each population.
The proposed model assumes a Markovian dependence among counts in neighboring age intervals. Such a dependence is included into the model through the Gaussian prior distributions elicited for $\bbeta_i=(\beta_{i,1},\dots,\beta_{i,A})$ and $\bmu_i=(\mu_{i,1},\dots,\mu_{i,A})$ for all $i$.
It is assumed that, given the associated precision parameters $\tau_{\beta}$ and $\tau_{\mu}$,  the regression parameters are independent among populations, that is, $\bbeta_1$, \dots,$\bbeta_n$ and $\bmu_1$, \dots,$\bmu_n$ are mutually independent for $i=1,...,n$.
The proposed dynamic Poisson model is hierarchically represented  as follows:
\begin{eqnarray}\label{eq:Poisson}
Y_{i,x}|\theta_{i,x} &\stackrel{ind}{\sim}& Poisson(E_{i,x}~\theta_{i,x}),~~for~~i=1,\dots,n;~x=1,...,A\\
\log(\theta_{i,x}) & = & \beta_{i,x} + \mu_{{i}}S_{x}\nonumber\\
\beta_{{i,0}}|\beta^0_{{i}}, \tau_{\beta} &\stackrel{ind}{\sim} & N\left(\beta^0_{{i}}, \tau_{\beta}\right)\nonumber\\
\beta_{{i,x}}|\beta_{{i,x-1}}, \tau_{\beta} &\stackrel{ind}{\sim} & N\left(\beta_{{i,x-1}}, \tau_{\beta}\right),~~~~for~~x=1,\dots,A\nonumber\\
\mu_{{i}}| \tau_{\mu} & \stackrel{iid}{\sim} & N\left(0, \tau_{\mu}\right),~~~~for~~x=1,\dots,A\nonumber\\
\tau_{\beta}, \tau_{\mu} &\stackrel{iid}{\sim} & Gamma(0.01,0.01)\nonumber\\
\beta^{{0}}_i &\stackrel{iid}{\sim} & N\left({0}, \tau=100 \right),\nonumber
\end{eqnarray}
where $\bbeta^{{0}}$ is a vector of initial values for the dynamic structure. The Normal distributions are parameterized in terms of mean and precision.
The model was implemented in package \textit{Nimble} \citep{deValpire2017} from software R \citep{softwareR}.
%

We propose to jointly model the $n$ mortality schedules of interest. 
By doing that,  a greater amount of sample information is guaranteed to estimate the precision parameters $\tau_{\beta}$ and $\tau_{\mu}$, which are shared by the $n$ populations.  We noted that such a strategy improve smoothness of the resulting mortality curves.

\subsection{TOPALS model from \cite{Gonzaga2016}}\label{sec:TOPALS}

\cite{Gonzaga2016} introduce a new method to estimate age-specific mortality rates in small populations.
It corresponds to a Poisson regression model based on TOPALS, a relational model
developed by \cite{DeBeer2012} for smoothing and projecting age-specific probabilities of death.
Their approach estimates a complete schedule of log-mortality rates  by adding a linear spline function to a pre-specified standard schedule.
The spline represent additive offsets in specific ages. 
As in Section \ref{sec:dynamicPoisson} denote by  $Y_{i,x}$ and $E_{i,x}$ the death count and the total exposure population in an population $i$ and age $x$, for $x=0,1,\dots,A$, respectively. The TOPALS model from \cite{Gonzaga2016} is defined as follows:
\begin{eqnarray}\label{eq:TOPALS}
Y_{i,x}|\theta_{i,x} &\stackrel{ind}{\sim}& Poisson(E_{i,x}\exp\{\theta_{i,x}\})\\
\theta_{i,x}|\balpha & = & S_{i,x} + \bB_{x}\balpha,\nonumber
\end{eqnarray}
where $S$ is a standard mortality schedule; $\bB_{x}$ is a $1\times 7$ vector of constants corresponding to the $x$th line of a $A\times 7$ matrix $\bB$ in which each column is a linear B-spline basis (for details, see \cite{Gonzaga2016} and their references); and $\balpha$ is a $7\times 1$ vector of parameters representing offsets to the standard schedule for specific knots $t_0,...,t_6$ at ages $(0, 1, 10, 20, 40, 70, 100)$, respectively. 
For ages $x \in {0, 1, ..., 99}$ and columns $k \in {0,...,6}$ the basis functions in $\bB$ are

{\Large $$B_{x,k}=\left\{\begin{array}{l} 
	\frac{x-t_{k-1}}{t_{k}-t_{k-1}}~~~~~\mbox{if}~~x\in\left[t_{k-1},t_{k}\right],\\
	\frac{t_{k+1}-x}{t_{k+1}-t_{k}}~~~~~\mbox{if}~~x\in\left[t_{k},t_{k+1}\right],\\
	0~~~~~~~~~~~~\mbox{otherwise.}
	\end{array}\right.$$}

The authors argue that under such a parameterization, the $\balpha=(\alpha_1,...,\alpha_7)$ values represent additive offsets $(\theta_{i,x}-S_{i,x})$ to the log-rate schedule at exact ages $(0, 1, 10, 20, 40, 70, 100)$ and offsets change linearly with age between those knots.
Parameters $\alpha_1,...,\alpha_7$ are estimated by maximizing a penalized Poisson likelihood function for age-specific deaths, conditional on age-specific exposures.   
The penalty term added to the log-likelihood function increases as the linear spline offsets become less smooth. \cite{Gonzaga2016} considered such approach in order to avoid implausible fitted schedules for very small populations with very low numbers of deaths. 
TOPALS is applied to estimate mortality schedules in Brazilian municipalities and an illustration of the method is provided in Figure 1 of \cite{Gonzaga2016}.

All datasets and functions related to TOPALS adjustment were made accessible on the supplementary website   {http://topals-mortality.schmert.net}. 
We  consider the available material to fit the TOPALS model in the data analysis presented in the following sections and we compare its performance with the proposed dynamic models.

\section{Simulation Experiments}\label{sec:simulations_cap3}

In this section, we consider simulated datasets to compare the performance of the dynamic Poisson model introduced in Section \ref{sec:dynamicPoisson}, the Gaussian DLM presented in Section\ref{sec:gaussianDLM} and the TOPALS model briefly reviewed in Section \ref{sec:TOPALS}. 
To generate a complete mortality schedule, we have to determine a true underlying mortality mechanism, which must include information about the mortality rate and the associated total exposure population by age. 
For this purpose, we consider the mortality schedule observed for the S\~ ao Paulo State (SP), Brazil, available from the case study presented in Section \ref{sec:application_mortality_curves}. 
As SP has a large population size, it provides a sufficiently smooth mortality curve to be used as a reference in our simulation study. In addition, the mortality curve of S\~ ao Paulo State characterizes a modern population and it  guarantees the simulation of mortality patterns potentially conformable to those we are interested in our application to Brazilian data. 

We simulate nine populations with sizes 1,000; 2,000; 5,000; 10,000; 20,000; 50,000; 100,000; 500,000 and 1,000,000. 
In each case, the total exposures per 1-year age interval, $E_{i,x}$, was proportionally calculated to mimic the population pattern of SP. 
By considering the mortality rates observed for S\~ ao Paulo as the true underlying risks, $\theta_{i,x}$, we generate the death counts $Y_{i,x}$ from a Poisson distribution such that $Y_{i,x}\sim Poisson(E_{i,x}\theta_{i,x})$, for $i=1,,\dots,9$ and $x=0,1,\dots,99$. 

The generated datasets were fitted under the three models and the estimates for the associated mortality rates, $\hat{\theta}_{i,x}$, were compared in terms of relative bias (RBias), square root of the mean squared error ($\sqrt{MSE}$) and mean absolute percentage error (MAPE) averaged over the $A=100$ age-mortality rates in each simulated population. 
Such measures are calculated, respectively, as $RBias=\frac{1}{A}\left[\sum_{x=1}^{A}\left(\frac{\theta_{i,x}-\hat{\theta}_{i,x}}{\theta_{i,x}}\right)\right]$;  
$MSE=\frac{1}{A}\left[\sum_{x=1}^{A}\left(\theta_{i,x}-\hat{\theta}_{i,x}\right)^2\right]$ and 
$MAPE=\frac{1}{A}\left[\sum_{x=1}^{A}\left\lvert\frac{\theta_{i,x}-\hat{\theta}_{i,x}}{\theta_{i,x}}\right\lvert\right]$.
Under the Poisson and Gaussian DLM models, for each generated dataset, two chains were run in the MCMC scheme considering a burn-in period of 100,000 iterations and a lag of 5,000 iterations was selected to avoid autocorrelated posterior samples, ending with a posterior sample of size 2,000 for each chain. The TOPALS model was fitted by using the functions available at {http://topals-mortality.schmert.net}.
The three models were fitted on an Intel (R) Core (TM) i7-8550U 1.80GHz CPU with 8GB RAM. Respectively, the computational effort to run the algorithm for the TOPALS, Poisson and Gaussian models was around 1, 16 and 13 minutes proportionally per population. 

\begin{figure}[htb!]
	\centering
	\subfigure
	{
		\includegraphics[scale=0.225]{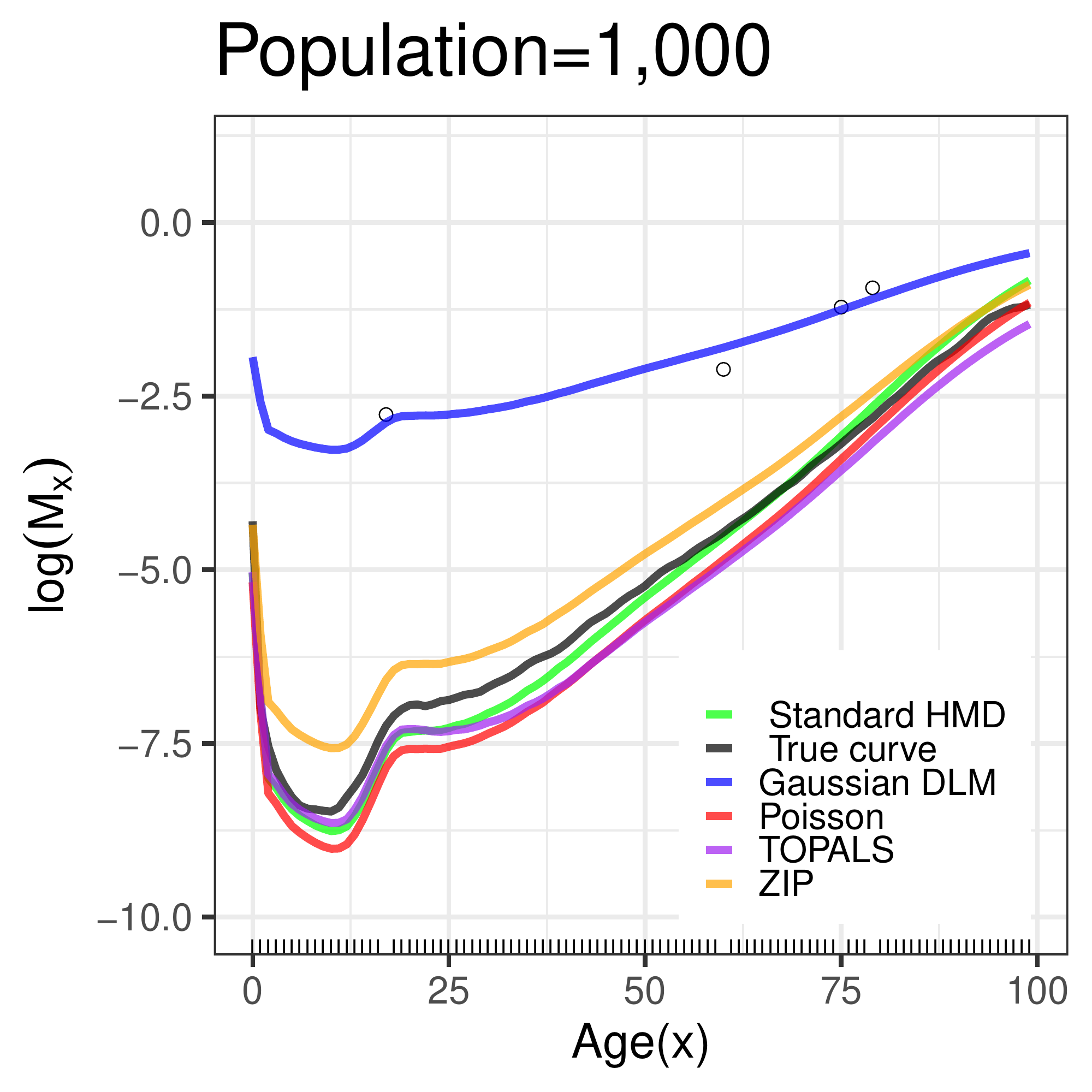}
	}
	\subfigure
	{
		\includegraphics[scale=0.225]{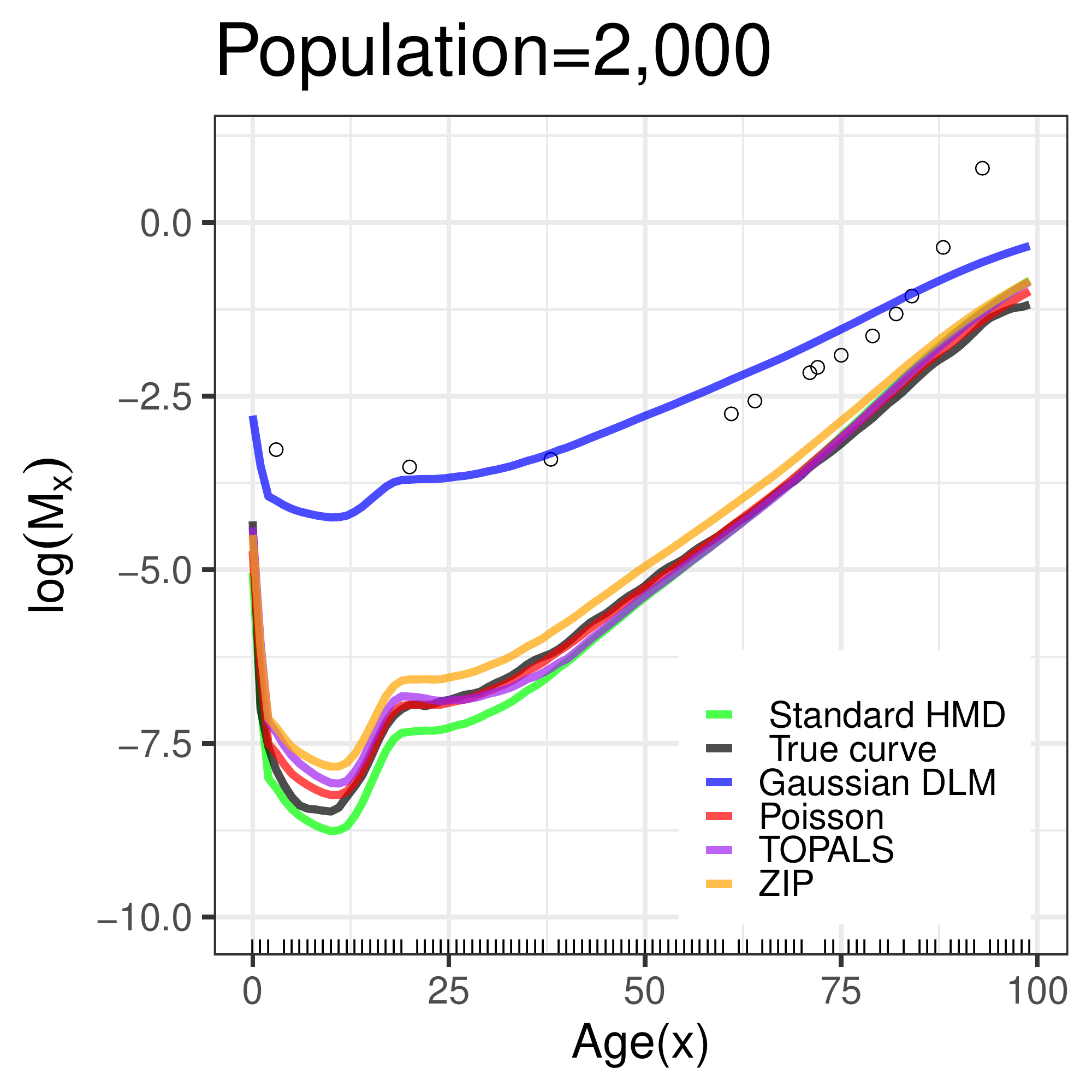}
	}\subfigure
	{
		\includegraphics[scale=0.225]{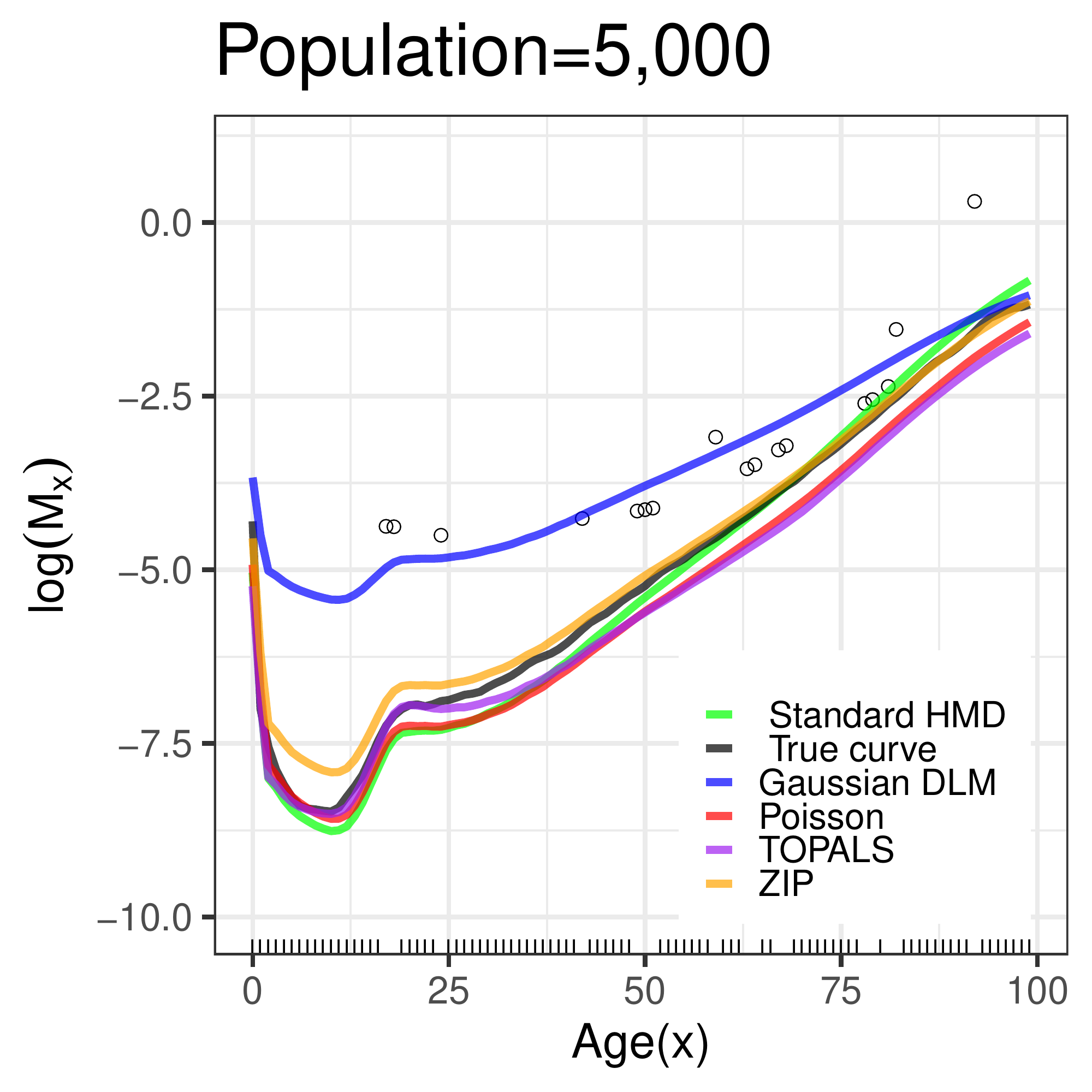}
	} \\
	\subfigure
	{
		\includegraphics[scale=0.225]{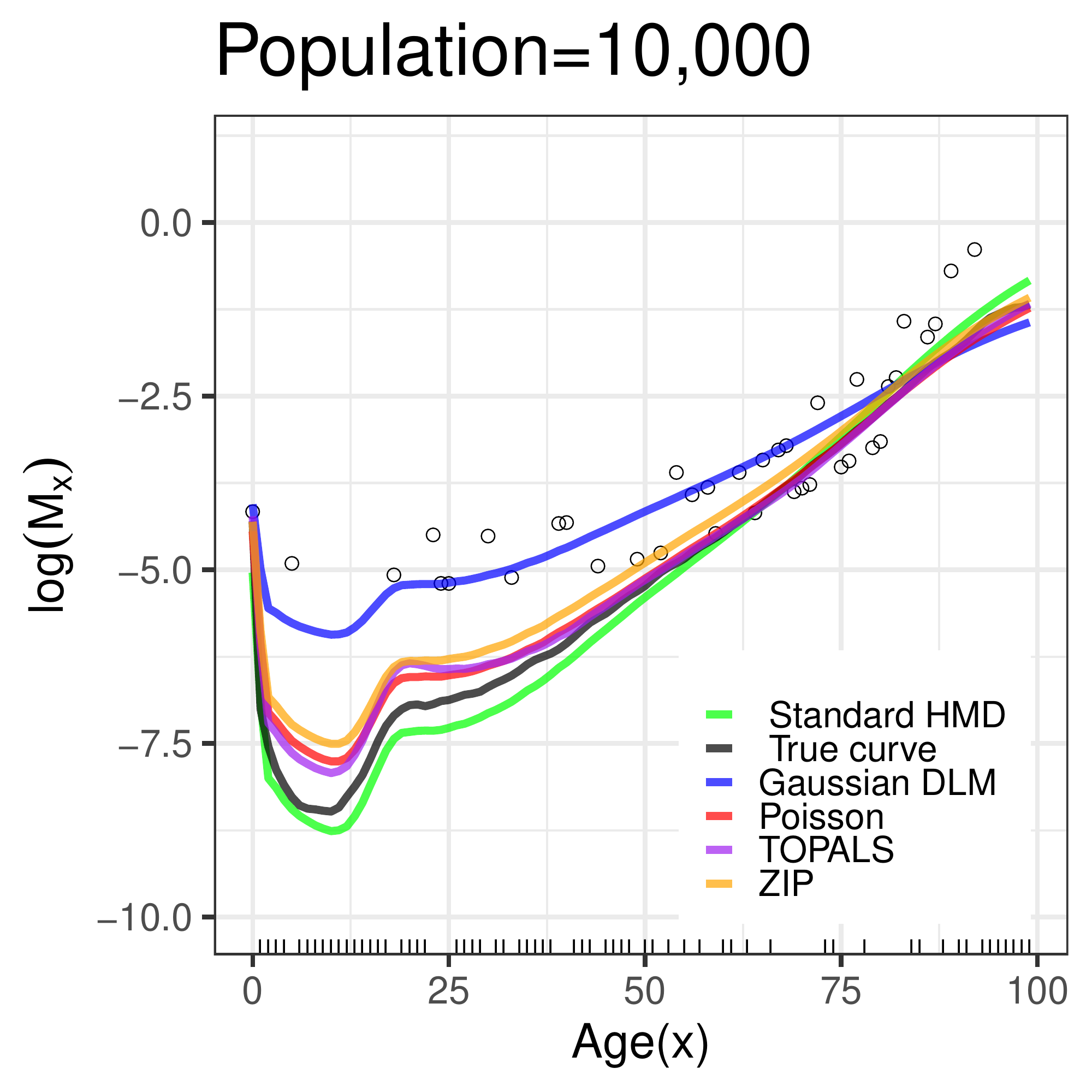}
	}
	\subfigure
	{
		\includegraphics[scale=0.225]{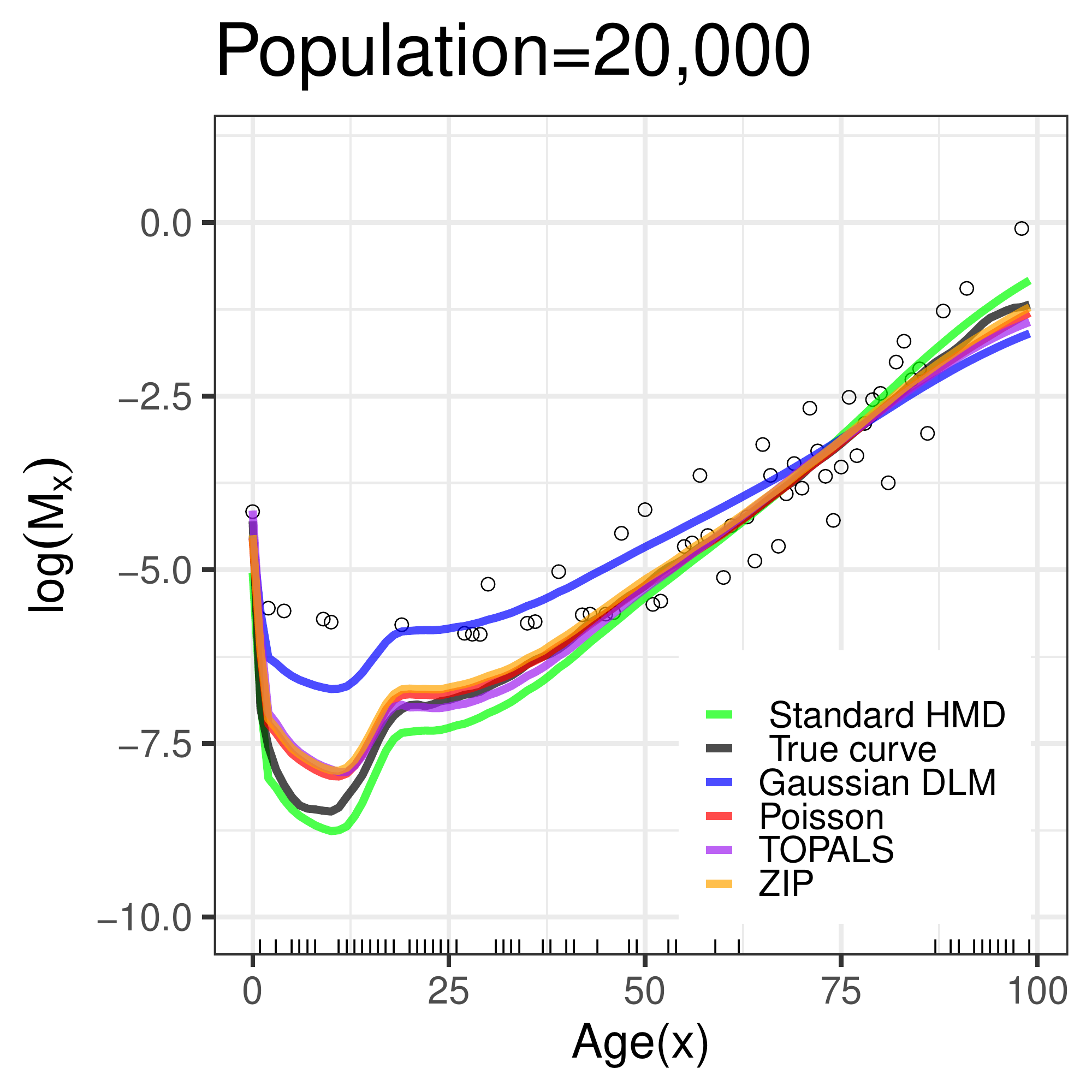}
	}\subfigure
	{
		\includegraphics[scale=0.225]{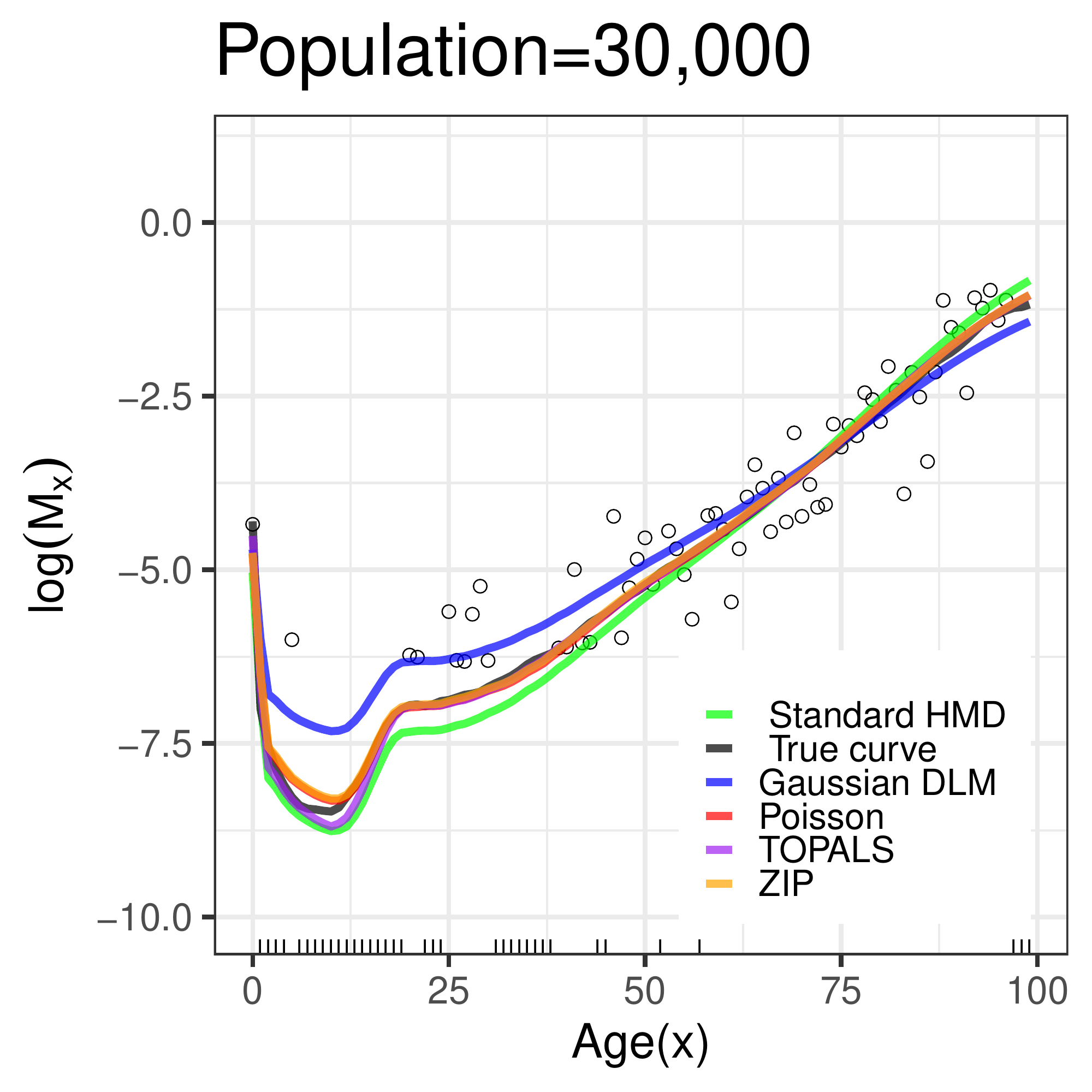}
	} \\
	\subfigure
	{
		\includegraphics[scale=0.225]{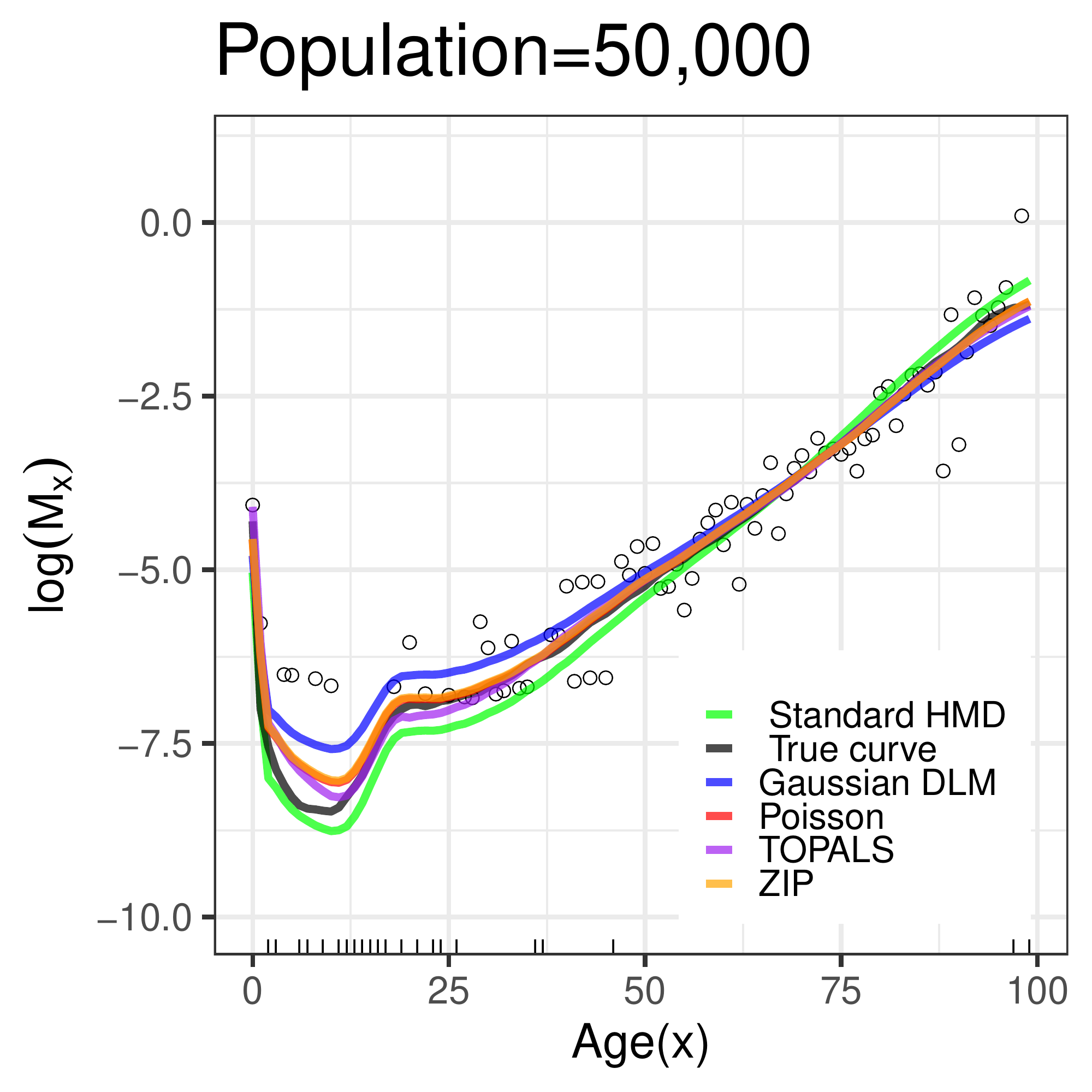}
	}
	\subfigure
	{
		\includegraphics[scale=0.225]{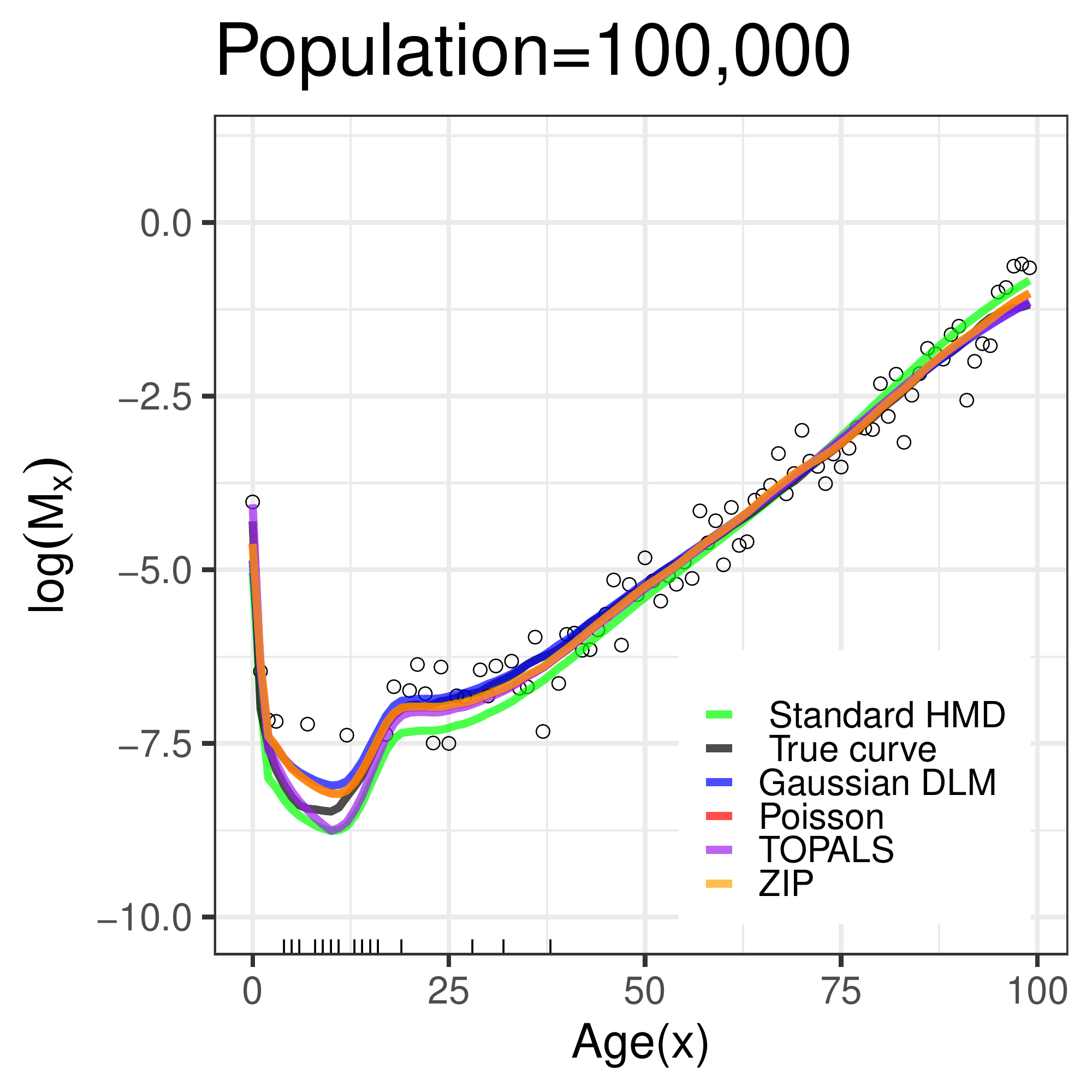}
	}\subfigure
	{
		\includegraphics[scale=0.225]{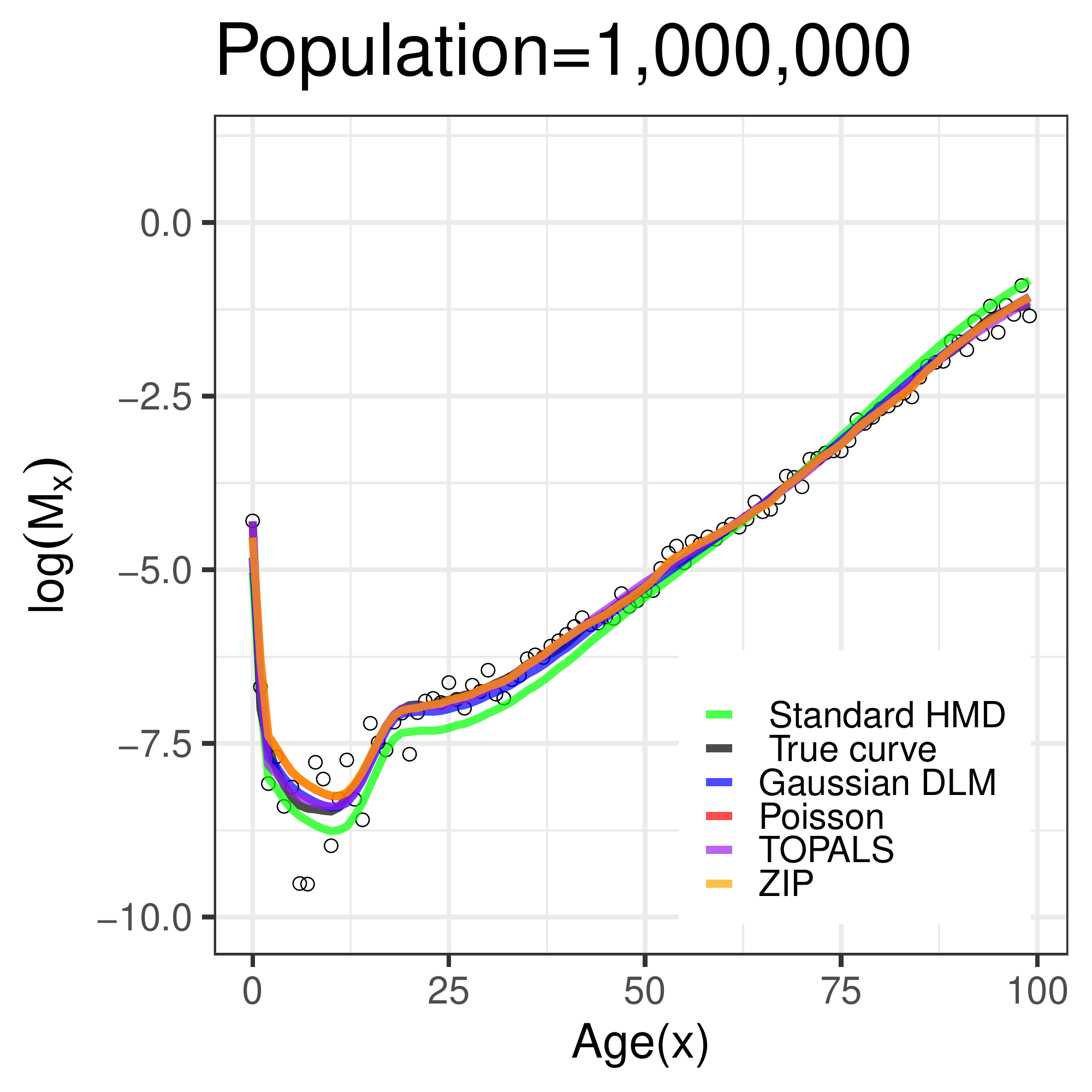}
	} \\
	\caption{Simulated mortality schedules (in the log scale) for selected population sizes. Open circles represent the log-mortality rate for each single-year of age. Tick marks on the horizontal axis represent ages with no observed deaths. Black curve represents the true underlying mortality rates considered to generate the datasets. The simulated datasets were fitted under the three models described in Section \ref{model:mortality_curves_models}: the dynamic Poisson model (red curve), the Gaussian DLM (blue curve) and the TOPALS model (purple curve). The green curve represents the standard mortality schedule assumed in all models (HMD, 2015). Orange curve represents the fit of a zero-inflated Poisson model not discussed in Section 2 yet.}
	\label{fig:mortality_curves_simulations}
\end{figure}

The log-mortality rates observed for each simulated population are showed in Figure  \ref{fig:mortality_curves_simulations} along with the estimates provided by the three models. 
The same standard mortality schedule obtained from the HMD is considered when fitting the models.
All models provided mortality schedules that evolves smoothly across ages, even for populations with a high frequency of null counts and  a high noise in the observed data.
The Gaussian DLM only provides a reasonable fit in large populations (greater than 100,000 inhabitants) with quite smooth mortality curves. 
In small populations, there are a great number of age intervals for which the number of deaths is zero, which are treated as missing data when fitting the Gaussian model.
Such data feature imposes a restriction on the Gaussian model when applied for small populations. 

%

Figure \ref{fig:mortality_curves_simulations} shows that, as expected, Poisson and TOPALS models perform better than the Gaussian DLM in small populations.
In general, these models tend to present similar shapes for the mortality schedules with more remarkable discrepancies in young ages, around the ''bathtub'' pattern, and in the oldest ages. 
Such discrepancies are more evident in populations sized 1,000 and 5,000 with a visually better performance of the TOPALS model in such cases.
For the population of size 2,000 the discrepancy between the two estimates is only evident in the latest ages, with a better performance of the Poisson model.
For populations with 10,000 and 100,000 inhabitants the results are less similar for ages around the ''bathtub'' pattern of the mortality schedules.
Poisson and TOPALS models provided almost the same fit for populations with sizes 20,000, 50,000 and 1,000,000 .

\clearpage
The Poisson model seems to be more influenced by outliers than TOPALS model, specially in small populations.
For instance, in the population with size 1,000 a single point was observed for younger ages. This point seems to be very influential, pulling the curve upwards in previous ages. 
Similarly, the mortality curve tends to be pulled downwards whenever a high frequency of null counts is observed in the latest ages.
The influence of the mortality standard seems to be stronger in TOPALS model, making smaller the influence of atypical observations. 
That is a point which worth further investigation.

In addition to the visual analysis available through Figure \ref{fig:mortality_curves_simulations}, we present in Table \ref{tab_simul_cap3} some evaluation metrics comparing the three fitted models. 
We highlighted in bold the model presenting the smallest value for each metric.
In general, Poisson and TOPALS models present the smallest values for the evaluation metrics.
TOPALS performs better in the most populations according to $Rbias$ and $\sqrt{MSE}$ whereas the Poisson model figures as the best fitting model according to $MAPE$. 

\begin{table}[htb!]
	\centering
	\caption{Relative bias ($RBias$), square root of the mean squared error ($\sqrt{MSE}$) and mean absolute percentage error ($MAPE$) for the estimated log-mortality schedules in each simulated populations.}\vspace{0.1cm}
	\tabcolsep=4.25pt 
	\begin{tabular}{cccccccccccc}
		\hline
		Population     & $RBias$ & $\sqrt{MSE}$ & $MAPE$ & & $RBias$ & $\sqrt{MSE}$ & $MAPE$& & $RBias$ & $\sqrt{MSE}$ & $MAPE$ \\
		\hline
		& \multicolumn{3}{c}{Poisson}	& & \multicolumn{3}{c}{TOPALS}	& & \multicolumn{3}{c}{Gaussian DLM} \\ 
		\cline{2-4} \cline{6-8} \cline{10-12}
		1,000   &-0.183 &0.640  &\textbf{0.078} &  &\textbf{-0.108}  &\textbf{0.437} &0.105 &  & 0.601 &3.304 &0.598\\
		2,000     &\textbf{-0.007} &\textbf{0.201}  &\textbf{0.012} &  &0.025  &0.238 &0.032  &  &0.503 &2.636 &0.485\\
		5,000     &-0.143 &0.512  &\textbf{0.063} &  &\textbf{-0.111}  &\textbf{0.390} &0.077  & &0.262 &1.684 &0.273\\
		10,000    &\textbf{0.004} &0.394   &0.049 &  &0.020  &\textbf{0.309} &\textbf{0.020}  &  &0.159 &1.344 &0.212\\
		20,000   &\textbf{-0.012} &\textbf{0.254}   &\textbf{0.011} & &-0.034  &0.269 &0.019  & &0.056 &0.856 &0.131\\
		30,000 	&0.006 &0.104   &\textbf{0.007} &  &\textbf{0.004}  &\textbf{0.091} &0.008  &  &0.029 &0.531 &0.075\\
		50,000 	&0.008 &0.179   &\textbf{0.010} &  &\textbf{-0.001}  &\textbf{0.174} &0.011  &  &0.013 &0.396 &0.050\\
		100,000   &0.007 &\textbf{0.133}   &\textbf{0.014} & &\textbf{-0.005}  &\textbf{0.133} &0.018  &  &0.008 &0.156 &0.009\\
		1,000,000   &0.004 &0.087   &\textbf{0.006} &  &\textbf{-0.001}  &\textbf{0.057} &\textbf{0.006}  &  &\textbf{-0.001} &0.103 & 0.014\\
		\hline
		\hline     
	\end{tabular}
	\label{tab_simul_cap3}
\end{table}

\section{Application to Brazilian Municipalities}\label{sec:application_mortality_curves}

In this section we analyze data from selected Brazilian municipalities over calendar years 2009-2011. 
Since some municipalities have small populations, we only consider Poisson and TOPALS models to fit the data.
In both models, we assume the same standard mortality schedule obtained from the HMD, as in the simulation study.
Population and deaths for the 5,565 Brazilian municipalities were obtained from the \cite{Gonzaga2016}'s project website   {http://topals-mortality.schmert.net}. 
The data is available by 100 single-year ages and separately by sex. 
Those authors collected the municipal data from the Demographic Census (2010) and from the Ministry of Health's Mortality Information System (SIM/Datasus), respectively. 
The 2010 census populations is used to estimate age- and sex-specific exposure populations over 2009-2011. 
It is worth noting that in the complete dataset, despite using three years of exposure, 49.2\% of the 1,113,000 combinations of municipality, age and sex cells are null in number of deaths. 

\begin{figure}[htb!]
	\centering
	\subfigure
	{
		\includegraphics[scale=0.225]{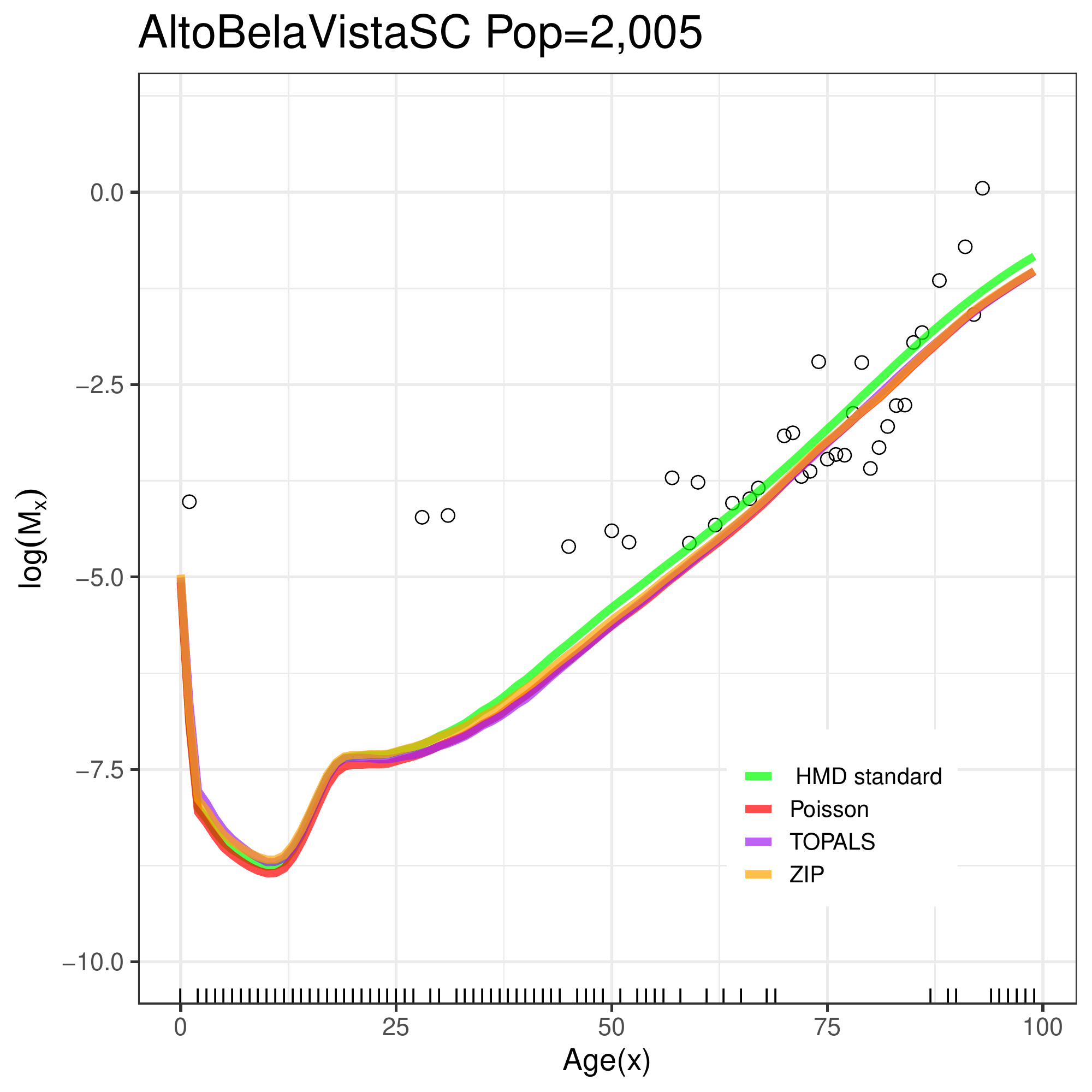}
	}
	\subfigure
	{
		\includegraphics[scale=0.225]{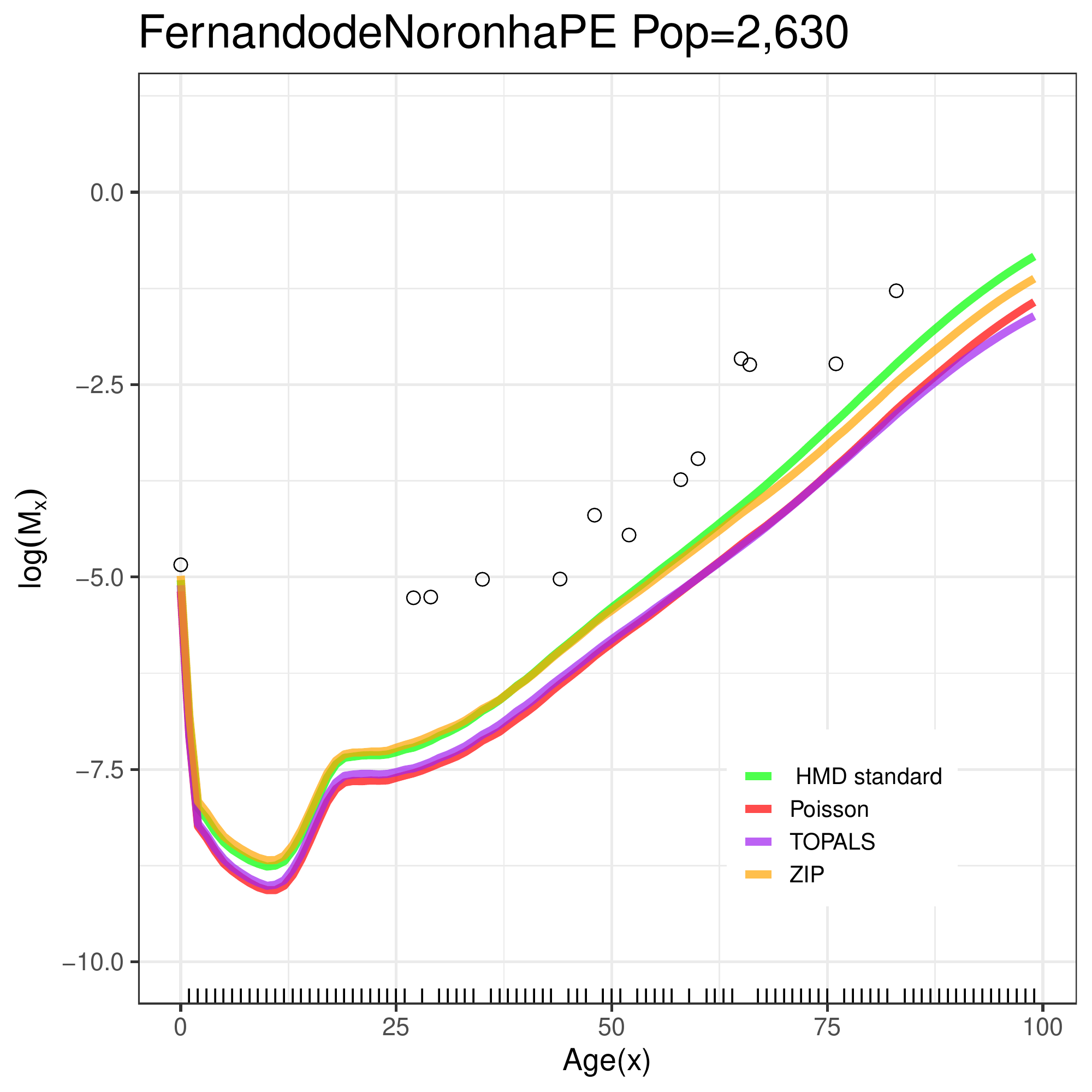}
	}\subfigure
	{
		\includegraphics[scale=0.225]{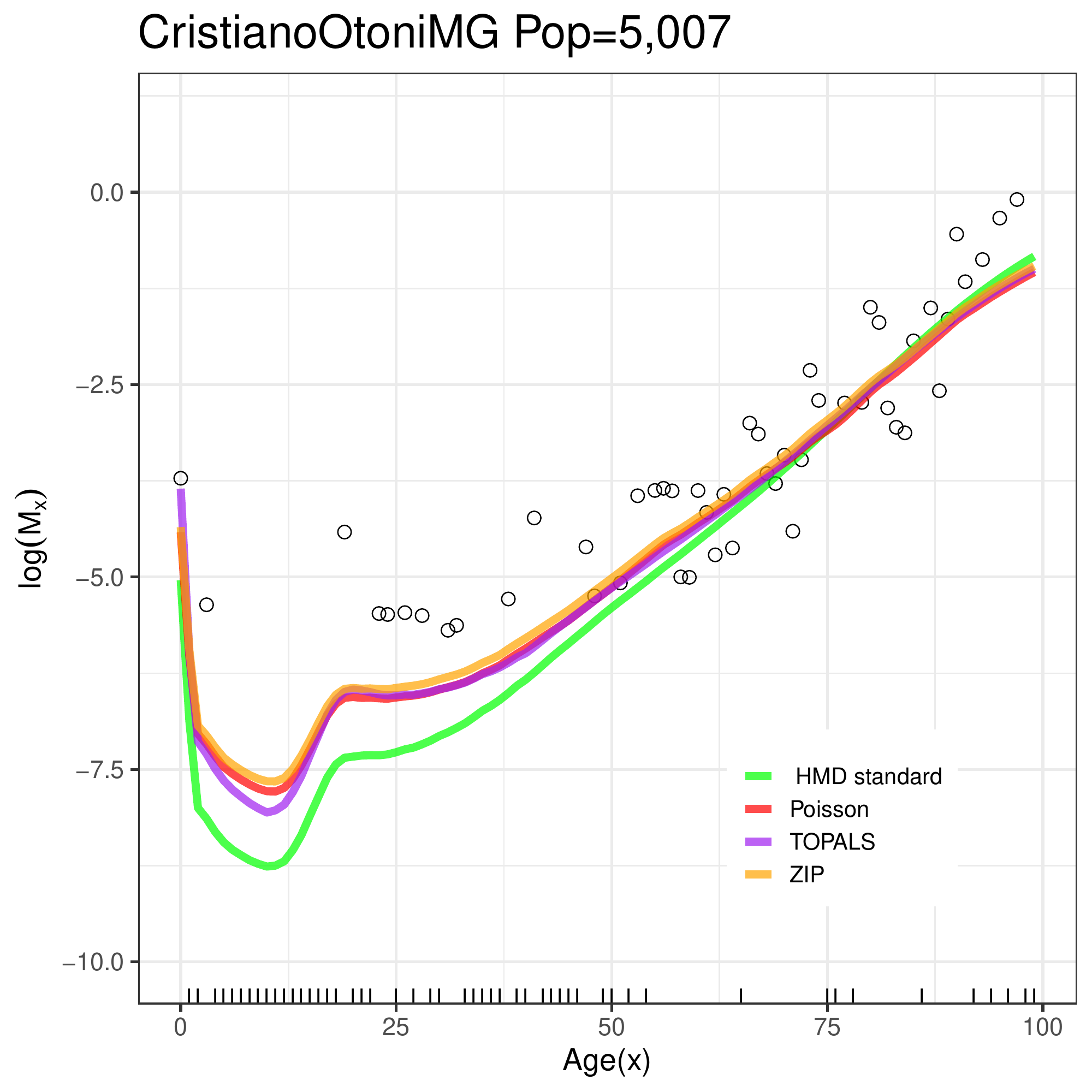}
	} \\
	\subfigure
	{
		\includegraphics[scale=0.225]{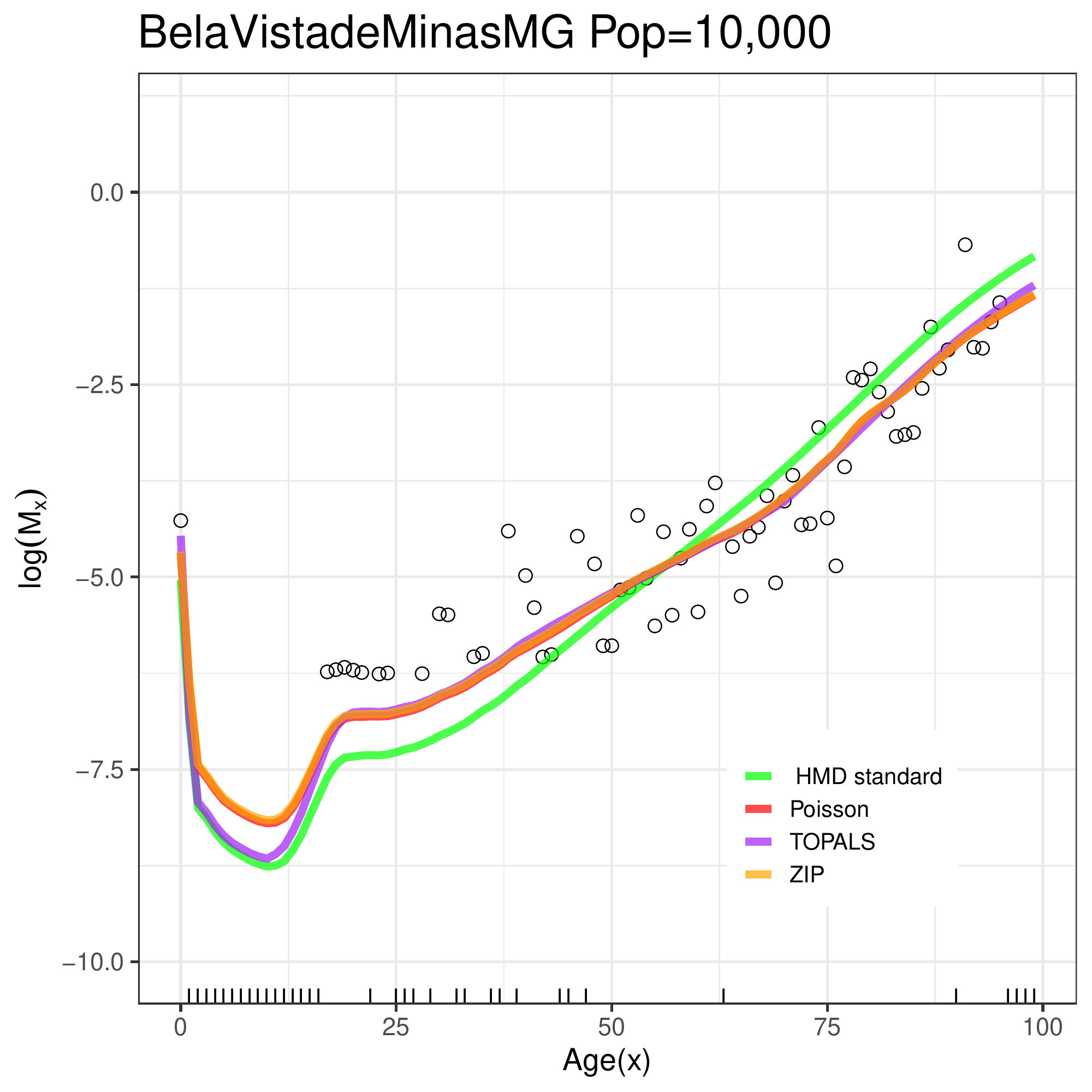}
	}
	\subfigure
	{
		\includegraphics[scale=0.225]{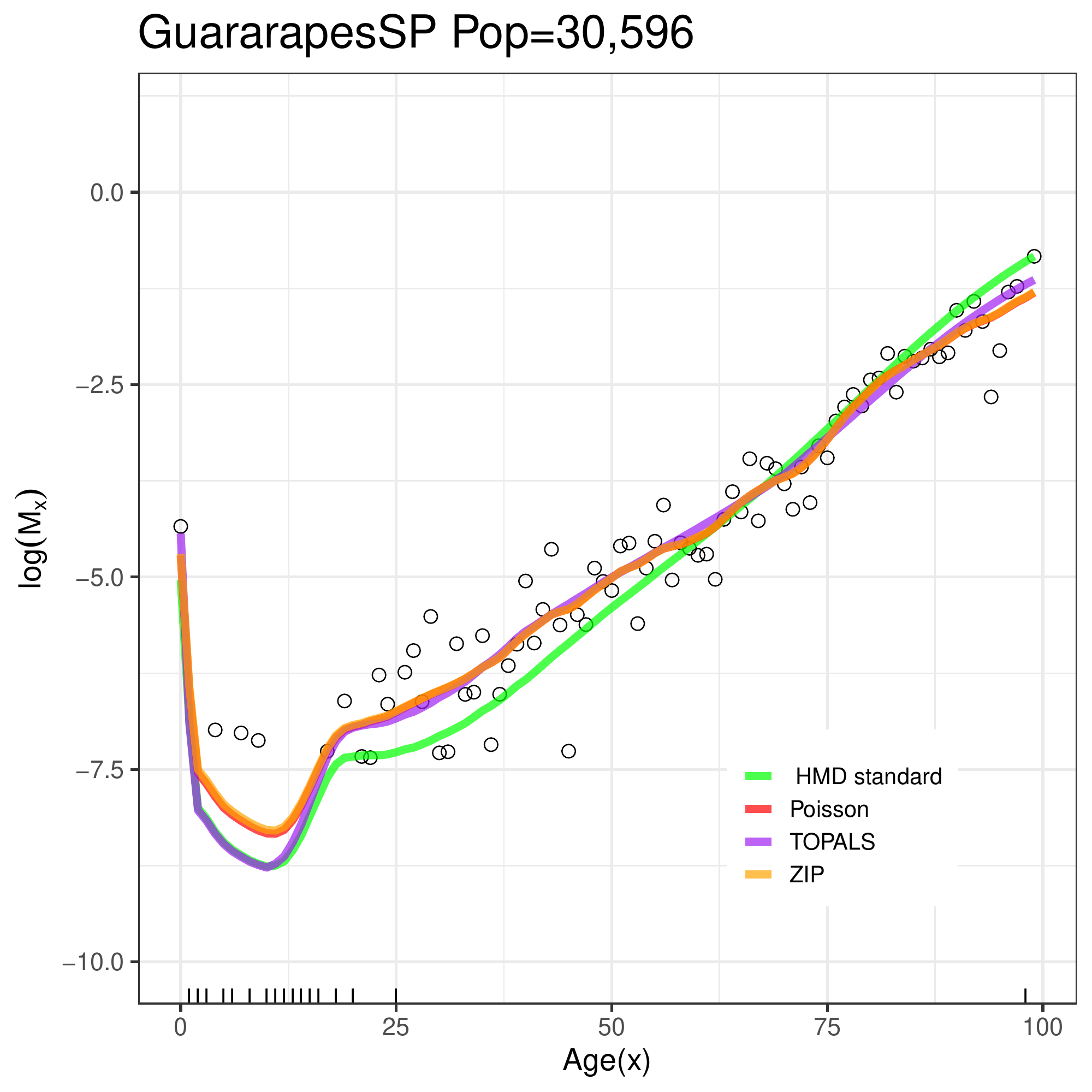}
	}\subfigure
	{
		\includegraphics[scale=0.225]{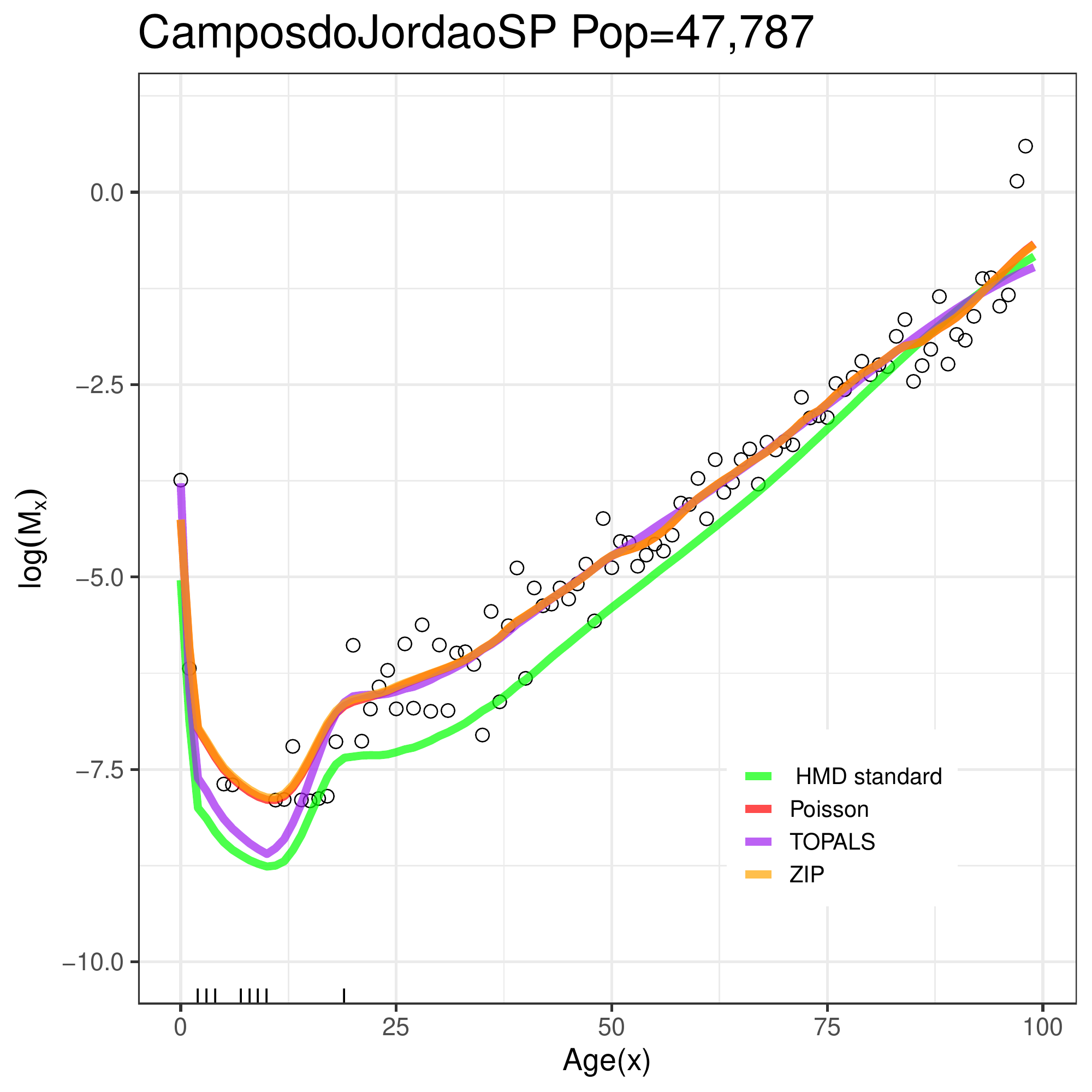}
	} \\
	\subfigure
	{
		\includegraphics[scale=0.25]{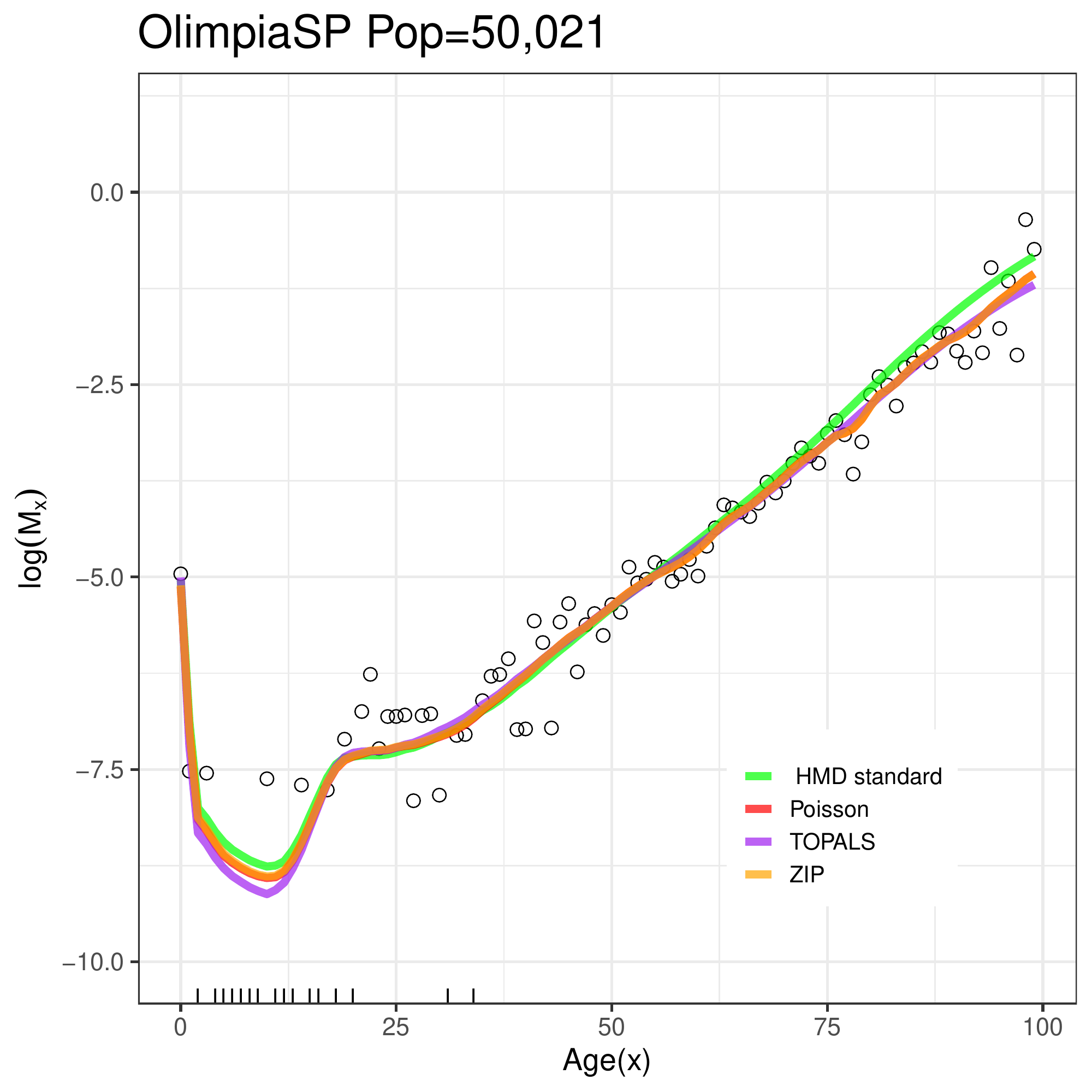}
	}
	\subfigure
	{
		\includegraphics[scale=0.225]{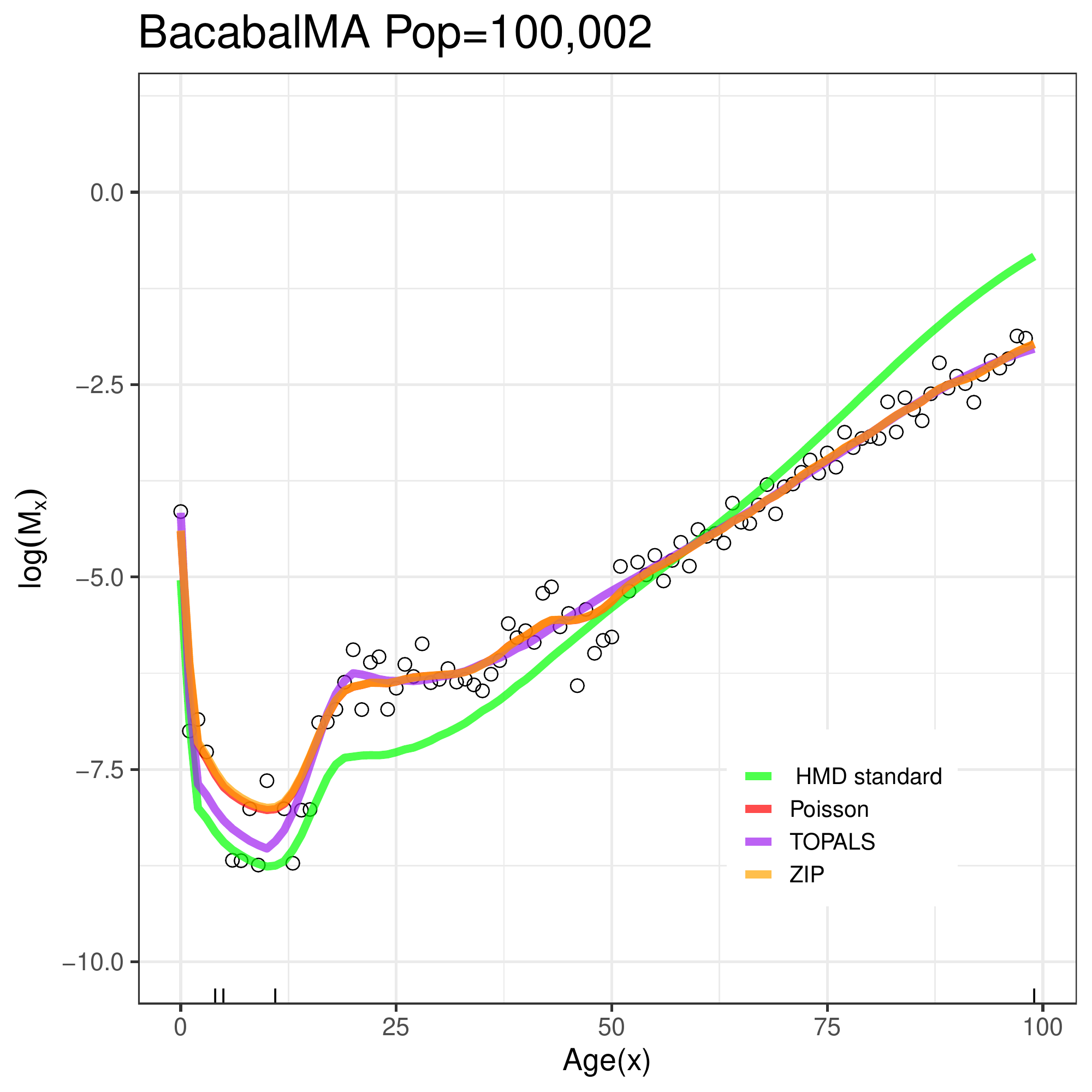}
	}\subfigure
	{
		\includegraphics[scale=0.225]{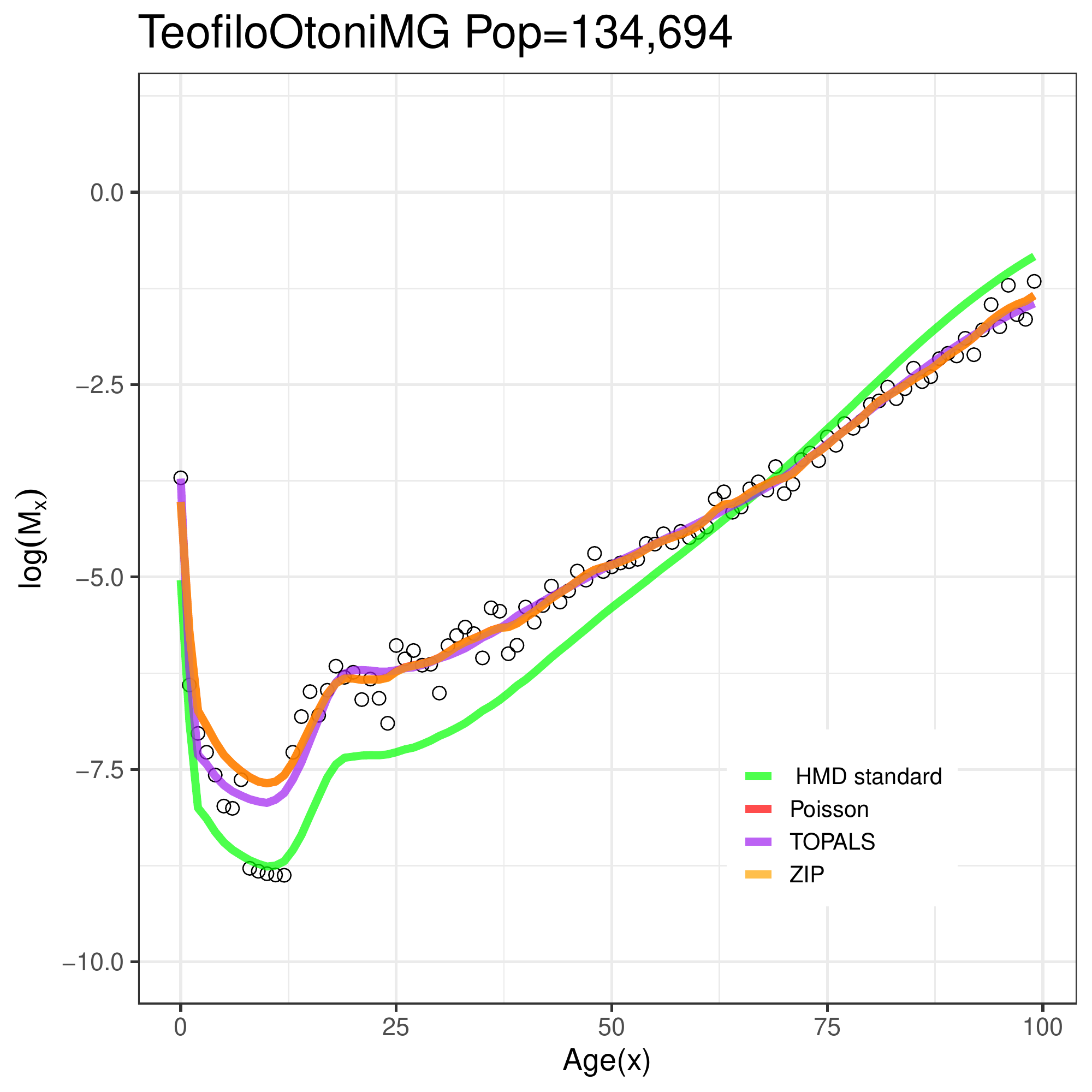}
	} \\	
	\subfigure
	{
		\includegraphics[scale=0.225]{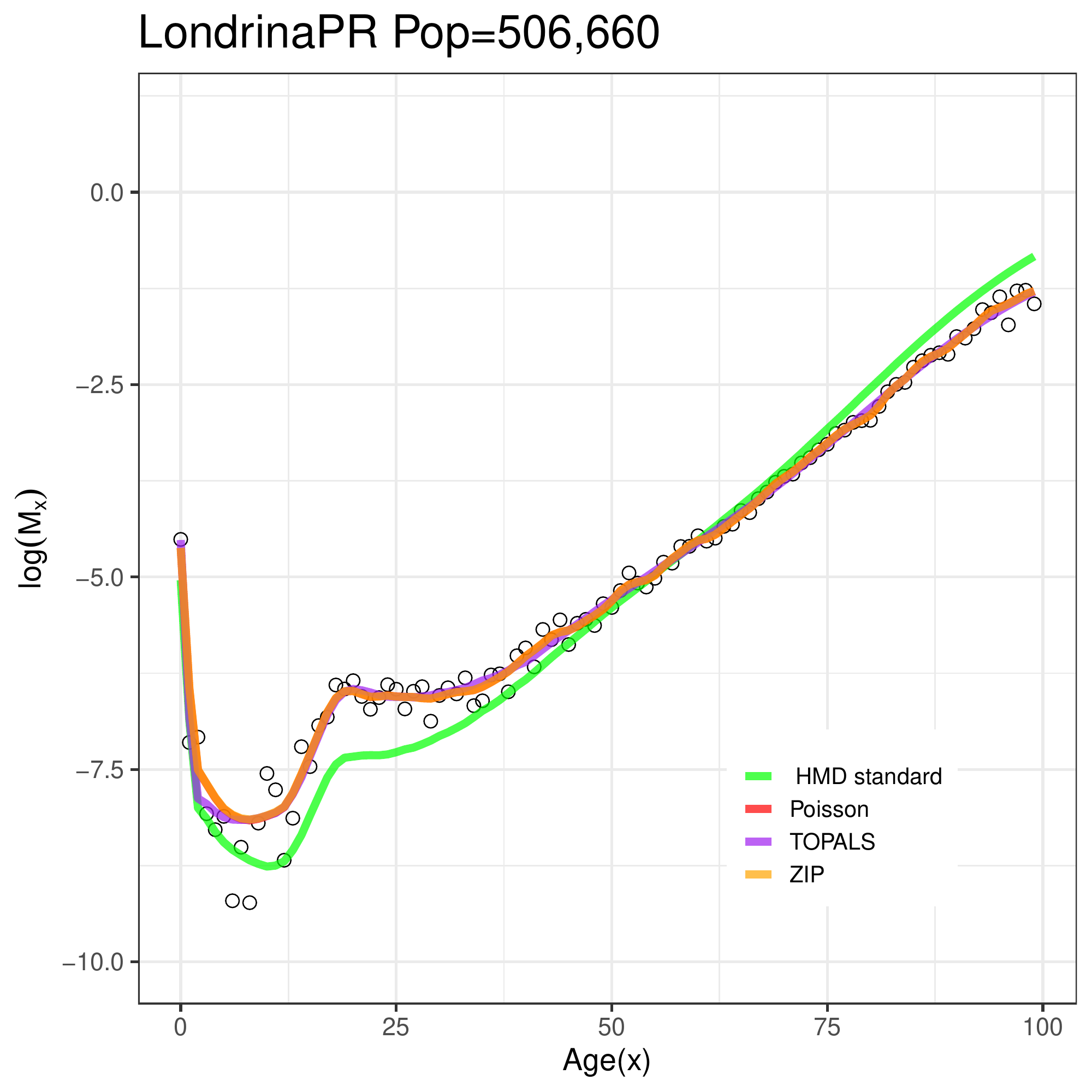}
	}
	\subfigure
	{
		\includegraphics[scale=0.225]{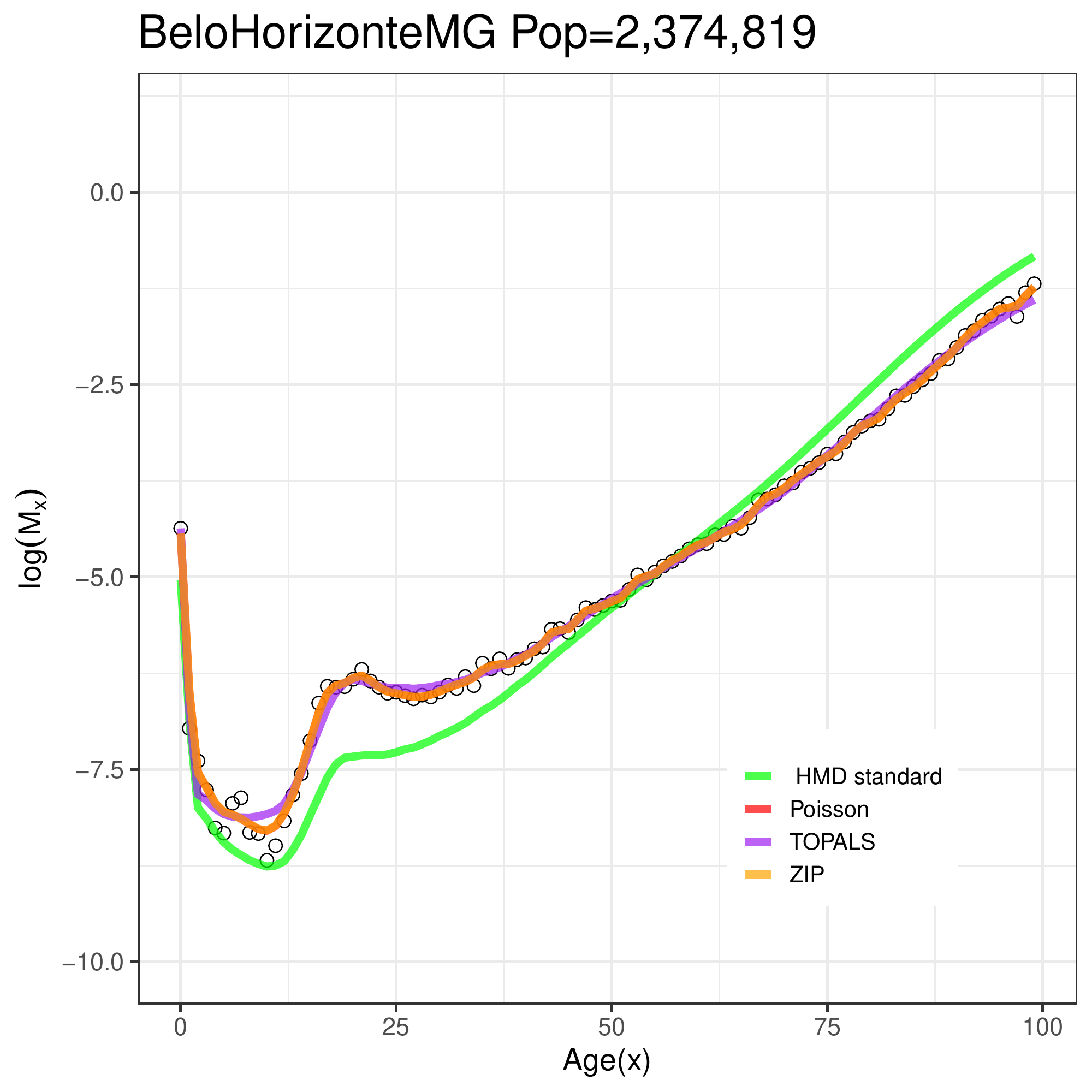}
	}\subfigure
	{
		\includegraphics[scale=0.225]{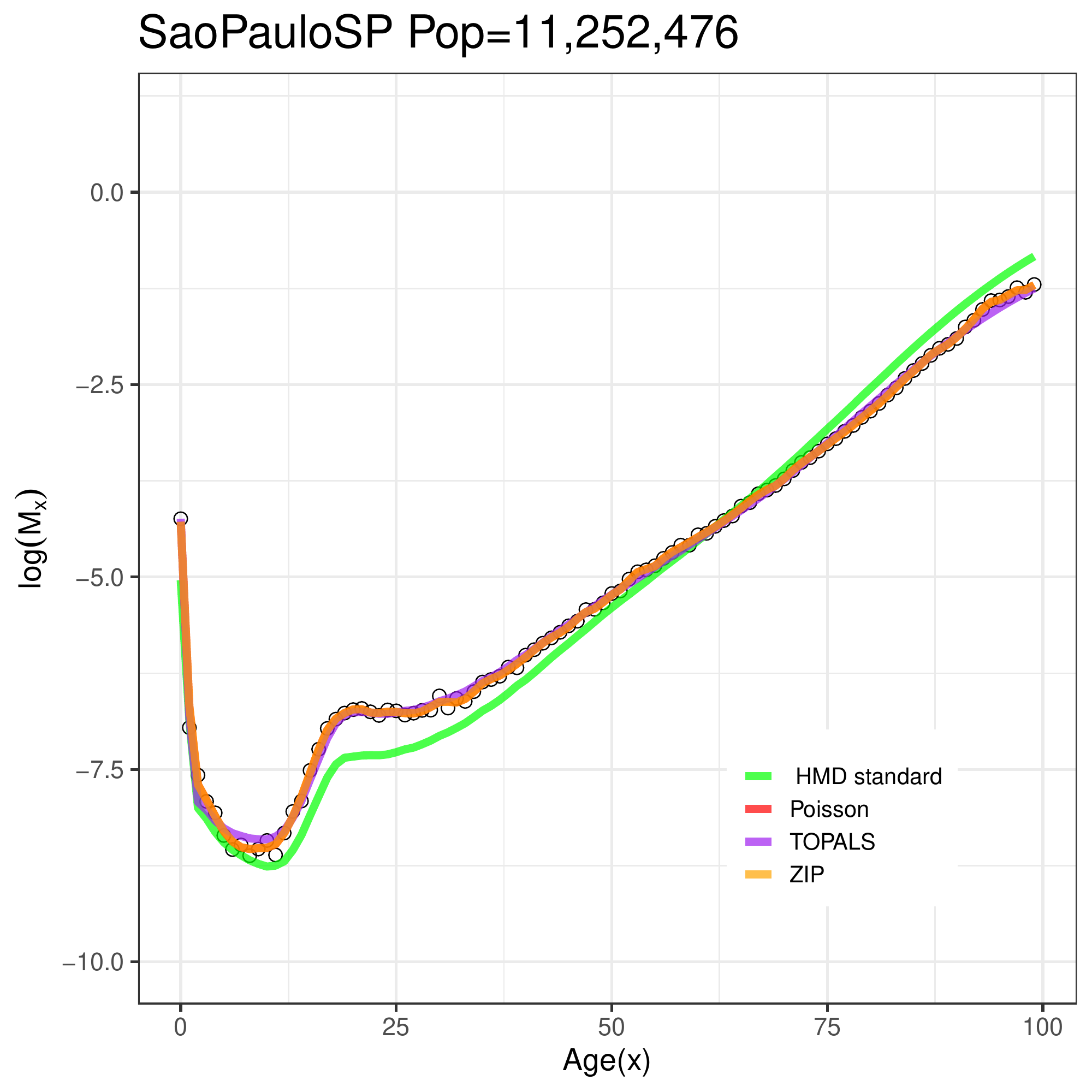}
	} \\
	\caption{Estimated mortality schedules (in the log scale) for selected Brazilian municipalities, both sexes. Open circles represent the observed log-mortality rate for each single-year of age. Tick marks on the horizontal axis represent ages with no observed deaths. The observed data were fitted under the dynamic Poisson model (red curve) and the TOPALS model (purple curve). Green curve represents the standard mortality schedule assumed in both models (HMD, 2015). Orange curve represents the fit of a zero-inflated Poisson model not discussed in Section 2 yet.}
	\label{fig:mortality_curves_application_both}
\end{figure}

\begin{figure}[htb!]
	\centering
	\subfigure
	{
		\includegraphics[scale=0.225]{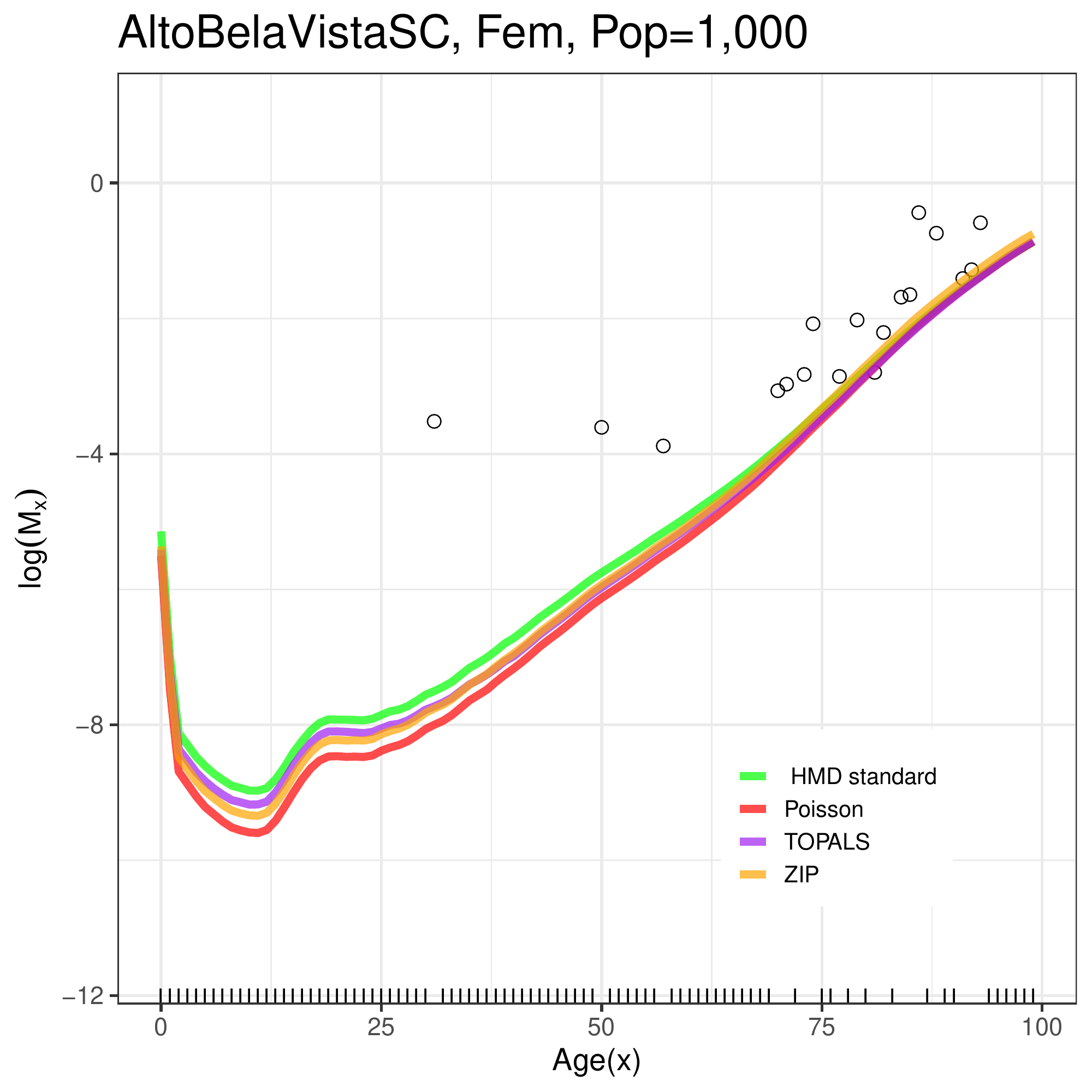}
	}
	\subfigure
	{
		\includegraphics[scale=0.225]{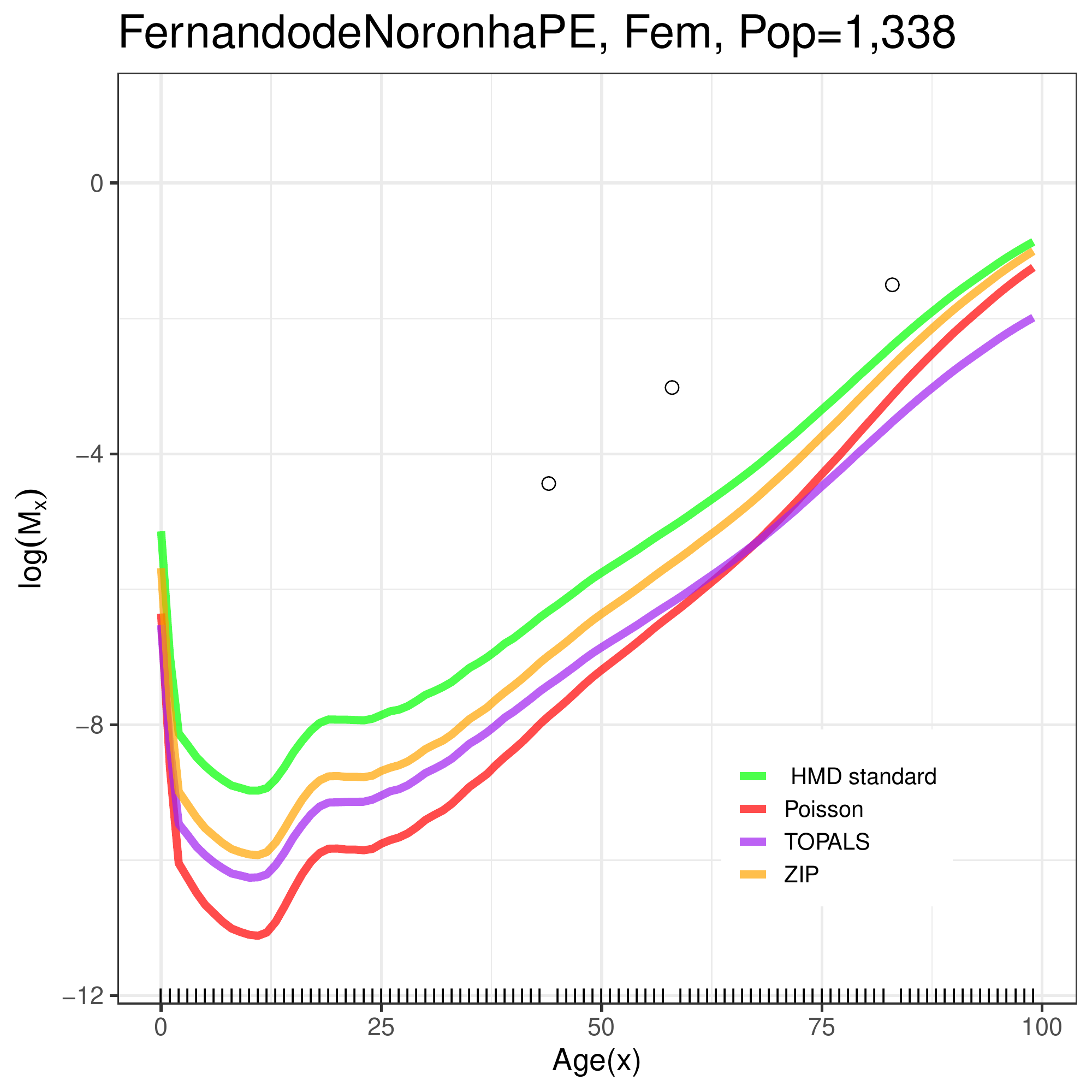}
	}\subfigure
	{
		\includegraphics[scale=0.225]{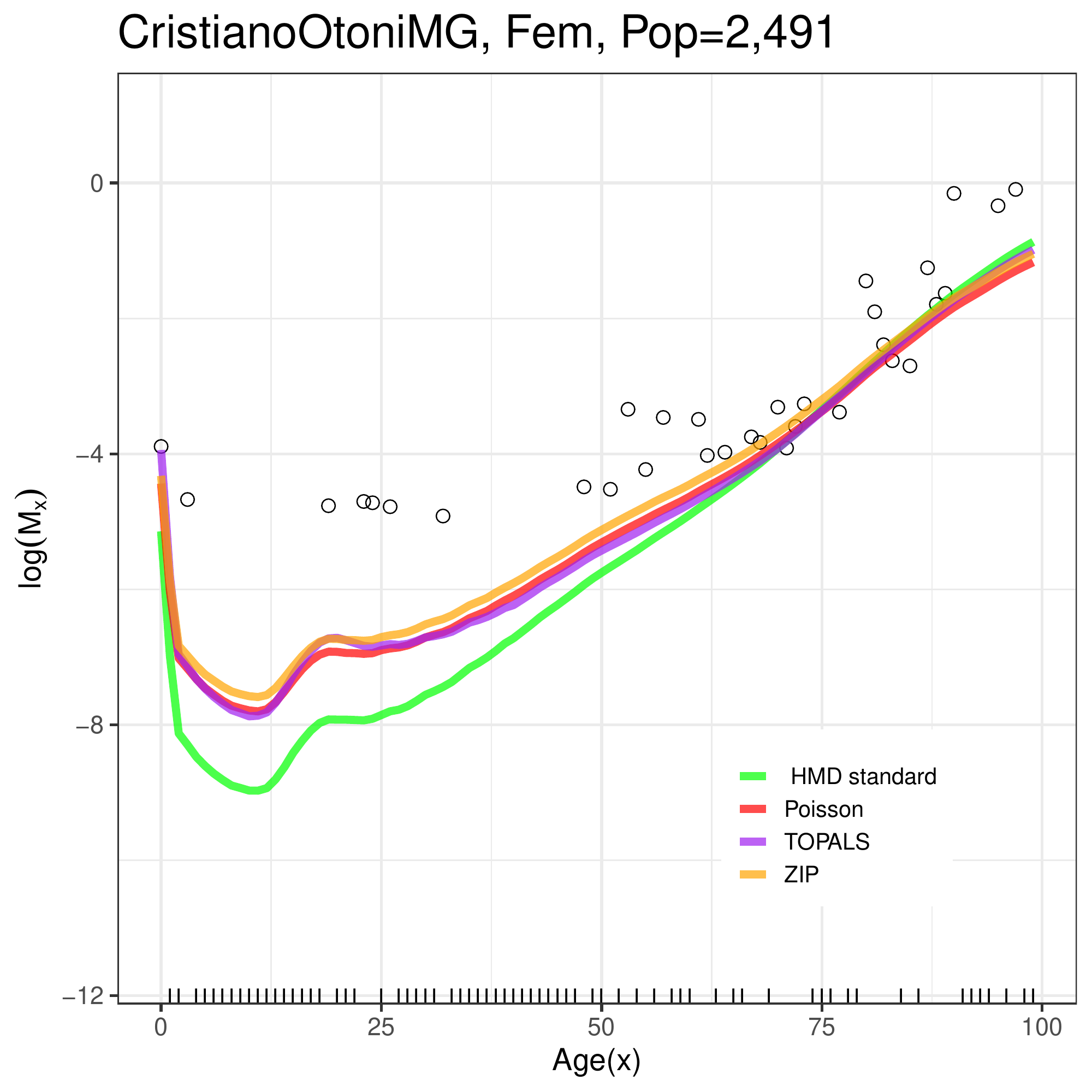}
	} \\
	\subfigure
	{
		\includegraphics[scale=0.225]{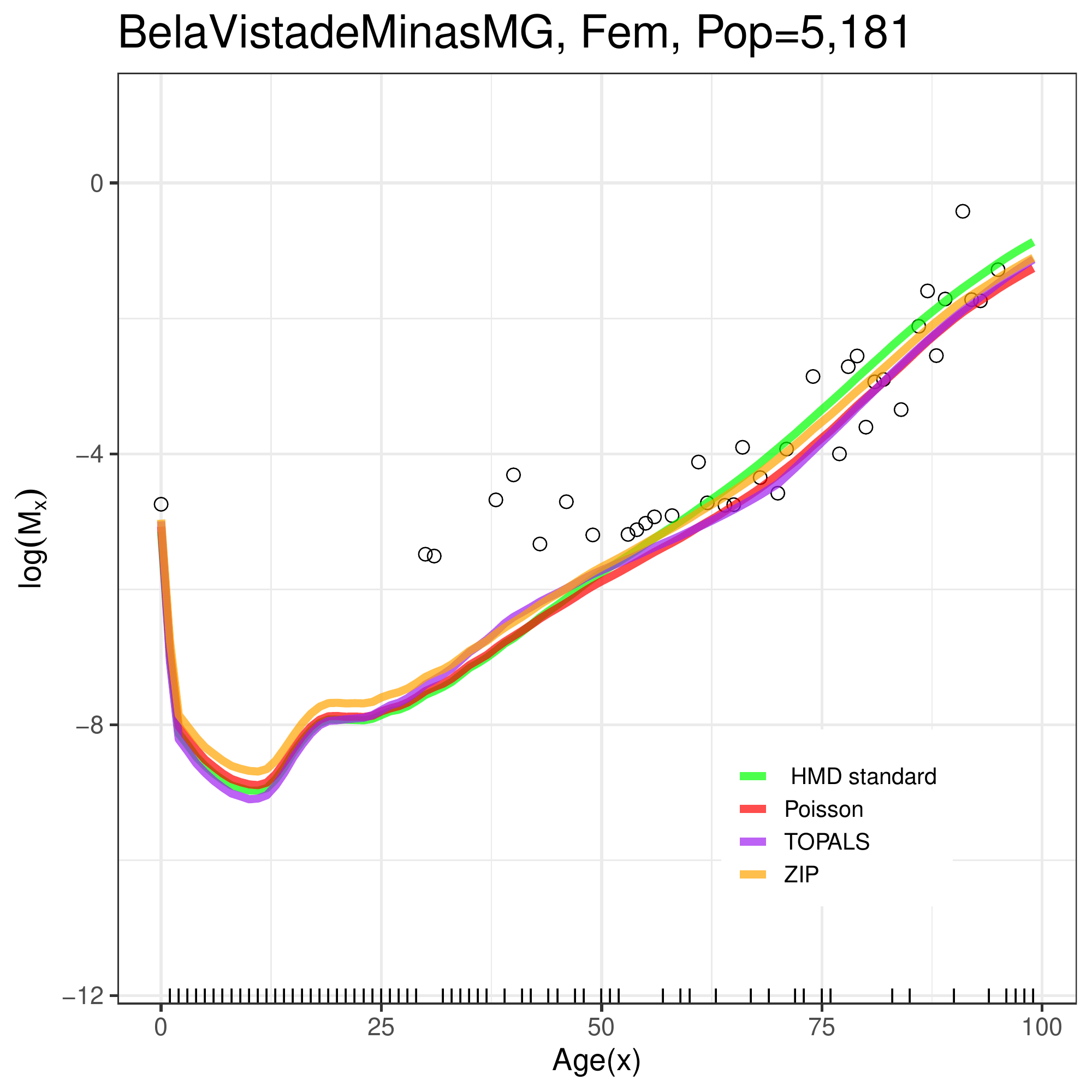}
	}
	\subfigure
	{
		\includegraphics[scale=0.225]{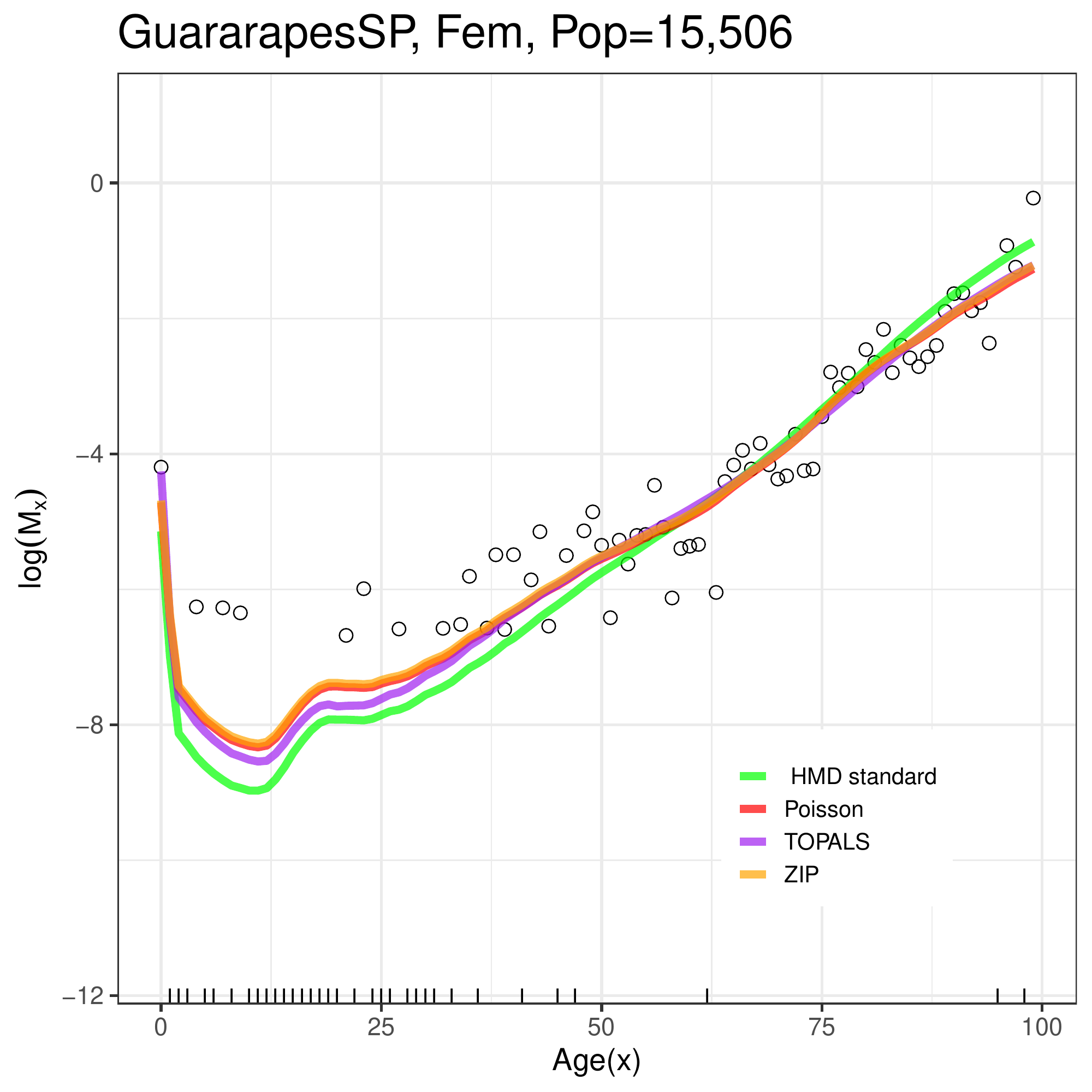}
	}\subfigure
	{
		\includegraphics[scale=0.225]{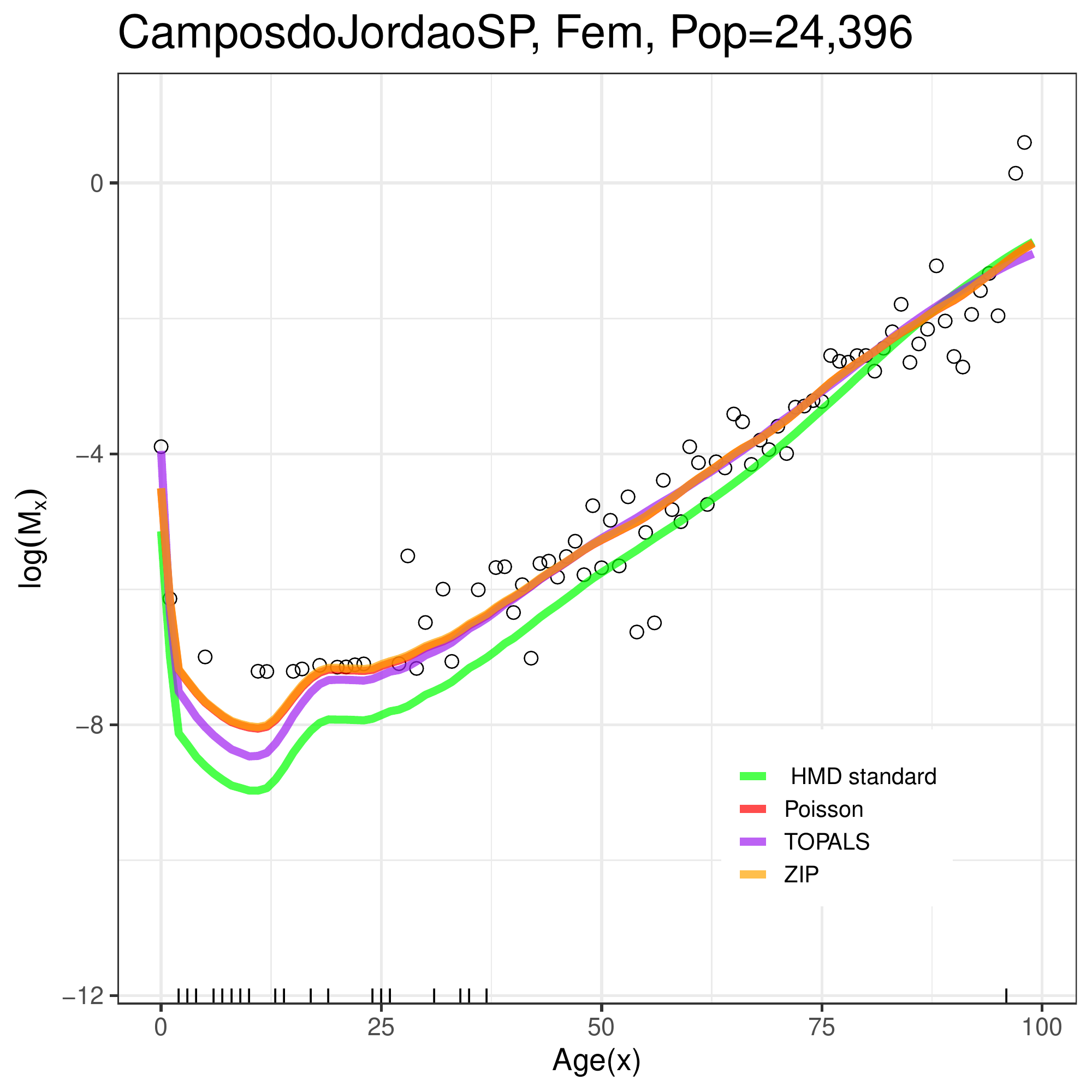}
	} \\
	\subfigure
	{
		\includegraphics[scale=0.225]{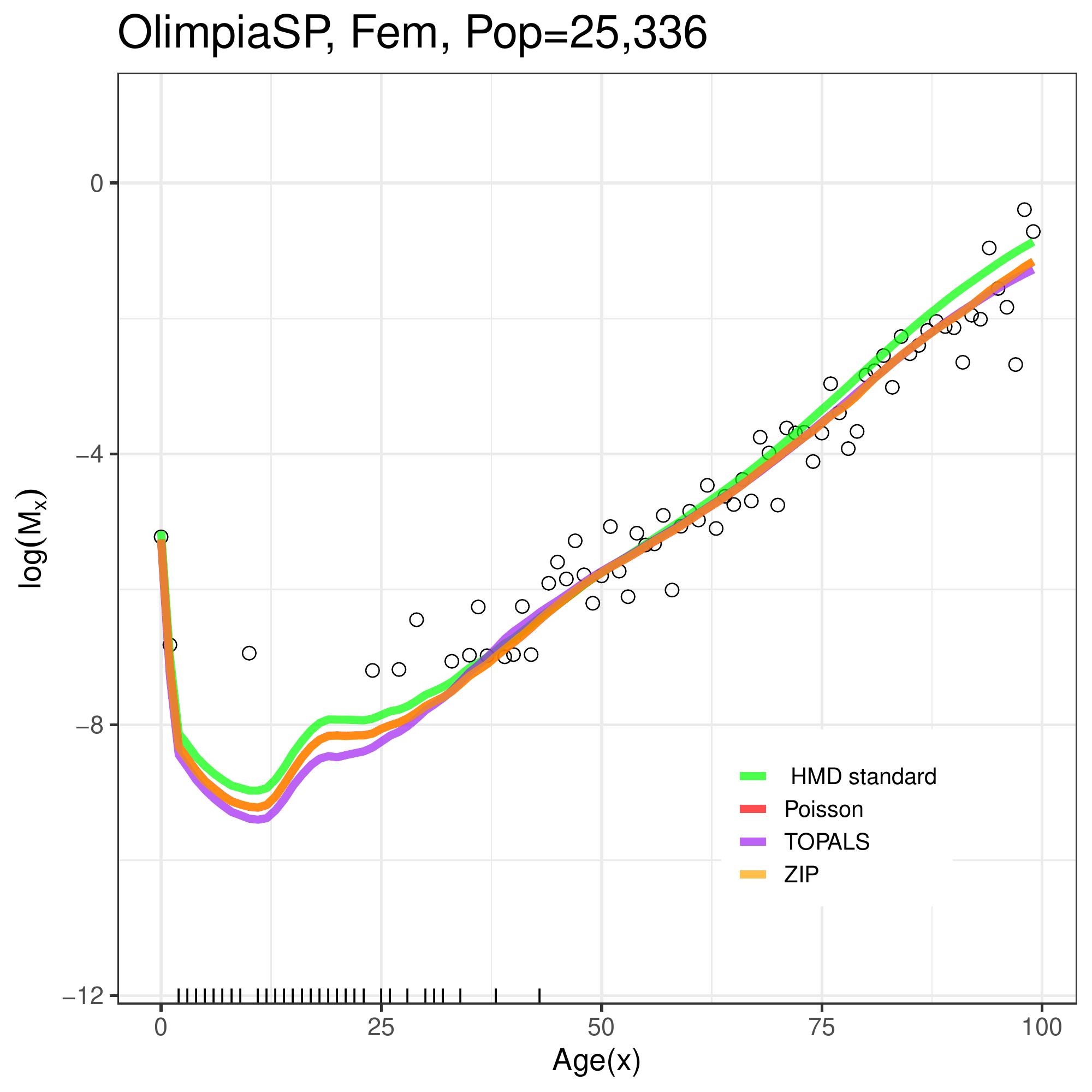}
	}
	\subfigure
	{
		\includegraphics[scale=0.225]{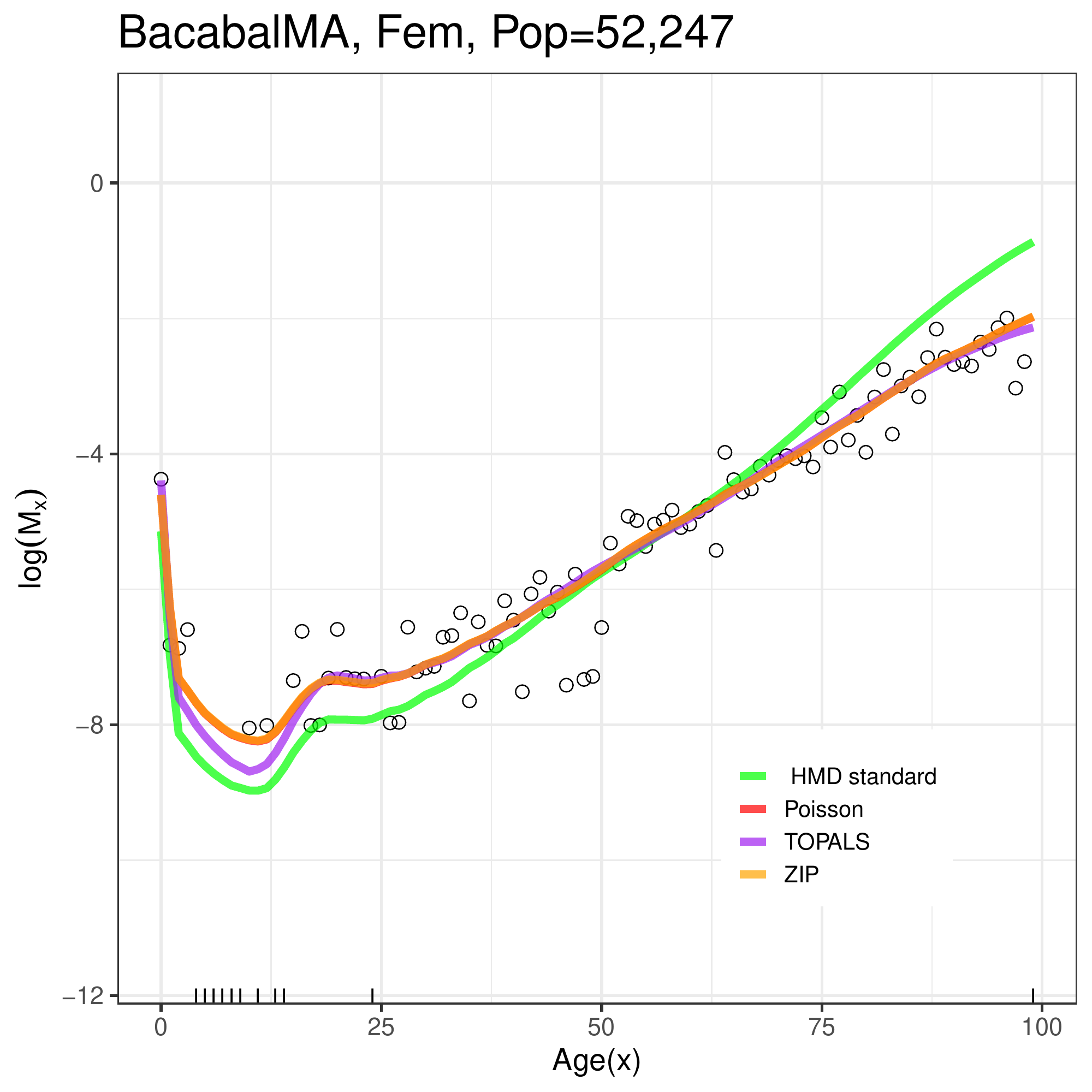}
	}\subfigure
	{
		\includegraphics[scale=0.225]{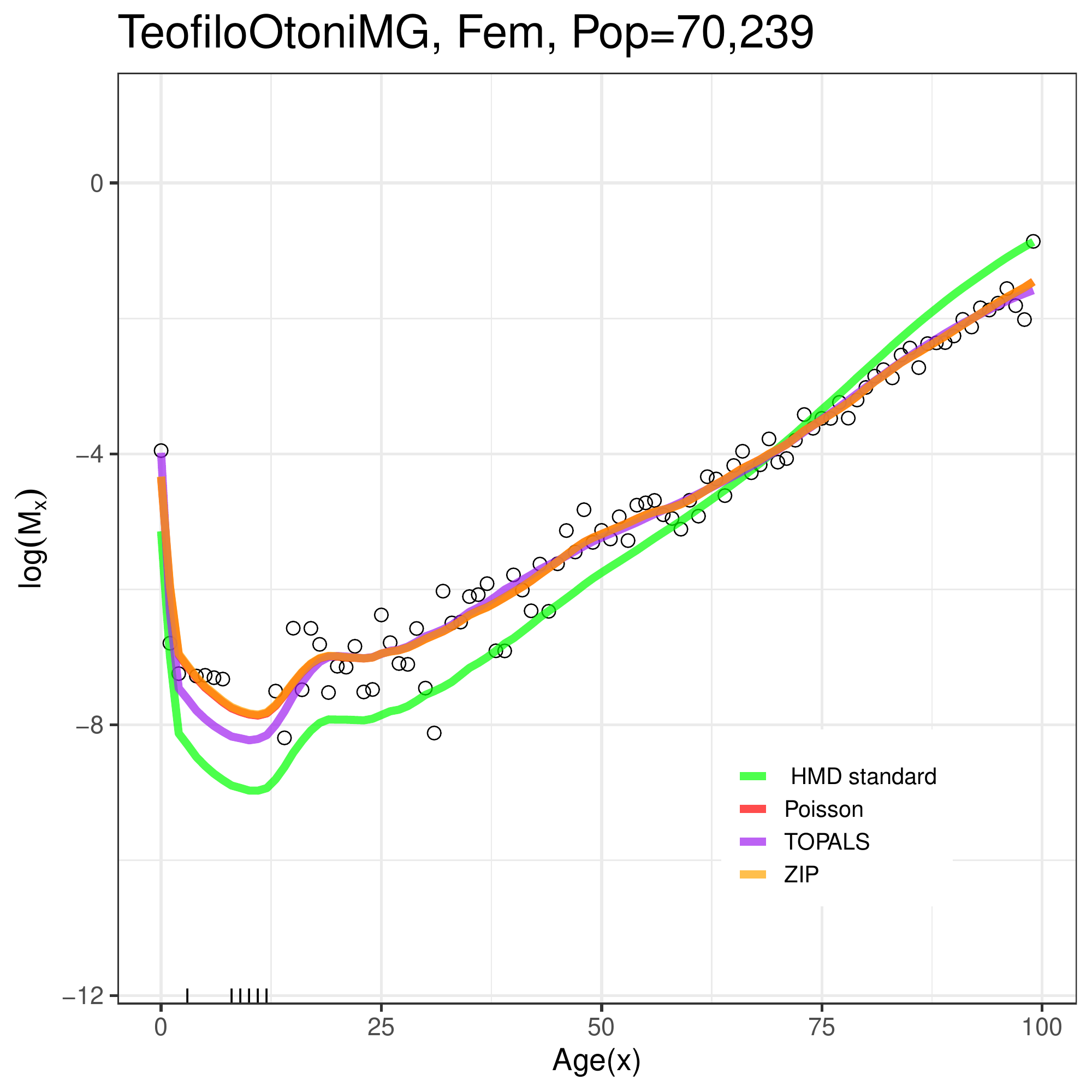}
	} \\	
	\subfigure
	{
		\includegraphics[scale=0.225]{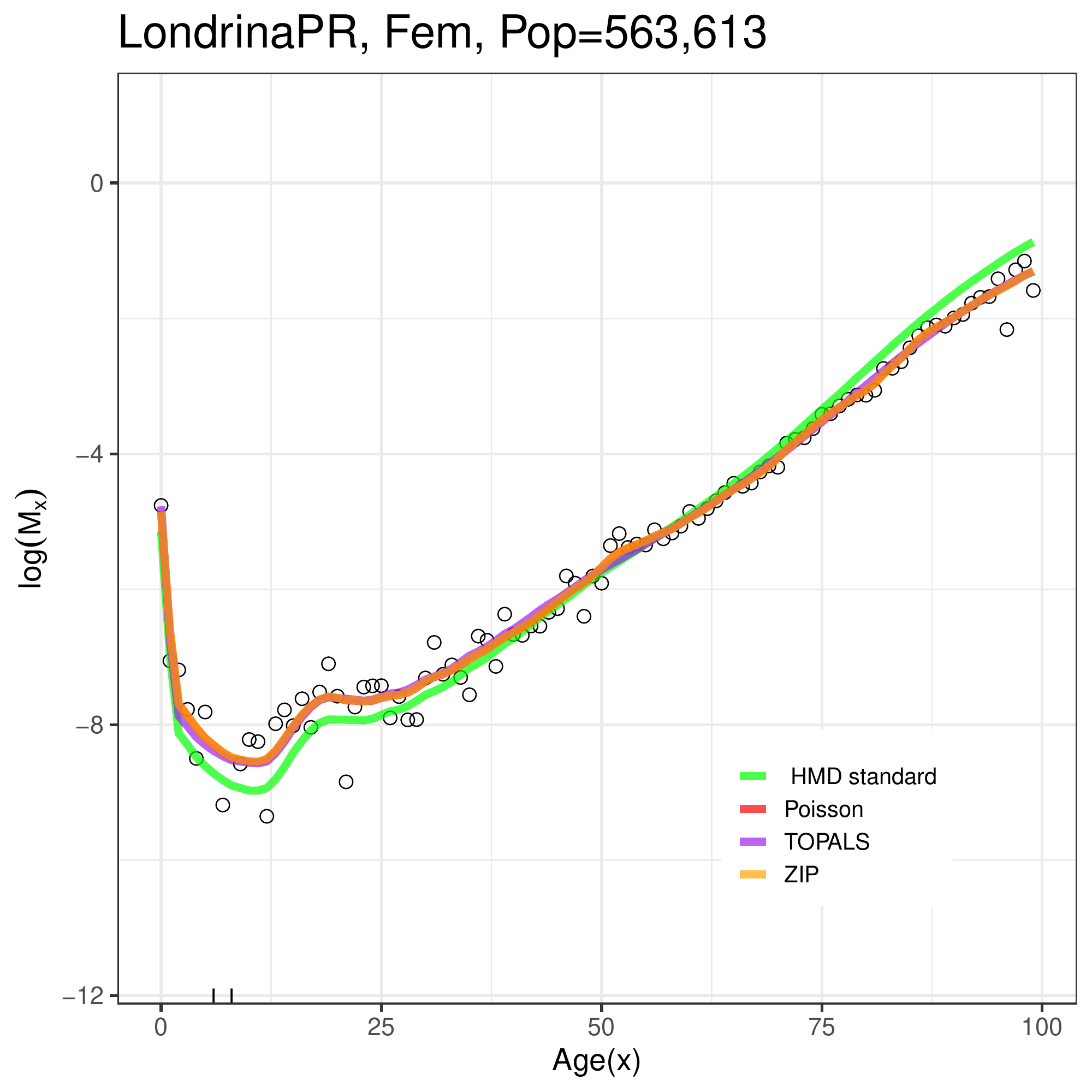}
	}
	\subfigure
	{
		\includegraphics[scale=0.225]{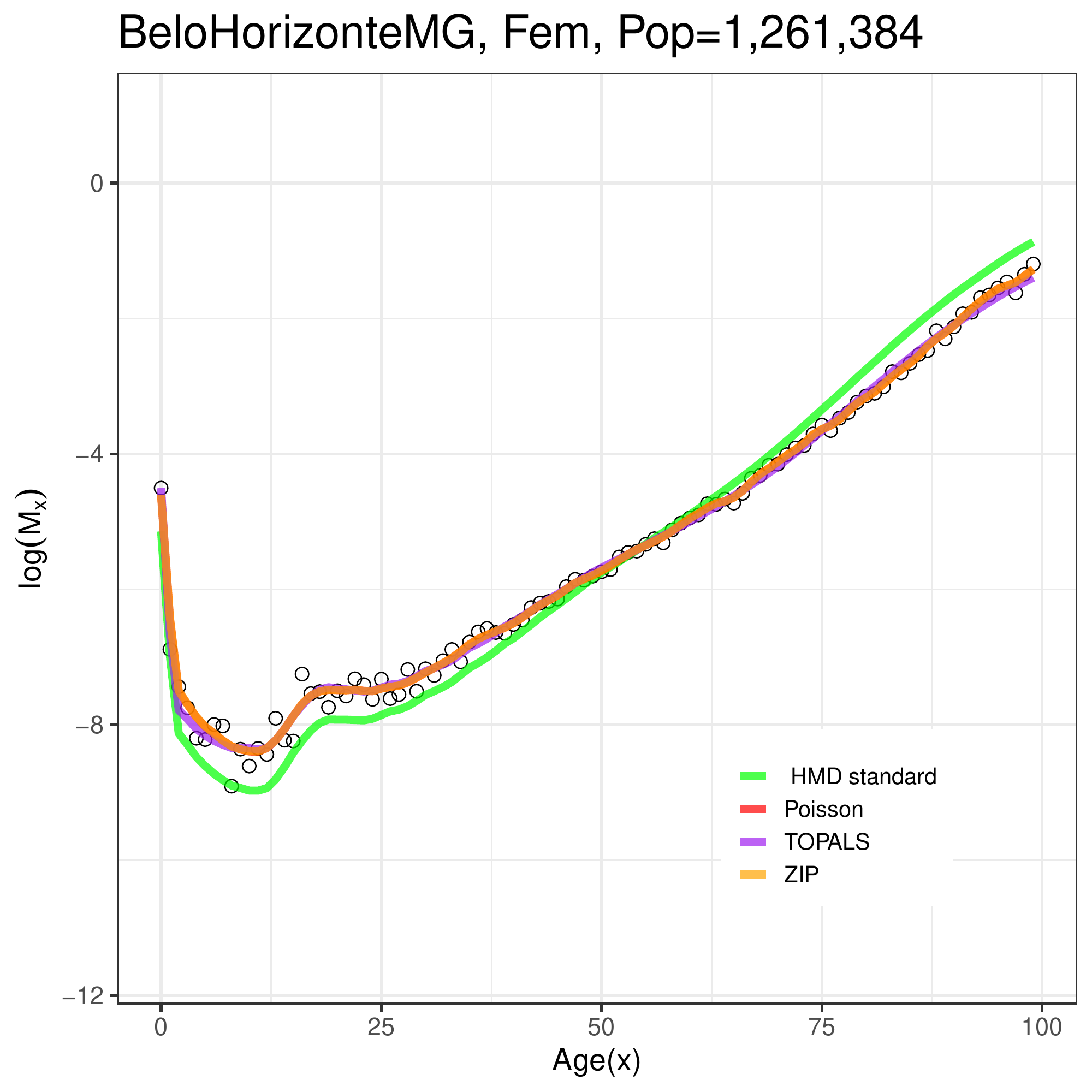}
	}\subfigure
	{
		\includegraphics[scale=0.225]{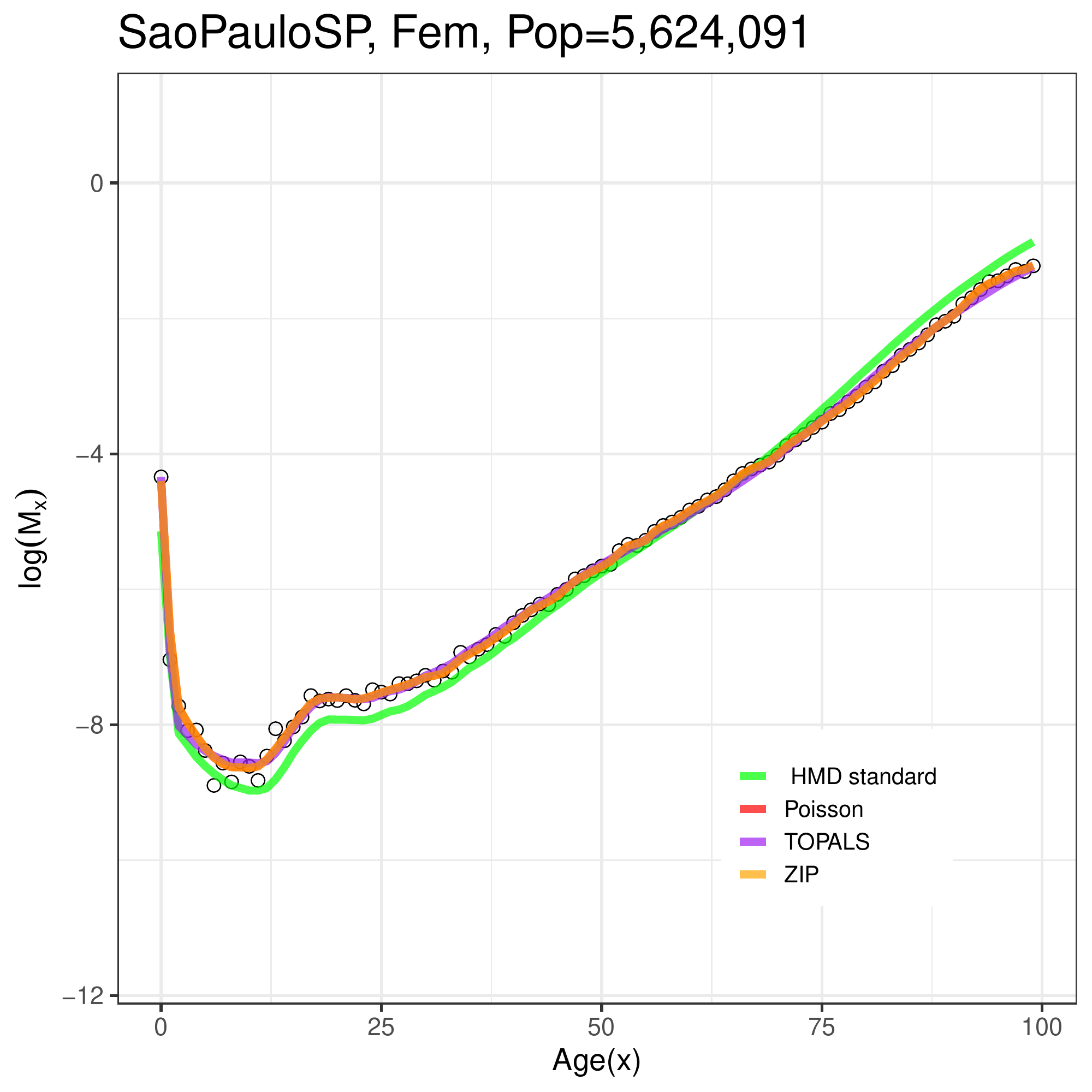}
	} \\
	\caption{Estimated mortality schedules (in the log scale) for selected Brazilian municipalities, females only. Open circles represent the observed log-mortality rate for each single-year of age. Tick marks on the horizontal axis represent ages with no observed deaths. The observed data were fitted under the dynamic Poisson model (red curve) and the TOPALS model (purple curve). Green curve represents the standard mortality schedule assumed in both models (HMD, 2015). Orange curve represents the fit of a zero-inflated Poisson model not discussed in Section 2 yet.}
	\label{fig:mortality_curves_application_female}
\end{figure}

\begin{figure}[htb!]
	\centering
	\subfigure
	{
		\includegraphics[scale=0.225]{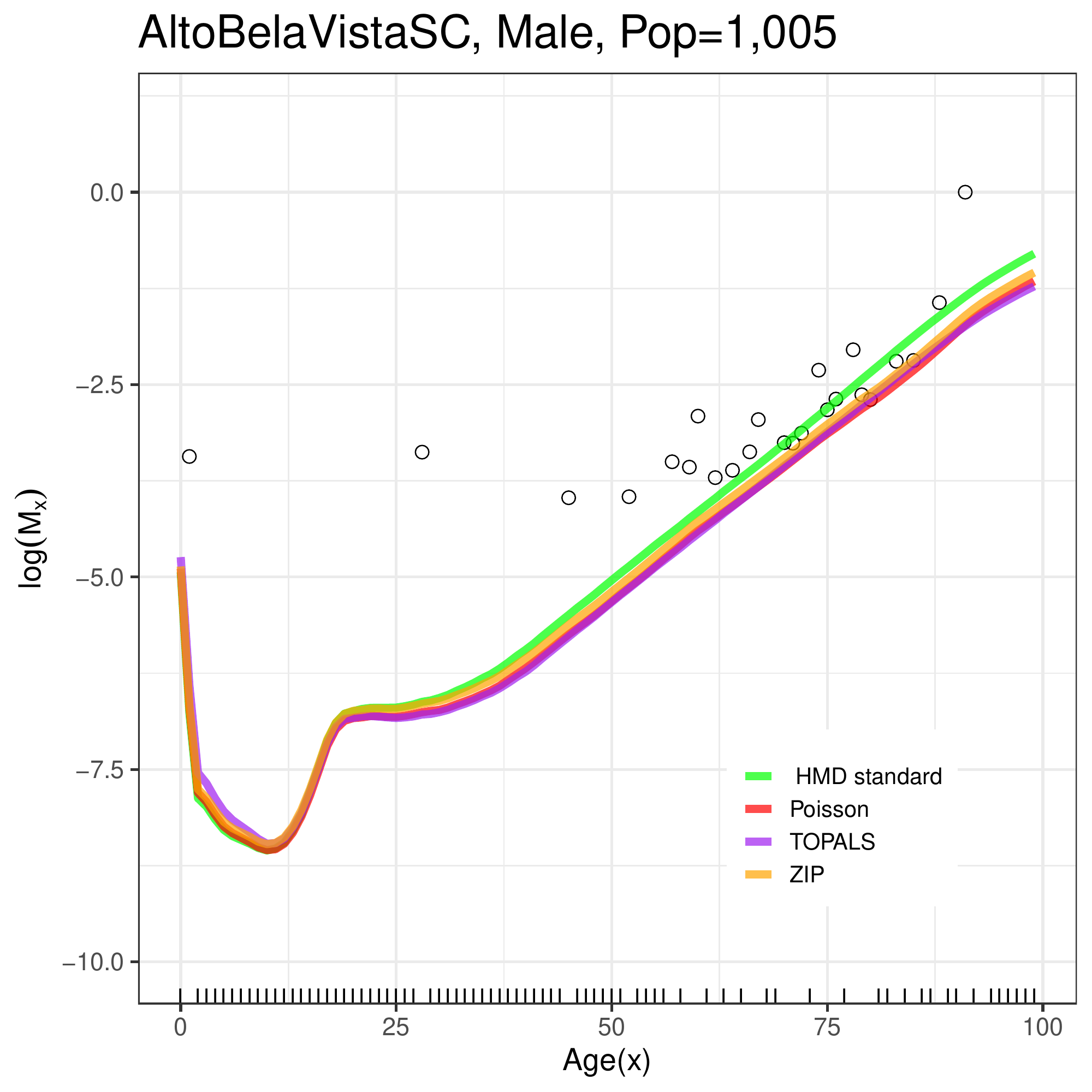}
	}
	\subfigure
	{
		\includegraphics[scale=0.225]{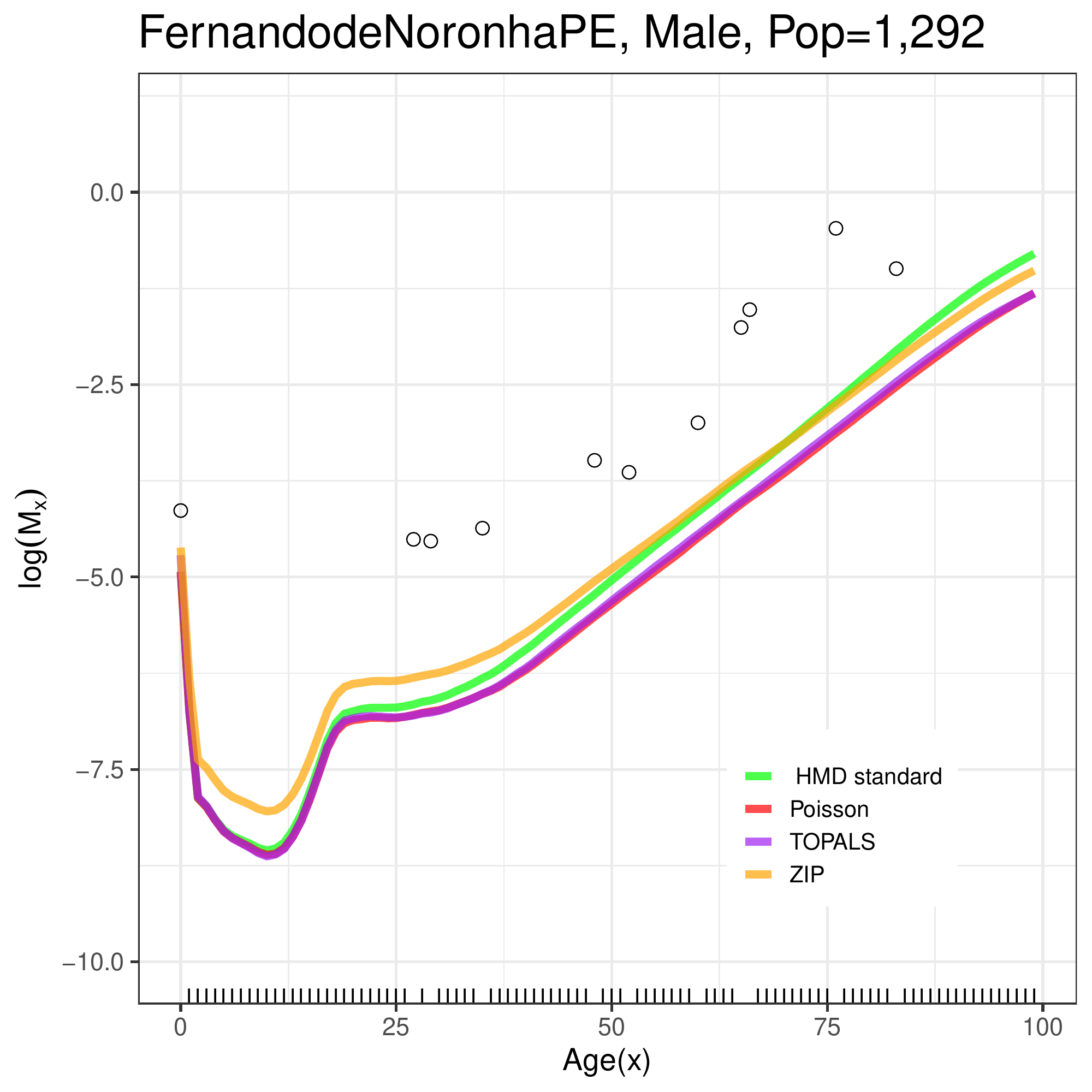}
	}\subfigure
	{
		\includegraphics[scale=0.225]{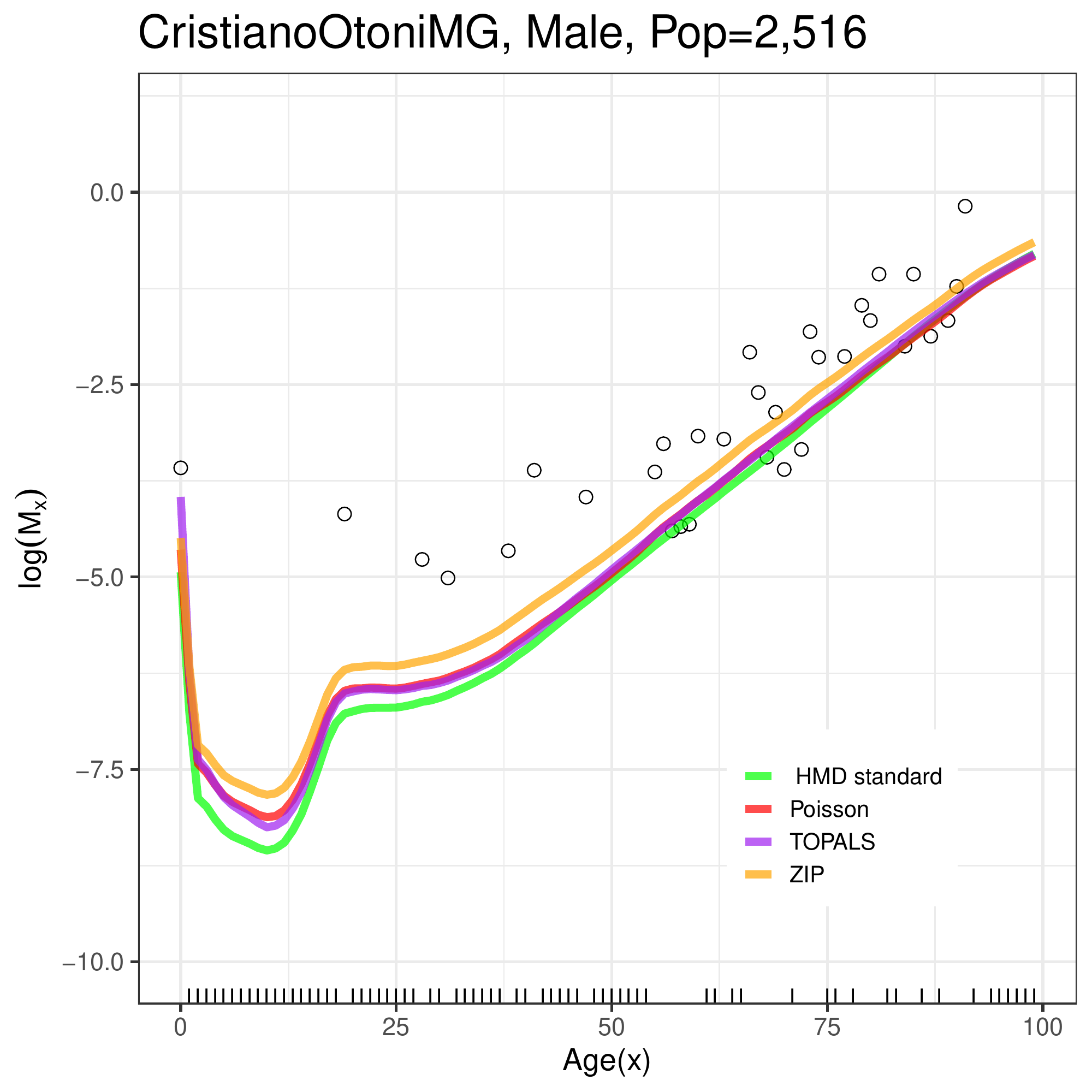}
	} \\
	\subfigure
	{
		\includegraphics[scale=0.225]{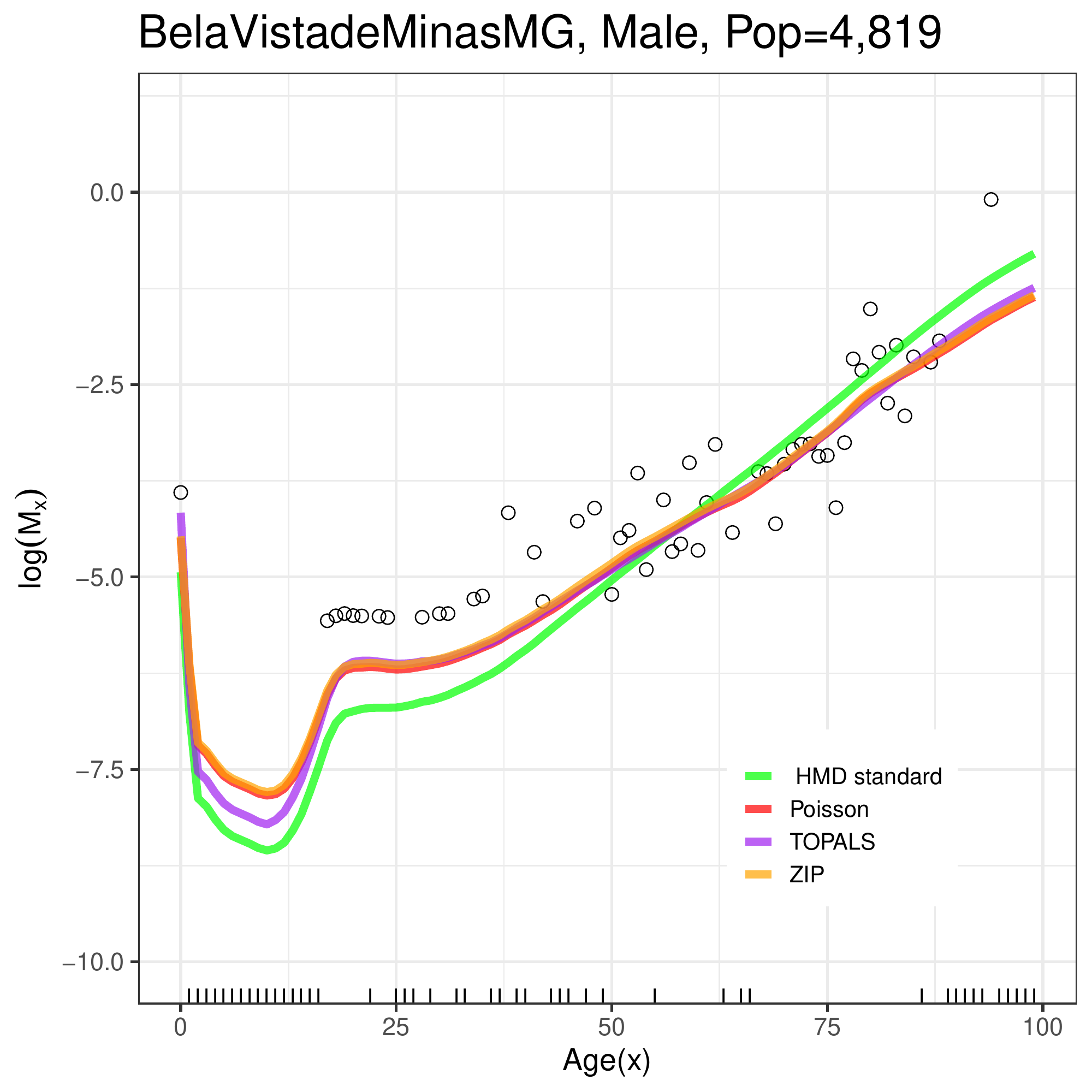}
	}
	\subfigure
	{
		\includegraphics[scale=0.225]{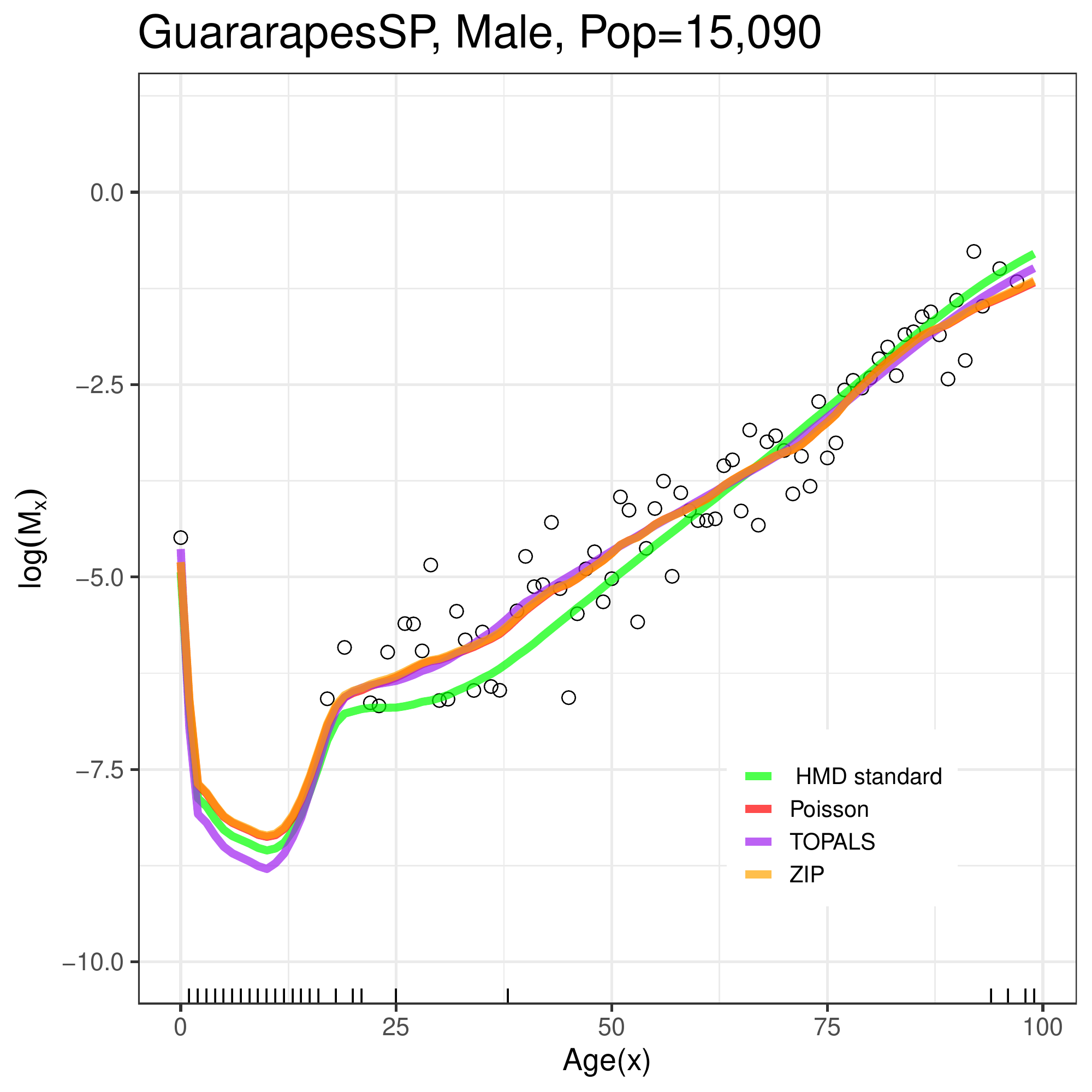}
	}\subfigure
	{
		\includegraphics[scale=0.225]{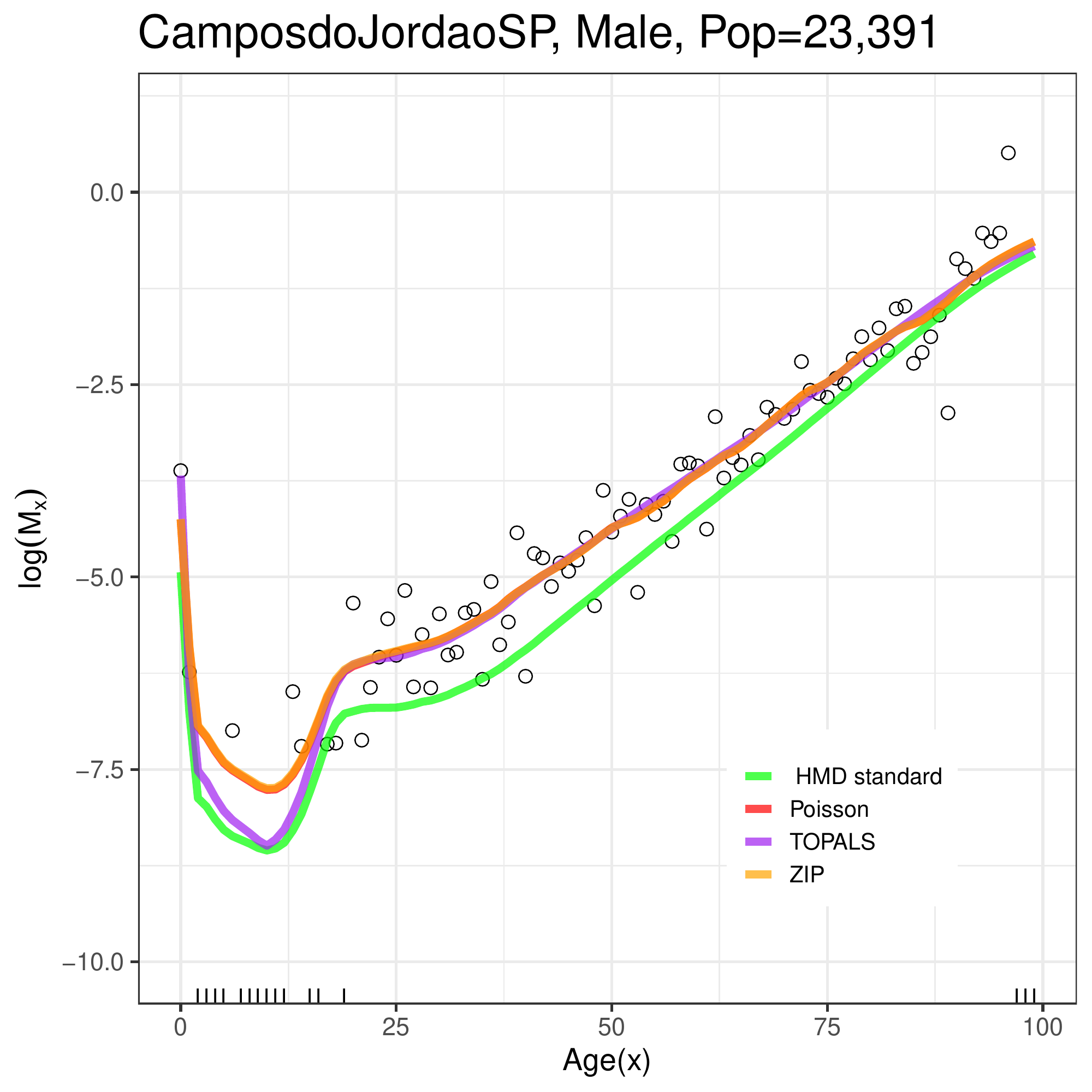}
	} \\
	\subfigure
	{
		\includegraphics[scale=0.225]{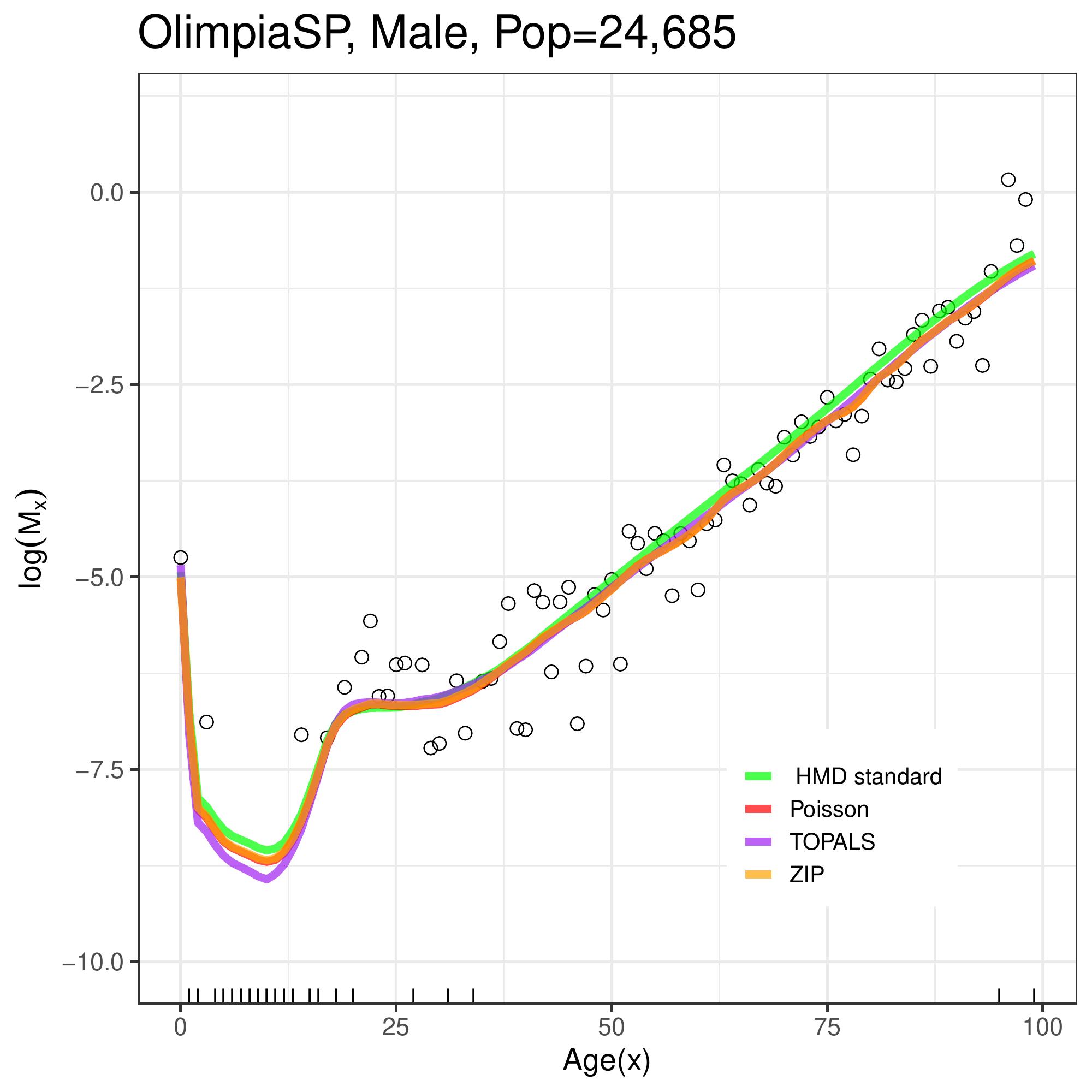}
	}
	\subfigure
	{
		\includegraphics[scale=0.225]{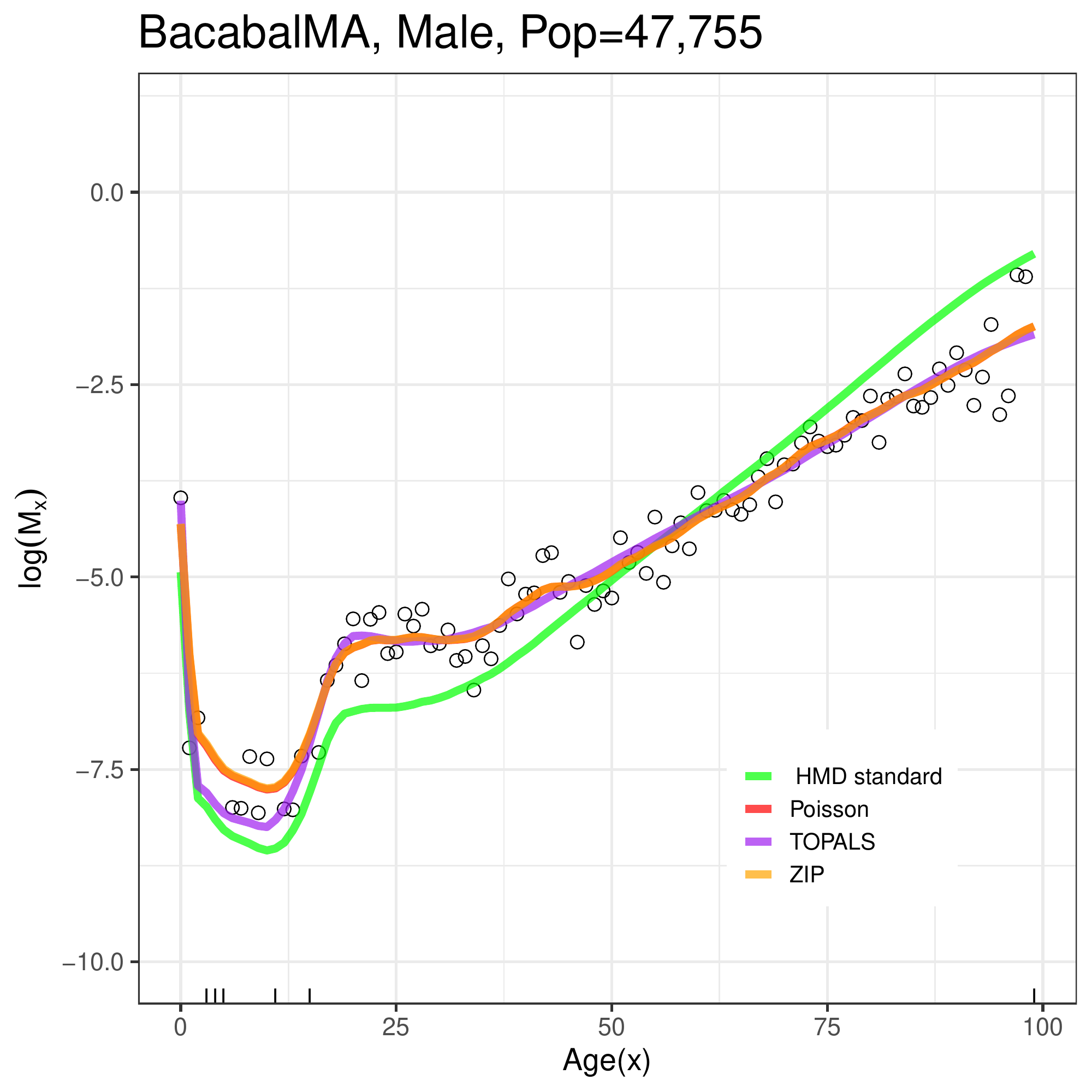}
	}\subfigure
	{
		\includegraphics[scale=0.225]{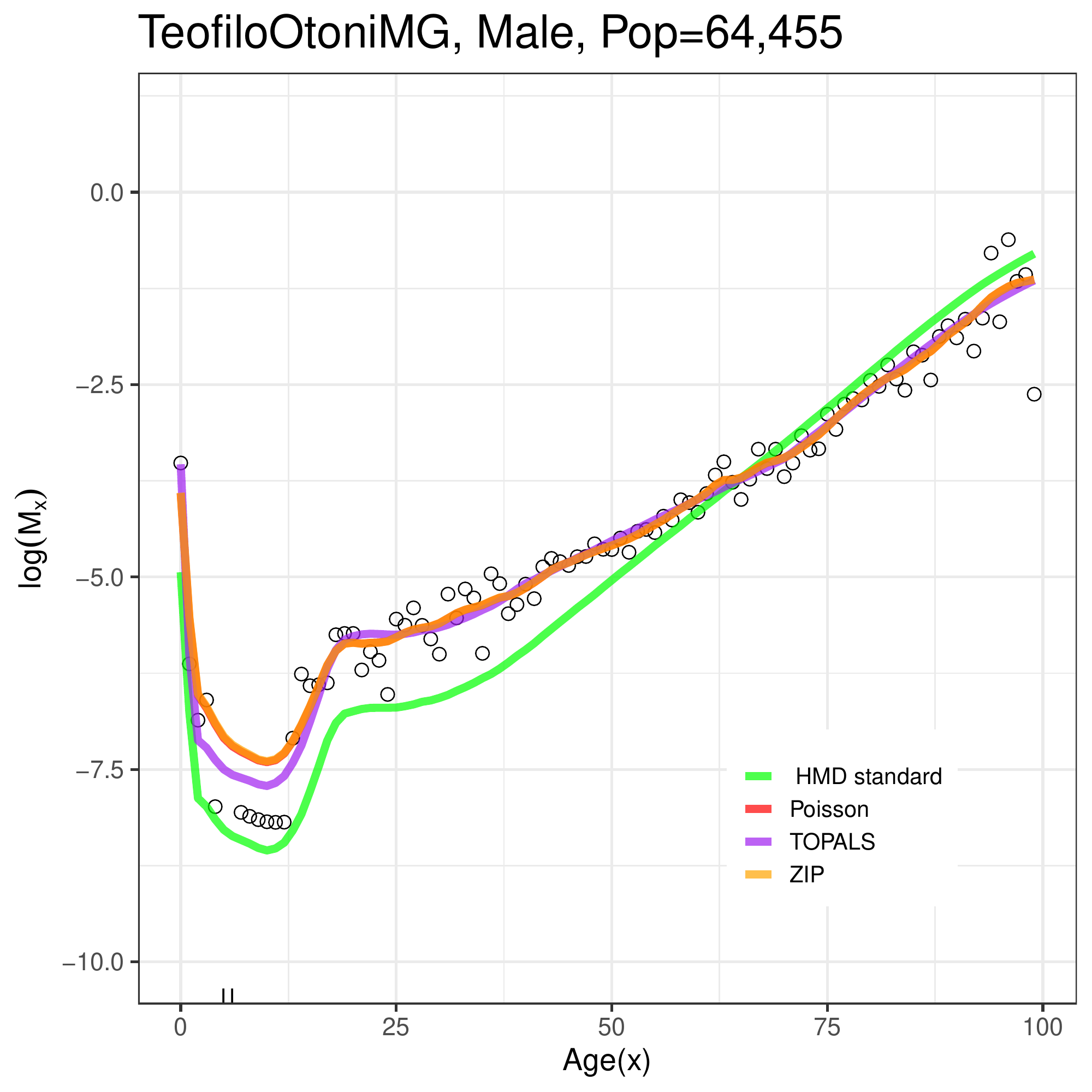}
	} \\	
	\subfigure
	{
		\includegraphics[scale=0.225]{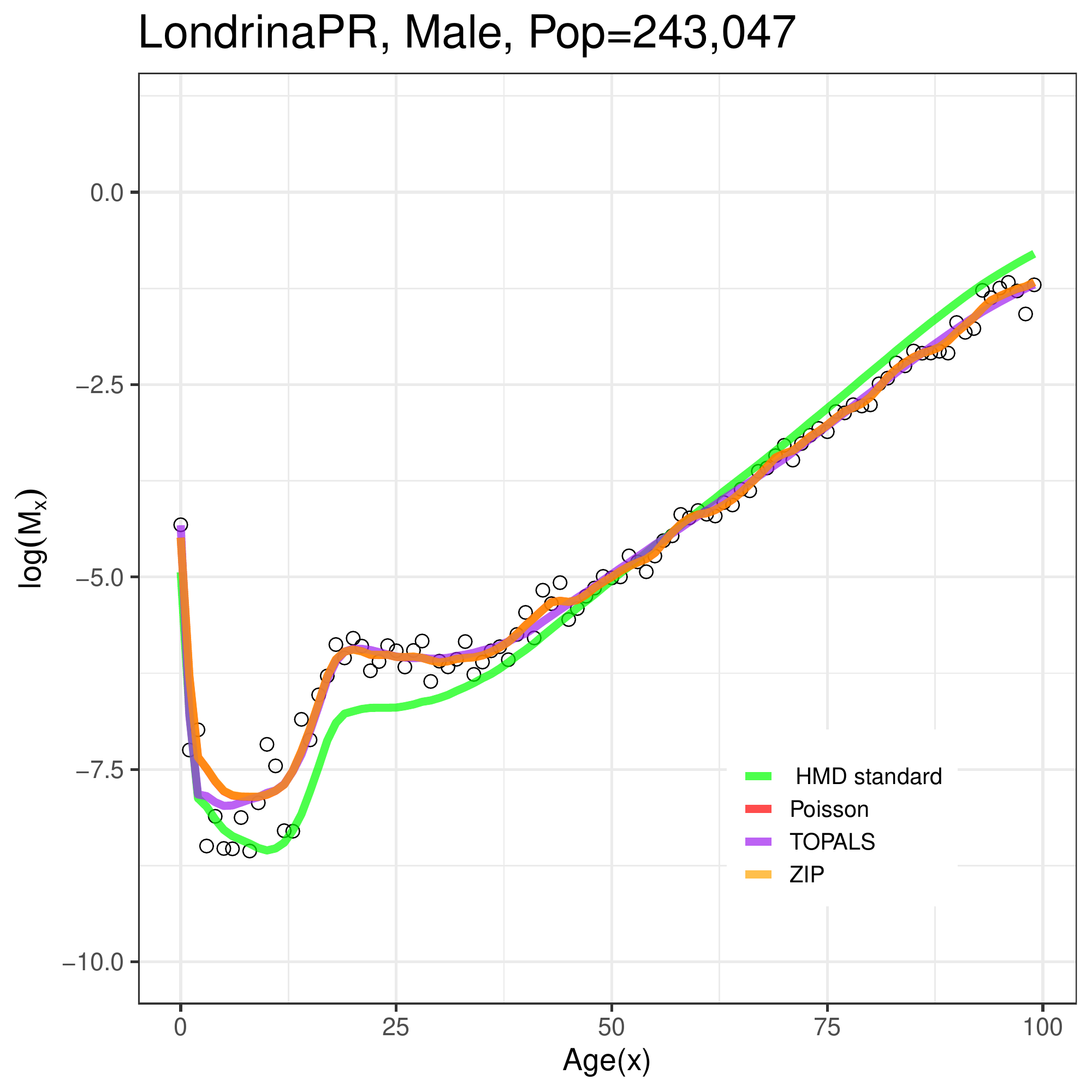}
	}
	\subfigure
	{
		\includegraphics[scale=0.225]{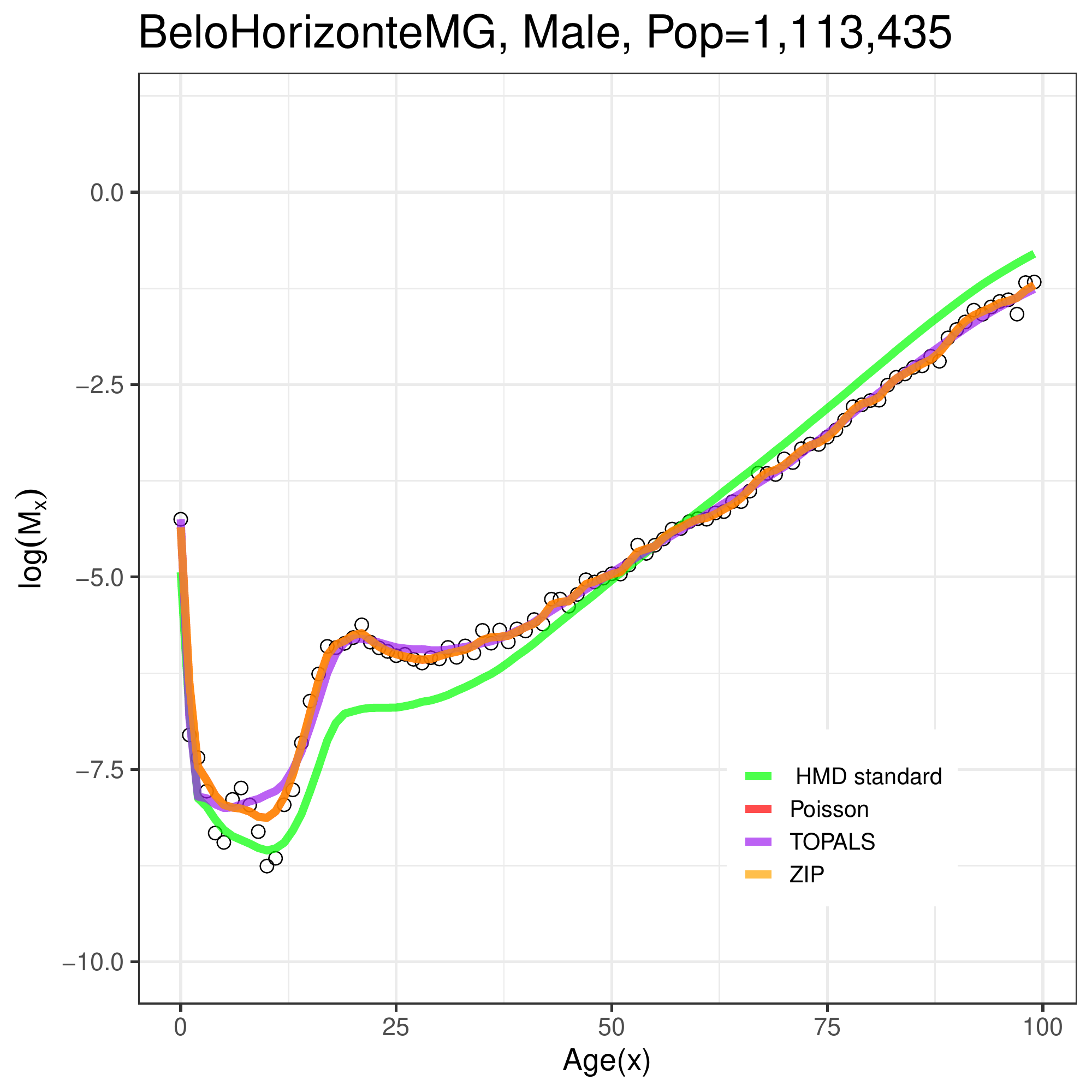}
	}\subfigure
	{
		\includegraphics[scale=0.225]{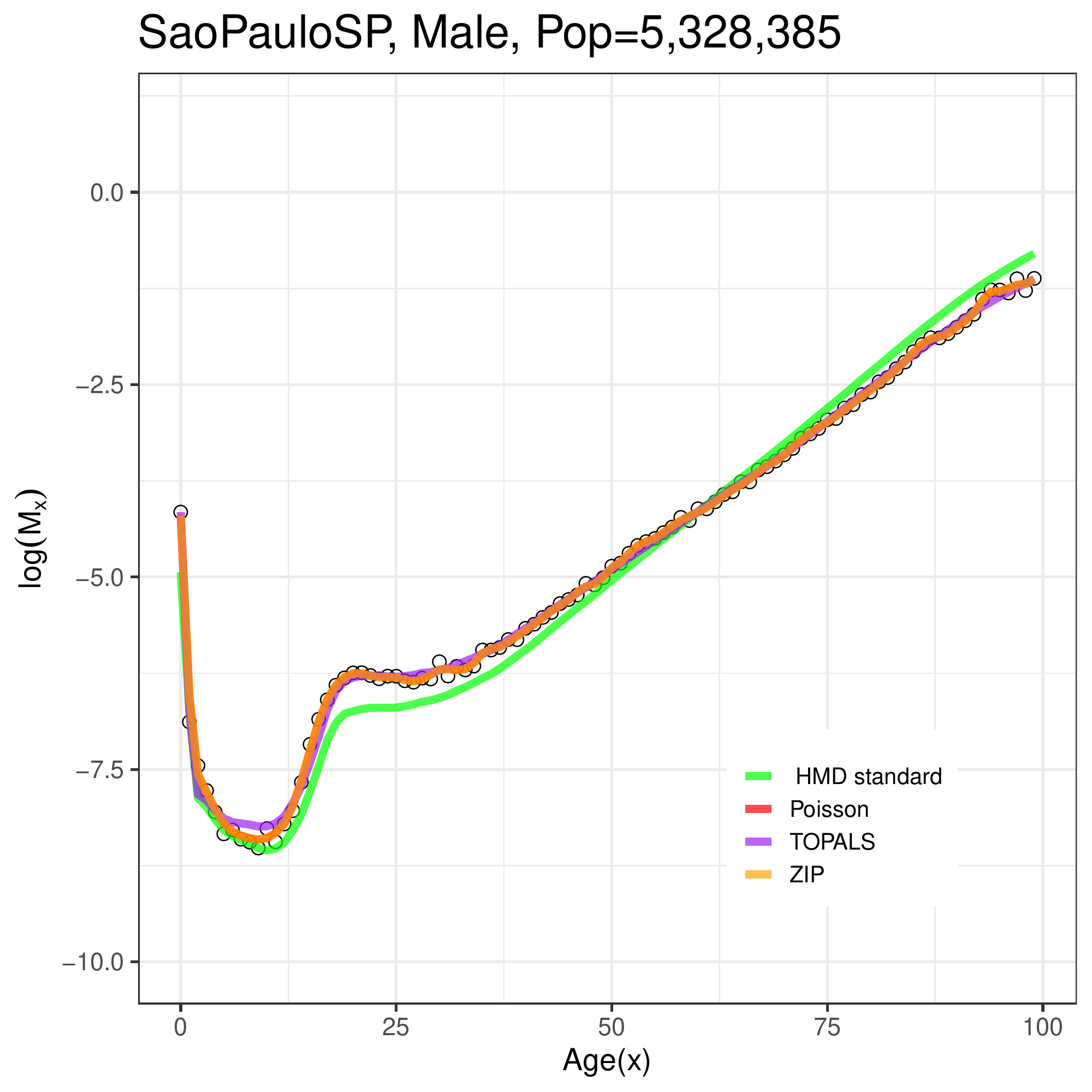}
	} \\
	\caption{Estimated mortality schedules (in the log scale) for selected Brazilian municipalities, males only. Open circles represent the observed log-mortality rate for each single-year of age. Tick marks on the horizontal axis represent ages with no observed deaths. The observed data were fitted under the dynamic Poisson model (red curve) and the TOPALS model (purple curve). Green curve represents the standard mortality schedule assumed in both models (HMD, 2015). Orange curve represents the fit of a zero-inflated Poisson model not discussed in Section 2 yet.}
	\label{fig:mortality_curves_application_male}
\end{figure}

We present the results in each selected municipality for both sexes together (Figure \ref{fig:mortality_curves_application_both}) and separately for females (Figure \ref{fig:mortality_curves_application_female}) and  males (Figure \ref{fig:mortality_curves_application_male}).
By comparing Figures \ref{fig:mortality_curves_application_female} and \ref{fig:mortality_curves_application_male} it can be noticed the usual differences between mortality patterns for females and males. 
As noted in the simulation study, Poisson and TOPALS models tends to present similar shapes for mortality schedules. 
In all cases, both models provided smooth mortality curves even  for population with highly erratic data.
For the three smallest selected populations, both models provided almost the same fit, except for females in municipalities Alto Bela Vista SC (population=1,000) and Fernando de Noronha PE (population=1,338). 
In the most cases, the biggest discrepancies between the two adjustments, when they can be visually noted, occur at latest ages or at ages around the "bathtub pattern'' of the mortality schedules.
In agreement with the findings in the simulation study, the Poisson model seems to be more influenced by outliers.
This can be specially noted for high observed values for the mortality rates at the end of the ages' scale, which apparently pulls the estimates upwards (see graphs for municipality Campos do Jord\~ ao SP, both sexes).
A similar effect is noted when occurs a high frequency of sequential null counts in the ages' range, which tends to pull the curve downwards (see graphs for females in municipalities Alto Bela Vista SC and Fernando de Noronha PE).

\clearpage
\section{Concluding Remarks}
\label{conclution:mortality_curves}


The reliable measurement and comparative analysis of mortality schedules for different populations helps to  highlight differences among groups of people and guide analysts to understand what drives health disparities.
For developing countries, specially in subnational geographic populations that do not have the resources to establish reliable death registration, this goal could be approached through the incorporation of empirical information to improve the estimates provided by usual parametric models.
That is a common approach in the called relational demographic models.
The problem with data coming from small and underdeveloped populations is the high occurrence  of low or null counts, which impairs the estimation of the true underlying mortality rates by usual methods.

In this work we proposed two regression models with dynamic parameters to estimate the complete mortality schedules per 1-year age intervals. Inference is made under the Bayesian paradigm. Since mortality curves generally have a specific pattern, to prevent unrealistic estimates, we use a standard schedule as a covariate in the regression model, similar to the idea of relational models which are common in demography. Dynamic evolution across ages is included in order to provide smoothed estimates for the mortality schedules. 

We consider a dynamic Poisson model to directly fit the observed mortality counts as well as the Gaussian dynamic linear model  to model the observed log-mortality rates.
The TOPALS model proposed by \cite{Gonzaga2016} is also fitted for comparison purpose. TOPALS  allows the derivation of complete schedules of age mortality rates via mathematical adjustments to a specified standard schedule through penalized splines function. 
A simulation study is performed leading to interesting initial visualizations of the models performance. 

In general, the Poisson and TOPALS models demonstrated to be more efficient as some of the analyzed data are sparse. 
Despite the Gaussian model showed to be a competitive model in large populations,  it is not appropriated for small areas. 
Poisson and TOPALS models  were applied to fit data from Brazilian municipalities.  
The Poisson model showed to be promising to estimate smoothed mortality schedules influenced to discrepant values in the observed mortality rates.  

We believe that an implementation of the model as a generalized dynamic linear model \citep{West1985} may improve estimates, for instance, with the inclusion of discount factors in the covariance matrix of the dynamic parameters throughout the well-known Kalman filtering and Kalman smoothing estimation algorithms \citep{Campagnoli2009} or appropriate Markov chain Monte Carlo techniques \citep{Gamerman1998,Schmidt2011}.  
In this context, the use of the generalized dynamic Poisson model proposed by \cite{Schmidt2011} to fit time series of count data in epidemiological studies can be investigated. Their model have a time-dependent parameter that captures possible extra-variation present in the data and also zero-inflated versions are proposed. As such, their approach may provide interesting results in the context of mortality schedules estimation.

The simulation study presented in Section \ref{sec:simulations_cap3} is no longer exhaustive to support the preference for one specific model in a general case. 
Thus, the limited results must not be generalized. 
A broader simulation study must be performed in order to provide more confident evidence about the models performances in different scenarios and to support the previous findings discussed in this work. It includes a Monte Carlo study with more populations sizes and particular characteristics. Also, the application to a greater range of municipalities and other geographic aggregation levels can be included. 
A sensitivity study on the standard schedule choice must be performed  as well. 
In this context, as addressed by \cite{Alexander2017}, we aim to investigate the use of multiple mortality patterns, in particular the consideration of principal components analysis with basis on a large set of trustful mortality curves such as those available in the HMD.
The definition of a multiple regression model based on the main principal components may increases flexibility in the estimates, potentially reducing the effect of outliers in the Poisson model evidenced by the studies developed so far.

Finally, we note that the excess of null death counts in some localities can indicate that the events is really rare or that there is a high level of underreporting of death counts.
Therefore, the appropriate use of a zero-inflated Poisson model (\cite{Lambert1992,Piancastelli2019,Jussiane2020}), or other models which account for overdispersion, can be investigated as in \cite{Lima2016} besides the consideration of underreporting bias correction. 
We intend to formulate a zero inflated Poisson model which, in addition to taking into account a standard mortality schedule, is capable of accounting for the occurrence of under-registration. 
In particular, we aim to explore the clustering model present in Chapter \ref{cap:paperA} \citep{Oliveira2020} within the problem of mortality schedule estimation. In the demography literature, it is known that the level of underreporting varies between ages intervals. Particularly, specialists argue that such a problem is worse in infant ages than in old ages which, in turn, tends to present a higher level of under-registration than young and adult ages (see e.g. \cite{Schmertmann2018} and references there in).
A clustering structure between subsequent ages intervals may be determined in order to apply the clustering model jointly with an adequate usage of a standard mortality schedule to ensure usual mortalities patterns.


\bibliographystyle{plainnat}

\end{document}